\documentclass[aps,prx,onecolumn,amsmath,nofootinbib,amssymb,superscriptaddress,11pt]{revtex4-1}
\pdfoutput=1
\usepackage{comment}
\usepackage{graphicx}
\usepackage{ulem}
\usepackage{soul}
\usepackage[colorlinks, linkcolor=blue]{hyperref}
\usepackage{verbatim}
\usepackage{esint}
\usepackage{amssymb}
\usepackage{amsmath}
\usepackage[compatibility=false,font=small,labelfont=bf, justification=justified,format=plain]{caption}
\usepackage{subcaption}
\DeclareMathOperator{\sign}{sign}
\renewcommand{\vec}[1]{\boldsymbol{#1}}
\def \k {{\vec k}}
\def \p {{\vec p}}

\def \ve {\varepsilon}
\def \r {{\vec r}}
\def \v {{\vec v}}
\def \el {\ell}

\def \q {{\vec q}}

\def \l {{\vec l}}
\def \ve {\varepsilon}

\def \C {{\cal{C}}}

\def \x {{\vec x}}

\def \J {{\vec J}}

\def \beq {\begin{eqnarray}}
\def \eeq {\end{eqnarray}}
\def \tn {\textnormal}

\def \la{\langle}
\def \ra{\rangle}

\def \tv{\tilde{v}_F}

\begin{document}
\title{Translationally invariant non-Fermi liquid metals with critical Fermi surfaces:\\ Solvable models} 
\author{Debanjan Chowdhury} \email{debch@mit.edu}
\affiliation{Department of Physics, Massachusetts Institute of Technology, Cambridge MA 02139, USA.}
\author{Yochai Werman} 
\affiliation{Department of Condensed Matter Physics, Weizmann Institute of Science, Rehovot-76100, Israel.}
\author{Erez Berg}
\affiliation{Department of Physics, University of Chicago, Chicago IL 60637, USA.}
\author{T. Senthil}
\affiliation{Department of Physics, Massachusetts Institute of Technology, Cambridge MA 02139, USA.}

%\vspace{1.6in}}
\begin{abstract}
We construct examples of translationally invariant solvable models of strongly-correlated metals, composed of lattices of Sachdev-Ye-Kitaev dots with identical local interactions. These models display crossovers as a function of temperature into regimes with local quantum criticality and marginal-Fermi liquid behavior. In the marginal Fermi liquid regime, the dc resistivity increases linearly with temperature over a broad range of temperatures.  
By generalizing the form of interactions, we also construct examples of non-Fermi liquids with critical Fermi-surfaces. The self-energy has a singular frequency dependence, but lacks momentum dependence, reminiscent of a dynamical mean field theory-like behavior but in dimensions $d<\infty$. In the low temperature and strong-coupling limit, a heavy Fermi liquid is formed. The critical Fermi-surface in the non-Fermi liquid regime gives rise to quantum oscillations in the magnetization as a function of an external magnetic field in the absence of quasiparticle excitations. We discuss the implications of these results for local quantum criticality and for  fundamental bounds on relaxation rates.  Drawing on the lessons from these models, we formulate conjectures on  coarse grained descriptions of a class of intermediate scale non-fermi liquid behavior in generic correlated  metals.
\end{abstract}

\maketitle

\setcounter{tocdepth}{2}
\tableofcontents
\section{Introduction}
A number of strongly correlated materials with a metallic parent state exhibit a variety of non-Fermi liquid (NFL) properties.  Some of the best known examples of such behavior occur in the ruthenates \cite{hussey,gegenwart,kapitulnik,allen}, cobaltates~\cite{Taillefer1,Ong1}, iron-based superconductors~\cite{Matsuda14} and heavy-fermion materials~\cite{Stewart}, amongst others. Some of these materials display striking non-Fermi liquid behavior over a broad range of temperatures above an emergent low energy scale but develop Fermi liquid-like properties and well defined Landau quasiparticles below this scale, while others remain non-Fermi liquid-like down to the lowest temperatures. Perhaps the most striking example of the latter behavior occurs in the ``strange-metal" regime~\cite{Takagi,Keimer15} of the cuprate superconductors and some quantum critical heavy-Fermion systems~\cite{rmpqcp,Stewart}. 
 
One of the most dramatic properties associated with many of these materials is a linear dependence of the dc resistivity on temperatures without any sign of saturation. In the cuprates, much of the phenomenology of the normal state   is apparently well described by the ``marginal Fermi liquid'' (MFL) model~\cite{Varma}, which postulates the existence of marginally defined quasiparticles, whose scattering rate is comparable to their energy. 

Broadly speaking, a few theoretical frameworks have been proposed to explain the phenomenology of strange metals: (i) Quantum critical fluctuations of a bosonic degree of freedom coupled to a Fermi-surface leading to a non-Fermi liquid ground state, which dominates the properties of the system in a range of temperatures above the critical point. Concrete examples of such theories involve the situation where an order-parameter field (such as a nematic or antiferromagnetic order-parameter) at its critical point couples to an electronic Fermi-surface \cite{rmpqcp}. Much progress has been made in understanding the properties of this class of  metallic quantum critical points in recent years~\cite{lee_review}. (ii) A distinct class of non-Fermi liquids arise at a critical point driven by electronic fluctuations associated with the destruction of the Fermi surface. Examples include a Kondo breakdown transition\footnote{The onset of antiferromagnetism (as a function of some tuning parameter) in several heavy Fermi liquids is known to have a striking non Hertz-Millis character and is accompanied by a dramatic change in the Fermi-surface volume.} in a heavy Fermi liquid~\cite{Coleman2000,Colemanetal,Si1,Si2,TSFL} and a Mott transition between a metal and a quantum disordered insulator~\cite{TS08,TSFL}. Such non-Fermi liquid quantum critical points have been argued~\cite{TS08,TSmott} to possess a critical Fermi surface - {\it i.e.}, the electronic excitations at the critical point are characterized by the presence of a sharply defined Fermi surface but with no sharp Landau quasiparticles.\footnote{Critical Fermi surfaces are also expected to occur at some quantum critical points driven by fluctuations of a Landau order parameter associated with ordering at zero momentum.} Currently known concrete low-energy theories for such quantum critical points involve fractionalized degrees of freedom and associated dynamical gauge fields. Theoretical progress has been possible on a few examples of such theories \cite{TSFL,TS08,TSmott,CFL_FL,OM,DChiggs}. 
While these concretely tractable examples are extremely useful, much more remains mysterious about the general theory of quantum critical points associated with the `death' of a Fermi surface.\footnote{In particular, in all the examples so far, there is a remnant `ghost' Fermi surface of fractionalized degrees of freedom once the electronic Fermi surface dies. It is not known if continuous quantum phase transitions can occur to phases where there is no such ghost.} (iii) Instead of appearing just at a critical point, a non-Fermi liquid can arise as a stable zero temperature phase, as has been observed for instance in numerical studies of lattice models \cite{mishmash}. A classic example of such non-Fermi liquid behavior occurs in a two-dimensional electron gas under high magnetic fields, when a compressible metallic phase is realized at a filling of $\nu=1/2$ \cite{HLR}. Indication of such non-Fermi liquid quantum phases have also been reported in correlated mixed-valence materials \cite{ybalb1,ybalb2}. (iv) Finally, in the limit of sufficiently strong interactions and at intermediate temperatures, it is possible that strange metal behavior arises generically without tuning to the vicinity of a quantum critical point.  However, the ground state is a Landau Fermi liquid or some other conventional state (e.g. a superconductor) and the strange metal regime appears only as a crossover at higher temperatures.

Despite all this progress in the theory of non-Fermi liquids, there is no clear mechanism that produces a linear in $T$ resistivity over a broad range of temperature in quantum critical or other non-Fermi liquids in translationally invariant models as a result of strong local electronic interactions. The phenomenological ``marginal Fermi liquid'' theory assumes coupling to a bosonic fluctuating mode that gives linear resistivity~\cite{Varma}; however, it is not clear how to derive such a bosonic spectrum from a microscopic model. The results of recent quantum Monte Carlo (QMC) simulations of an Ising nematic transition~\cite{Lederer2017} are consistent with a linear behavior of the resistivity at the quantum critical point.\footnote{These results are subject to uncertainties associated with analytical continuation from imaginary to real time. From the imaginary time data, one can extract ``resistivity proxies'' that coincide with the dc resistivity under certain assumptions, such as the absence of sharp features in the frequency-dependent conductivity over a scale $\omega \lesssim T$. The validity of these assumptions is hard to assess from imaginary-time data, and has to be checked independently.} There is currently no theoretical understanding of these results.

Empirically, it is likely that these different routes to non-Fermi liquid physics are realized in different materials. Our focus in this paper is on  route (iv) above. In a number of different systems (for example, in some cobaltates~\cite{Taillefer1,Ong1} and  ruthenates~\cite{Tyler1998,Bruin13}) it is indeed seen that there is a wide intermediate temperature $T_{UV} \gg T \gg T_{\tn{coh}}$ where strange metallic transport is observed, including 
non-Fermi liquid temperature dependent resistivity with values exceeding the Mott-Ioffe-Regel limit. As the temperature drops below a low `coherence scale' $T_{\tn{coh}}$ there is a crossover to more conventional behavior. Importantly, it does not appear that $T_{\tn{coh}}$ can be pushed close to zero by tuning some parameter,\footnote{It is worth pointing out that this is likely {\it not} the situation  for the cuprate strange metal and in some heavy electron materials like YbRh$_2$Si$_2$ \cite{rmpqcp}. In both these systems by tuning one parameter it has been possible to stabilize the NFL physics to  ultra-low $T$  suggesting that $T_{\tn{coh}}$ can, in principle, be tuned to zero.} suggesting that it may be fundamentally impossible to stabilize such NFL states at zero temperature. In other words, the intermediate-$T$ NFL physics of these systems may not in principle be controlled by $T = 0$ Infra-Red (IR) fixed points with a finite number of relevant perturbations.  
We call such intermediate-$T$ non-Fermi liquid states as examples of ``IR-incomplete" states of matter (see  Ref. \cite{tskitp2011} for a possibly useful exposition). By themselves, they cannot be the deep IR theory of any state of matter and hence require IR-completion.   
\newpage
Examples include electron-phonon systems above their Debye temperature~\cite{ziman}, lattice models with bounded kinetic energy at high $T$~\cite{lindner,Oganesyan}, spin-incoherent Luttinger liquids~\cite{sill}, electrons coupled to a lattice of bound-states \cite{PC98}, holographic non-Fermi liquids~\cite{Liu1,Liu2}, and some states found in DMFT calculations at finite temperature~\cite{DMFT,kotliar}.  Common to many of these examples of IR-incomplete theories is that they have extensive residual low-$T$ entropy ({\it i.e.} the entropy extrapolated to $T = 0$ from the regime in which the theory applies  is non-zero) which is then relieved below $T_{\tn{coh}}$ leading to a crossover to a conventional state.

Progress in understanding strongly interacting  IR-incomplete non-Fermi liquids has been hindered by the lack of suitable controlled theoretical techniques. The Sachdev-Ye-Kitaev (SYK) model~\cite{SY,kitaev_talk,Parcollet1,Parcollet2,FuSS,SS15,Maldacena_syk,kitaevsuh,Altman17}, consisting of a large number of degrees of freedom coupled via a random all-to-all interaction, provides a window into the behavior of strongly coupled systems with no quasiparticles. The model is $(0+1)-$dimensional, and thus it does not contain information about transport. Higher dimensional generalizations of the model
have been considered~\cite{Parcollet1,Burdin2002,Gu17,SS17,DVK17,Balents,hongyao}. Refs.~\cite{Parcollet1,Burdin2002} studied lattice models of itinerant fermions coupled to spins with a long-ranged all-to-all interactions. Refs.~\cite{Gu17,Balents,SS17} considered lattice models with an SYK dot placed in every site, with a random short ranged inter-site coupling. The charge and thermal transport properties have been computed.
The solution of these models have many appealing characteristics, such as a locally quantum critical, non-Fermi liquid crossover regime where the resistivity is linear in temperature and quasi-particles are destroyed.

In all of the above models, translational symmetry is strongly broken, raising a number of questions: (i) Does quenched disorder play an essential role in the behavior of strange metals as suggested in Ref. \cite{subir_kitptalk}, or could it be realized even in a perfectly crystalline system?  (ii) Can a non (or marginal-)Fermi liquid with a critical Fermi-surface (to be defined below) appear within this class of models, and what are its transport and other related properties? (iii) Does  a non-Fermi liquid with a critical Fermi surface show quantum oscillations in an external applied magnetic field?

In order to address these questions, in this work we construct a set of {\it translationally invariant} models that can be solved exactly in the large $N$ limit, where $N$ is the number of fermion flavors (or ``orbitals'') per site, coupled by a frustrated on-site interaction. Our construction is therefore different from other constructions of higher-dimensional generalizations of SYK-type models at a fundamental level. The crucial new ingredient, namely the exact translation symmetry (instead of a statistical symmetry) at the level of each realization will allow us to address many interesting questions beyond the scope of previous works. Specifically, we will address questions related to the possibility of obtaining non-Fermi liquid behavior in models without disorder, the existence of a sharp Fermi surface (or lack thereof) in translation invariant non-Fermi liquids, the fate of quantum oscillations due to critical Fermi surfaces beyond semiclassical quantization of quasiparticle-based theories and other related issues.  Our paper will also lead to new insights into a class of non-Fermi liquid metals, namely the ``IR-incomplete" NFLs (of which there are numerous examples, as highlighted later), and will potentially be useful for future developments in the field.

Within our construction, if there is a single band of bandwidth $W$, and the typical interaction strength is $U$, we find that the system crosses over at a temperature $T\sim W^2/U (\equiv\Omega^*)$ from a low-temperature Landau Fermi liquid ground state to locally quantum critical non-Fermi liquid state, where the Fermi surface is completely destroyed, but there still is a well-defined Fermi energy. The resistivity crosses over from $\rho \sim T^2$ at $T\ll \Omega^*$ to $\rho \sim T$ at $T\gg \Omega^*$; the value of the resistivity at the crossover scale $(T\sim\Omega^*)$ is $\rho\approx h/Ne^2$. In addition, the two salient features of the one band model are as follows: (i) At strong coupling (i.e. $U\gg W$) and at low temperatures compared to $\Omega^*$, the momentum dependence of the electron self-energy becomes parametrically smaller in $(W/U)$ compared to the frequency dependence. The resulting Fermi liquid has a sharp Fermi surface but the self-energy is momentum independent. At temperatures higher than $\Omega^*$, this sharp Fermi surface is lost and the electronic excitations become incoherent. (ii) In the incoherent regime, even though the system is translationally invariant, as a result of the locally critical structure of the correlation functions and strong momentum dissipation on the lattice, the previously established mechanism for incoherent transport in disordered SYK-like models \cite{Parcollet1,Balents} continues to be applicable to our one-band model. In the Fermi liquid regime, the resistivity is finite and arises from umklapp scattering. Our results for the translationally invariant one-band model shed interesting light on the validity of `locally critical' theories in a microscopic setting, where the self-energy is allowed to be momentum dependent apriori but becomes unimportant in the large$-N$ and strong coupling regime.

If there are multiple bands with parametrically different bandwidths (or an itinerant band coupled to localized electrons, as in a Kondo lattice), a richer behavior is observed. In addition to the low temperature Fermi liquid and the high temperature incoherent regime, we find an intermediate range of temperatures where the correlations in the narrow band are locally quantum critical, while the band with the larger bandwidth forms a marginal Fermi liquid, with a single particle inverse lifetime proportional to $\max(\varepsilon, T)$, where $\varepsilon$ is the energy. This region realizes the marginal Fermi liquid phenomenological model proposed in Ref.~\cite{Varma}, with the density (or flavor) fluctuations of the narrow, incoherent band (which have SYK like correlations) playing the role of the critical bosonic degree of freedom. Importantly, our microscopic electronic model defined on the lattice has only {\it local} interactions and preserves translational symmetry. Moreover, even though the light electrons have a feedback on the heavy electrons, there remains a parametrically broad regime of temperatures where the SYK form of the correlations in the heavy band survives. In the regime where the heavy electrons becomes incoherent, there can be strong momentum dissipation in the lattice model leading to a finite $T-$linear resistivity.

Within the multi-band setup, we also consider models where the on-site interactions for one of the bands involves $q>4$-body terms, which allows us to obtain non Fermi liquids with a singular self-energy and a critical Fermi-surface. Interestingly, upon applying a magnetic field, both the marginal Fermi liquid and the non-Fermi liquid regimes are characterized by quantum oscillations of the magnetization as a function of the inverse of the field. The period of the oscillations is the same as that of an ordinary Fermi liquid, but the temperature dependence of their amplitude is different from that of a Fermi liquid.

It has been proposed that transport in the strange metal regime \cite{Bruin13} can be understood in terms of the conjectured ``Planckian'' bound on relaxation rates, $1/\tau \lesssim k_B T / \hbar$~\cite{QPT, Zaanen04}. 
It is interesting to examine our results in the context of this proposal; however, there is no unique definition for a ``transport scattering rate". One can naively choose to define it from the dc conductivity by fitting it to a `Drude-like' form $\sigma = ne^2 \tau_{\tn{dc}}/m^*$, where $m^*$ is the effective mass of the low-temperature Fermi liquid state, and expect a bound on $\tau_{\tn{dc}}~ (\sim 1/T)$.\footnote{This is the definition used in Ref.~\cite{Bruin13}. One we may alternatively define a scattering rate by expressing $\sigma \propto \kappa v_F^{*2} \tau_{\tn{d}}$ where $\kappa$ is the compressibility, or $\sigma \propto \omega_p^2 \tau_\tn{p}$ where $\omega_p$ is the plasma frequency.} 
In the two-band non-Fermi liquid state described in Sec.~\ref{nfl}, we find that $1/\tau_{\tn{dc}}$ has a non-Planckian form: $1/\tau_{\mathrm{dc}} \sim T^\alpha$ with $\alpha<1$. Alternatively, a natural way of defining the transport scattering rate is to use the temperature dependent crossover frequency scale, $1/\tau_{\tn{opt}}$, across which the optical conductivity crosses over from its high frequency regime to the dc limit. For the models considered below, we find that $\tau_{\tn{opt}}$ satisfies a Planckian-type bound with $\tau_{\tn{opt}}^{-1} \le  a k_BT/\hbar$, where $a$ is an $O(1)$ number.\footnote{There are examples of models that violate this bound on $\tau_{\mathrm{opt}}$, however. See, e.g., Ref.~\cite{Werman}.} Thus, the question of the existence of a bound requires a sharp definition of what one means by the ``scattering rate.''

In light of the phenomenologically appealing features of the solution of these models, it is interesting to ask about lessons we might learn and  apply to real  correlated materials described by some generic model. Restricting to  IR-incomplete non-Fermi liquids, it is interesting to consider the structure of a coarse-grained description.  We expect that there will be a few distinct universality classes of such non-Fermi liquids with different coarse grained descriptions. The models studied in this paper suggest one possible universal route to non-Fermi liquid behavior.  Specifically we propose that in a class of generic systems that show intermediate-$T$ NFL physics, there is an emergent  large length scale $\ell  \gg a$ (the microscopic scale) such that within patches of size $\ell$ the system is maximally chaotic (in the sense that it obeys the chaos bound of Ref.~\cite{Maldacena2016}; see Appendix \ref{chaos} for details) though globally, {\it i.e.} at longer scales it may not be so. Further we expect that the assumption of maximal chaos severely restricts the structure of correlators within such a patch.  A coarse-grained description of the macroscopic physics - appropriate at scales much longer than $\ell$ - can then be built by coupling together maximally chaotic bubbles with generic interactions.  Note that  the $(0+1)$-dimensional SYK models are well known to be maximally chaotic. Thus the models we study may be viewed as a concrete  example of such a coarse grained effective model.  In general the appropriate description of a maximally chaotic bubble in such a metal will not likely  be an SYK-like model, and will in the future have to be replaced by a better theory that takes into account spatial locality within each bubble. Nevertheless these solvable models  point to the  importance of  maximally chaotic intermediate scale bubbles as a possible universal route to  a class of non-Fermi liquids. 

The rest of this paper is organized as follows: we introduce our model of a strongly interacting translationally invariant one-band metal in section \ref{mod} and compute the fermion Green's function,  thermodynamic and transport properties in sections \ref{spec}, \ref{thermo} and \ref{trans1} respectively. 
In section \ref{scaling} we provide a very simple qualitative understanding of these one-band models which demystifies their properties and provides a complementary approach to analyzing the key features of the model. We introduce an additional band with a parametrically smaller bandwidth and study the effect of inter-band interactions in section \ref{sec:MFL}. We compute the fermion Green's function in section \ref{MFLgreen} and find a regime with a marginal Fermi liquid behavior. We explore the thermodynamic and transport properties associated with the MFL in sections \ref{thermomfl} and \ref{transmfl} respectively. The two band model is generalized in section \ref{nfl}, where we find a regime with non-Fermi liquid behavior and a singular self-energy with a variable exponent; the thermodynamic and transport behavior are discussed in section \ref{thermonfl} and \ref{transnfl}. For the generalized model, we explore the ``$2K_F$" singularities and quantum oscillations in the magnetization as a function of an external magnetic field as a result of the presence of the critical Fermi surface in section \ref{2kfnfl} and \ref{qo}, respectively. On the basis of our study of all the models with locally critical degrees of freedom, we propose some general constraints on models with local quantum criticality in section \ref{LC}. Finally, in section \ref{sec:discussion} we conclude with a summary of our results and their relation to other recent works. In section \ref{conj} we also present our conjectures for intermediate scale non Fermi liquid  physics in generic strongly correlated models and explore their consequences for the phenomenology of a wide variety of non-Fermi liquid metals. We study the toy problem with $q=2$ (i.e. a random-matrix) in the presence of uniform hopping terms as an interesting exercise, which can be solved exactly, in Appendix~\ref{syk2} in order to shed some light on issues related to transport. A number of accompanying technical details appear in the appendices.

\section{One-Band Model} 
\label{mod}
\begin{figure}
\begin{center}
\includegraphics[width=1.0\columnwidth]{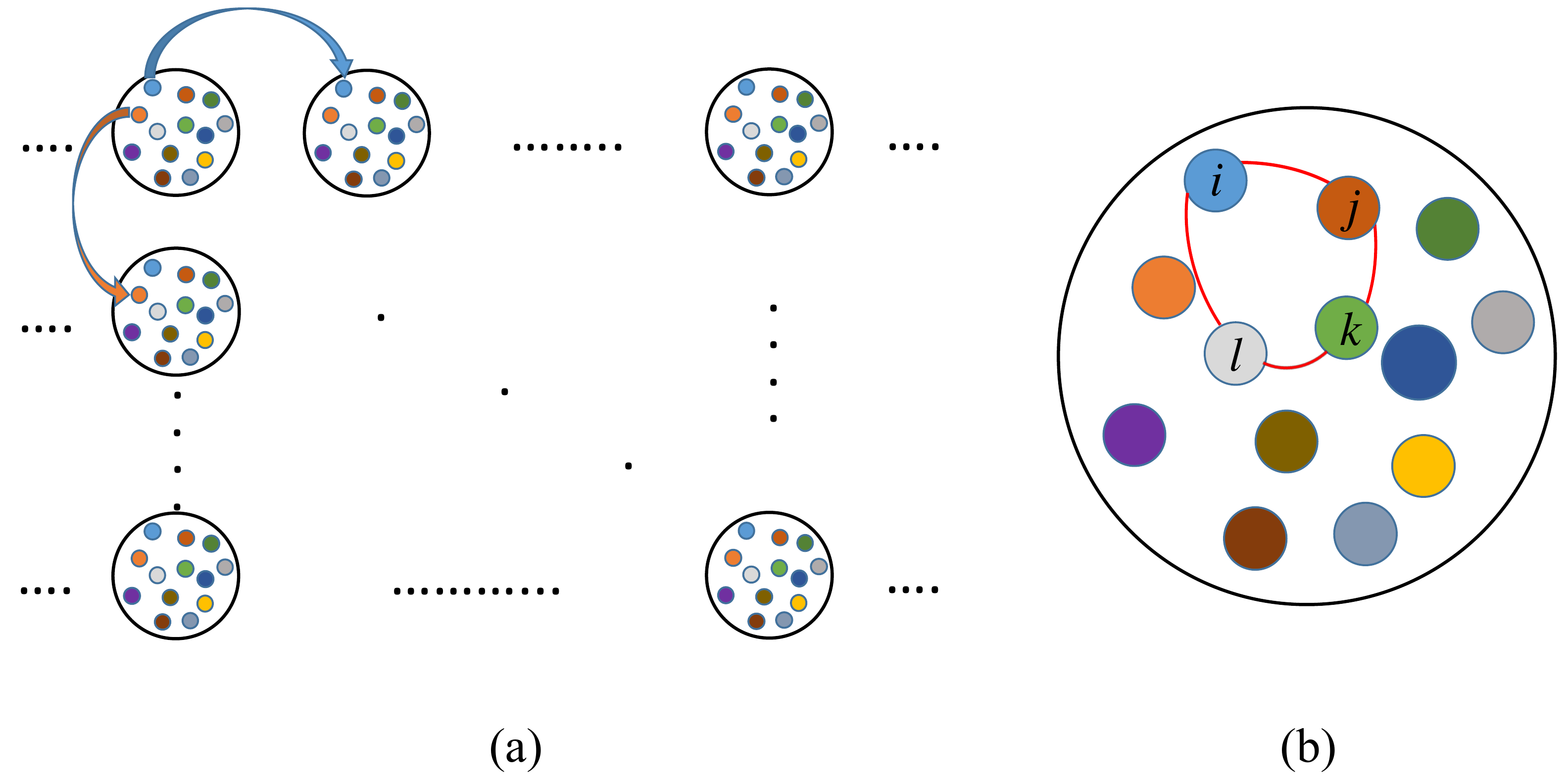}
\end{center}
\caption{(a) A two-dimensional lattice where each site contains $N$ orbitals (represented by different colors). The hoppings, $t^c_{\r\r'}$, between any neighboring sites (colored arrows) are diagonal in orbital-index. Each site is identical and the system is translationally invariant. (b) The internal structure of a single site with $N$ orbitals. The on-site interactions, $U^c_{ijk\el}$, are quartic in the fermion operators, with all orbital indices unequal.}
\label{model}
\end{figure}

Let us begin with a microscopic model in $d-$dimensions on a hypercubic lattice ($d=2$ will be of primary interest) with $N$ orbitals per site and fermionic operators defined by, $c^{\dagger}_{\r,\el}$, $c_{\r,\el}$, ($\el=1,...,N$). The fermions satisfy usual anti-commutation algebra $\{c_{\r,\el}, c_{\r',\el'}^{\dagger}\} = \delta_{\el\el'} \delta_{\r\r'}$. We assume that there is a global $U(1)$ symmetry corresponding to a {\it single} conserved density ($V\equiv$volume), $Q_c = \sum_{\r,\el} \la c_{\r\el}^{\dagger} c_{\r\el}\ra/(NV)$. The value of $0<Q_c<1$ can be tuned by a chemical potential $\mu_c$.
The Hamiltonian is given by
\beq
H_c = \sum_{\r,\r'} \sum_{\el}(-t^c_{\r,\r'}  - \mu_c \delta_{\r\r'}) c_{\r \el}^\dagger c_{\r'\el} + \frac{1}{(2N)^{3/2}}\sum_{\r}\sum_{ijk\el} U^c_{ijk\el} c^\dagger_{\r i}c^\dagger_{\r j} c_{\r k} c_{\r \el}, %\nonumber\\
\label{hc1}
\eeq
where the hopping terms between sites $\r$ and $\r'$, $t^c_{\r\r'}$, are diagonal in the orbital subspace and depend only on $|\r-\r'|$ (assumed to be identical for all orbitals). The interaction term, $U^c_{ijk\el}$, is purely on-site and is properly antisymmetrized with $U^c_{ijk\el} = -U^c_{jik\el} = -U^c_{ij\el k}$ and $U^c_{ijkl} = U^{c}_{klij}$. The values of $U^c_{ijk \el}$ are assumed to be independent of the site-label, $\r$ (see Fig.~\ref{model}(a) for a caricature of the model; Fig.~\ref{model}(b) elucidates the structure of interactions within each site). The model can be viewed as a lattice of Sachdev-Ye-Kitaev (SYK)~\cite{SY,kitaev_talk,Parcollet1,Parcollet2,FuSS,SS15} quantum dots with identical on-site interactions, connected by orbital-diagonal, translationally invariant hopping matrix elements.\footnote{A one-dimensional field theory with similar translationally-invariant interactions has been considered in Ref.~\cite{berkooz2017comments}.} 

The model (\ref{hc1}) is difficult to solve. However, just as in the SYK model, if we consider the interaction terms $U^c_{ijk\el}$ to be random, independent variables with a zero mean, and take the limit $N\rightarrow \infty$, then it is possible to compute properties of the model {\it averaged over realizations of $U^c_{ijk\el}$}. 
It is important to note that we are not only assuming that the coupling constants on different sites have the same distribution; rather, in every realization they are identical to each other, and hence the Hamiltonian defined in Eq.~\ref{hc1} is translationally invariant. For convenience, we set the distribution of the coupling constants to be Gaussian. The distribution satisfies $\overline{U^c_{ijk\el}}=0$ and $\overline{(U^c_{ijk\el})^2} \equiv U^2_c$, where $U_c$ characterizes the strength of the interactions. The other energy scale in our problem is the free electrons' bandwidth, which we denote by $W_c$.

It is believed that the properties of the SYK model are self-averaging, in the sense that the correlation functions of a typical realization are close to those of the mean, up to $1/N$ corrections. In Appendix \ref{app:selfaverage}, we demonstrate that the standard deviations and higher cumulants of the correlation functions in our model are suppressed by powers of $1/N$. We therefore expect that the correlation functions in our model are self-averaging in the large $N$ limit, as in the single-site SYK model. 

\subsection{Fermion Green's Function} 
\label{spec}
\begin{figure}
\begin{center}
\includegraphics[width=0.6\columnwidth]{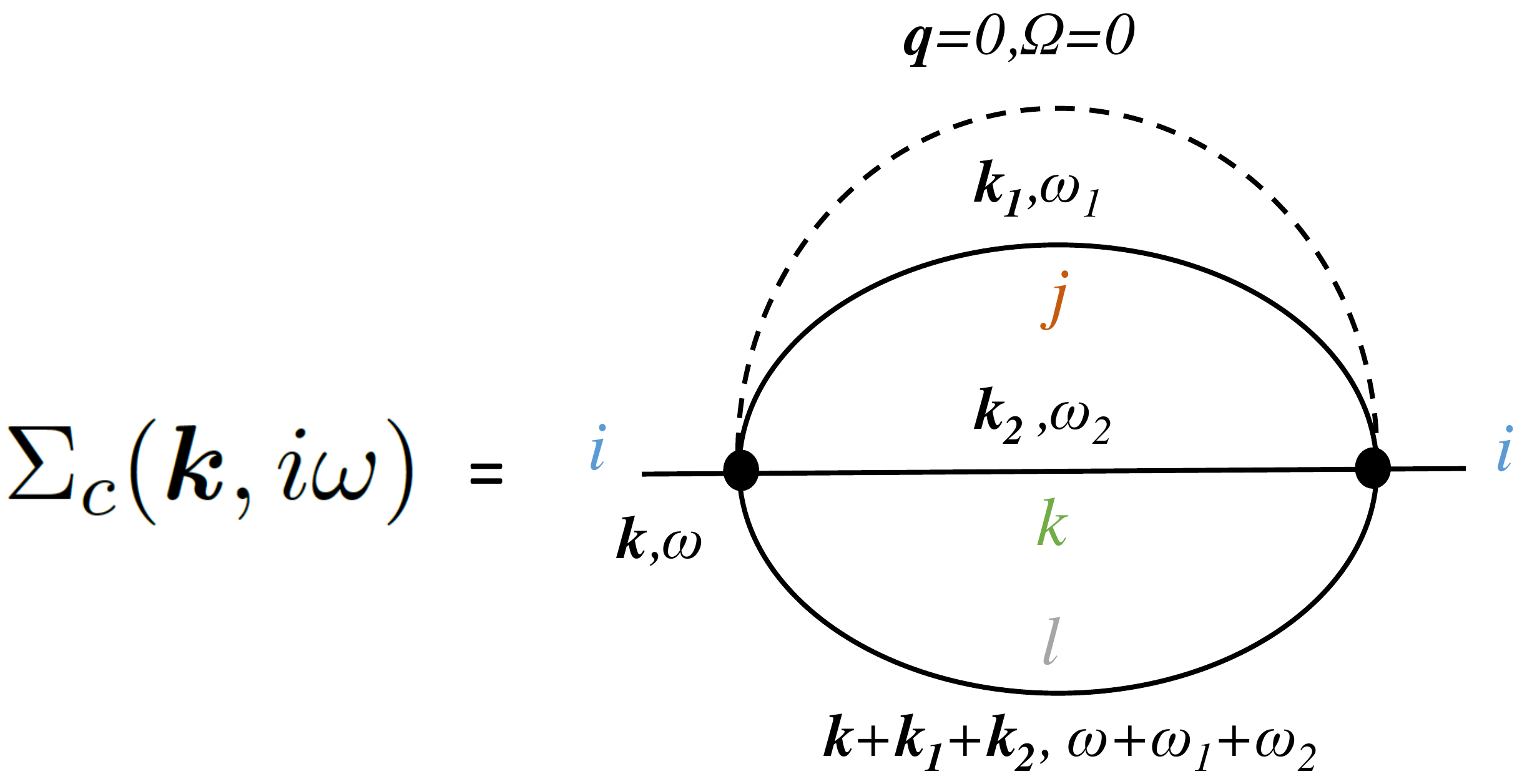}
\end{center}
\caption{The self-energy diagram, $\Sigma_c$,  for $c-$fermions with orbital index $i$ in the single-band model due to $U_c$. The solid black lines represent fully dressed Green's functions, $G_c(\k,\omega)$; see Eq.~(\ref{SP1}). The dashed line corresponds to $U_c^2$ contraction and carries no frequency/momentum. }
\label{se}
\end{figure}

The fermion Green's function can be analyzed diagrammatically, such that the large-$N$ saddle-point solution reduces to studying the following set of equations self-consistently,
\begin{subequations}
\begin{eqnarray}
\label{SP1}
G_c(\k,i\omega) &=& \frac{1}{i\omega - \ve_\k - \Sigma_c(\k,i\omega)}, \label{SPc_a}\\
 \Sigma_c(\k,i\omega) &=& - U_c^2 \int_{\k_1} \int_{\omega_1} G_c(\k_1,i\omega_1)~\Pi_c(\k+\k_1,i\omega+i\omega_1), \label{SPc_b}\label{selfenergyc} \\
\Pi_c(\q,i\Omega) &=& \int_{\k} \int_{\omega} G_c(\k,i\omega)~G_c(\k+\q,i\omega+i\Omega),
\label{SPc_c}
\end{eqnarray}
\end{subequations}
where $\int_\k\equiv\int d^d\k/(2\pi)^d$ and $\ve_\k$ is the dispersion for the $c-$band. Formally, the above set of equations corresponds to resumming an infinite class of `watermelon-diagrams', as shown in Fig.~\ref{se}. One can arrive at the same set of saddle-point equations by starting from the path-integral formulation, as described in Appendix \ref{pif}. In  Sec.  \ref{scaling}, we provide a simple alternate derivation of the results for the one band model using scaling-type arguments which provide much physical insight. 

As we shall now show, the fermionic spectral function has qualitatively different behavior at different temperatures. When the temperature is much lower than the characteristic crossover scale $\Omega^*_c \equiv W_c^2 / U_c$, the spectral function has a Fermi-liquid like form. In the interesting case $U_c \gg W_c$, there is a second regime defined by $\Omega^*_c \ll T \ll U_c$, 
where the spectral function has an incoherent, local form without any remnant of a Fermi-surface. To make this statement more precise, we can take the limit of $U_c\rightarrow\infty$ keeping $W_c$ finite (such that $\Omega_c^*$ collapses to zero), and then take the limit of $T\rightarrow0$, thus obtaining a compressible phase of electronic matter without quasiparticle-excitations in a clean system, lacking any sharp momentum-space structure. We refer to this state as a local incoherent critical metal (LICM). 

To analyze the equations~(\ref{SPc_a}-\ref{SPc_c}), we focus on the two extreme limits of $T$ (or $\omega$) that are either much larger or much smaller than $\Omega^*_c$. In the limit $T\ll \Omega^*_c$, we find that the system follows Fermi liquid behavior at sufficiently low frequencies. To show this, let us use a Fermi liquid-like {\it ansatz} for the fermionic self energy. At low frequencies we assume that $\Sigma_c$ has the following form near the Fermi surface:
\beq
\Sigma_c(\k,i\omega) = -i(Z^{-1} - 1) \omega + (\tv - v_F)k + \dots,
\label{Sigma_c}
\eeq
where $Z$ is the quasiparticle residue, to be determined self-consistently, $k=|\k-\k_F|$  ($\k_F$ is the Fermi momentum), $\tv~(v_F)$ are the renormalized (bare) Fermi-velocities with the renormalization $\tv/v_F=A$ to be determined self-consistently, and the $\dots$ denote higher power terms in an expansion in $\omega,~k$. We stress that $\tilde{v}_F$ is different from the effective Fermi velocity $v^*_F = Z \tilde{v}_F$, which is the physical speed with which quasi-particles propagate.  
For simplicity, we have dropped the constant term, which can be absorbed in the chemical potential. Inserting this form into the self-consistency equations~(\ref{SPc_a}-\ref{SPc_c}), we obtain after a standard computation (see Appendix {\ref{M1A}} for details)
\beq
\Pi_c(\q, i\Omega) = Z \nu_0 \left(1 - \frac{|\Omega|}{\sqrt{(Z \tv q)^2 + \Omega^2}} +O(q^2)\right).
\label{Pi_c}
\eeq
Here, $\nu_0 \sim k_F^d/W_c$ is the bare density of states at the Fermi energy. (We set the units of length such that the lattice spacing $a=1$.)
In Eq.~(\ref{Pi_c}) we have taken into account the contribution of the quasi-particle poles of the Green's functions at $i\omega = Z \tilde\varepsilon_\k$, and ignored the additional branch cut singularities, that turn out not to change the final result qualitatively. Next, we feed Eq.~(\ref{Pi_c}) back into~(\ref{selfenergyc}), giving 
\beq
\Sigma_c(\k,i\omega) = \nu_0^2 U_c^2 \left[ i Z \omega + i \alpha \nu_0 |\omega|^2 \ln\bigg(\frac{Z \tilde{v}_F k_F}{|\omega|} \bigg) \mathrm{sgn}(\omega) - Z^2 \zeta v_F k \right], \label{eq:sigmac}
\eeq
where $\alpha,~\zeta$ are numerical factors of order unity that depends on the geometry of the Fermi surface (Appendix~\ref{M1A}). The factor of $\ln\left(\frac{Z\tilde{v}_F k_F}{|\omega|}\right)$ in (\ref{eq:sigmac}) is special to $d=2$; it is absent in higher dimensions. Equating this to Eq.~(\ref{Sigma_c}), we get that
\beq
\nu_0^2 U_c^2 Z &=& (Z^{-1} - 1),\\
\nu_0^2 U_c^2 Z^2 \zeta &=& (A - 1).
\label{Z}
\eeq
In particular, in the weak coupling limit, $\nu_0 U_c \ll 1$, we get that $Z \approx 1 - (\nu_0 U_c)^2 $. In the opposite limit, $\nu_0 U_c \gg 1$, we get to logarithmic accuracy that $Z = 1/(\nu_0 U_c)$, and $A$ is $O(1)$. In this strong coupling limit, even though the electronic self-energy is allowed to be apriori momentum dependent, the frequency dependence is parametrically larger in $(U_c/W_c)$ compared to the momentum dependence.  Hence, the ground state is a Fermi liquid for any coupling strength; in the strong coupling limit, the quasi-particle weight becomes small, and the effective mass increases as $m^* = m/Z \approx m\nu_0 U_c$, where $m$ is the bare mass while the momentum dependence of the self-energy is independent of $U_c$. This state is therefore a {\it heavy} Fermi-liquid. Moreover, since the self-energy is only weakly dependent on the momentum but strongly frequency dependent, the resulting state is reminiscent of a DMFT description \cite{DMFT} of a heavily renormalized Fermi liquid. Note, however, that while DMFT is exact in the limit of infinite dimension, in our case $d$ is finite; instead, we have to take the large $N$ and strong coupling limits.

Next, we turn to the behavior of $\Sigma_c(\omega)$ at high frequencies. We focus on the strong coupling limit, $\nu_0 U_c \gg 1$. In this regime, $\Sigma_c(\omega)$ exceeds the Fermi energy for sufficiently large $\omega$. Extrapolating $\Sigma_c(\omega)$ from Eq.~(\ref{Sigma_c}) with $Z = 1/(\nu_0 U_c)$, we get that this occurs at frequencies larger then $\Omega^*_c = W_c^2/U_c$.  Then, to zeroth order, we can neglect $\varepsilon_\k$ relative to $\Sigma_c(\omega)$ in Eq.~(\ref{SPc_a}). In this limit, the self-consistent equations~(\ref{SPc_a}-\ref{SPc_b}) reduce to those of the single site SYK model~\cite{SY,kitaev_talk,Parcollet1,Parcollet2,FuSS,SS15,Maldacena_syk}. In particular, we get that at frequencies smaller than $U_c$, $\Sigma_c(\omega) \sim i\mathrm{sgn}(\omega)\sqrt{U_c|\omega|}$~\cite{SY,Parcollet1,Parcollet2}. Extrapolating $\Sigma_c(\omega)$ from high to intermediate frequencies, we reproduce the result that $\Sigma_c(\omega) \gg W_c$ for $\omega\gg W_c^2/U_c$, consistent with the extrapolation from low frequencies. 

To find the residual momentum dependence of the Green's function in the strong coupling incoherent regime, we expand the self-consistent equations~\ref{SPc_c} in powers of $\varepsilon_\k$\footnote{The results below are also readily obtained by simply calculating the Green's function in perturbation theory  in the hopping $t_c$ along the lines of Sec. \ref{scaling}.}. To leading order, we get that $G_c(\k,\omega)-G_{0}(\omega) \sim \varepsilon_\k/[\Sigma_0(\omega)]^2$, where $G_{0}(\omega)$ and $\Sigma_0(\omega)$ are the Green's function and the self-energy of the single site SYK model, respectively (see Appendix \ref{licm} for details). Importantly, we see that although the momentum dependence of the Green's function decreases with increasing frequency, the correlation length over which $G_c(\r,\omega)$ decays (obtained by taking the fourier transform of $G_c(\k,\omega)$) remains frequency-independent and is determined by the spatial extent of the hopping parameters, $t^c_{\r\r'}$.

To summarize, we get that for strong coupling, $G_c(\k,\omega)$ has the following form in the two extreme frequency limits:
\begin{equation}
G_c(\k,i\omega) \sim 
\begin{cases}
\frac{Z}{i\omega - Z \tilde\varepsilon_\k + i \alpha \nu_0^2 U_c|\omega|^2 \ln(\frac{1}{|\omega|})\mathrm{sgn}(\omega)}, &\omega \ll W_c^2/U_c,\\
\frac{i\mathrm{sgn}(\omega)}{\sqrt{U_c |\omega|}} -  B(\omega) \frac{\varepsilon_{\k}}{U_c |\omega|}, &W_c^2/U_c \ll \omega \ll U_c,
\end{cases}
\label{limits}
\end{equation}
where $Z \sim 1/(\nu_0 U_c)$, and $\alpha$ is a number of order unity. $B(\omega)$ is a constant independent of frequency for both $\omega > 0$ and $\omega < 0$ though its precise value is different for the two signs of $\omega$.  Indeed it is a direct descendant of the ``spectral asymmetry" that characterizes the Green's function of a single SYK island \cite{SY,Parcollet1}.

At low frequencies, there is a Fermi surface with well-defined, albeit strongly renormalized quasiparticles. The renormalized bandwidth is $W_c^* \sim \Omega^*_c = W_c^2/U_c$. The $\omega^2$ term in the denominator of $G_c$ becomes the imaginary part of the self-energy after an analytic continuation to real frequency. It can be written in a revealing form: $\Sigma''(\omega) \sim \omega^2 \ln\left(\frac{W_c^*}{|\omega|}\right)/W^*_c$. At finite temperatures, the zero-frequency imaginary part is $\Sigma''(0,T)\sim \pi^2 T^2 \ln\left(\frac{W_c^*}{T}\right) /W_c^*$. Note that, upon extrapolating this form to the  crossover scale, $\Sigma''(0,T\sim \Omega^*_c) \sim W_c^*$, {\it i.e.} at this scale, the scattering rate of quasiparticles is comparable to the effective bandwidth, and we expect the quasi-particle picture to break down.

At energies much higher than the renormalized bandwidth, the Fermi surface is destroyed, and the single-particle spectral function has no sharp features in momentum space. Instead, it is well approximated by $A_c(\k,\omega)\sim 1/\sqrt{U_c\, \mathrm{max}(|\omega|,T)}$. 
This is the LICM regime.

\subsection{Thermodynamic Properties}
\label{thermo}
We now turn to discuss the thermodynamic properties of the one-band model. As we saw in the previous subsection, at sufficiently low temperatures, $T\ll\Omega_c^*$, the system is well described by Fermi-liquid theory. This implies, in particular, that the entropy per unit cell follows a linear temperature dependence, $S(T\ll \Omega_c^*) = N\gamma T$, where $\gamma \propto m^* \sim U_c/W_c^2$. At temperatures much higher than $\Omega_c^*$, we can calculate the thermodynamic properties perturbatively in the inter-site hopping\footnote{Such a perturbative expansion breaks down at sufficiently low temperatures, since the hopping is a relevant perturbation.}. Then, the entropy is given by that of a single SYK dot, up to a correction of the order of $(W_c/U_c)^2$. The entropy takes the form $S(T\gg \Omega_c^*) = N (S_0 + \gamma_0 T)$, where $S_0$ and $\gamma_0$ are known constants~\cite{Parcollet1}. At temperatures of the order of $\Omega^*_c$, we expect the entropy to interpolate between these two behaviors. Based on our analysis of the saddle point equations in this section, as well as our simpler understanding using scaling in Sec.~\ref{scaling} below where we study the perturbative effects of the relevant hopping terms as a function of decreasing energy starting from the decoupled SYK-like regime, we find a strong indication of a single crossover separating the two regimes at the coherence-scale $\Omega_c^*$.  All of the thermodynamic quantities, as well as the frequency dependent self-energies, evolve smoothly through this crossover without any associated phase transitions; we have checked this explicitly by solving the saddle-point equations numerically for small system sizes (results not shown). These aspects of our results are thus qualitatively similar to the results reported in Ref.~\cite{Balents} for the disordered version of the one-band model.

Next, we turn to discuss the compressibility, given by $N\kappa = (\partial n/\partial \mu)$, where $n = \sum_{\r,\ell} c_{\r,\ell}^\dagger c_{\r,\ell}$ is the total density for all the orbitals. We begin by noting that each site ({\it i.e.} SYK island) has a finite compressibility which is given by $\kappa_0 \sim 1/U_c$ \cite{SS17,Balents}. As a result of the finite hopping and bandwidth, there is a correction to this result and at strong coupling we obtain
\beq
\kappa = \frac{c_0}{U_c}\bigg[1+ O\bigg(\frac{W_c^2}{U_c^2}\bigg)\bigg],
\eeq
where $c_0$ is a constant of order unity. As discussed earlier, in this regime the mass enhancement factor $m^*/m =Z^{-1} \approx U_c/W_c$. This can be reconciled within the Fermi-liquid description of the state if one introduces a large dimensionless `Landau-parameter', $F_0\sim(U_c/W_c)^2$.

\subsection{Transport}
\label{trans1}
\begin{figure}
\begin{center}
\includegraphics[width=0.8\columnwidth]{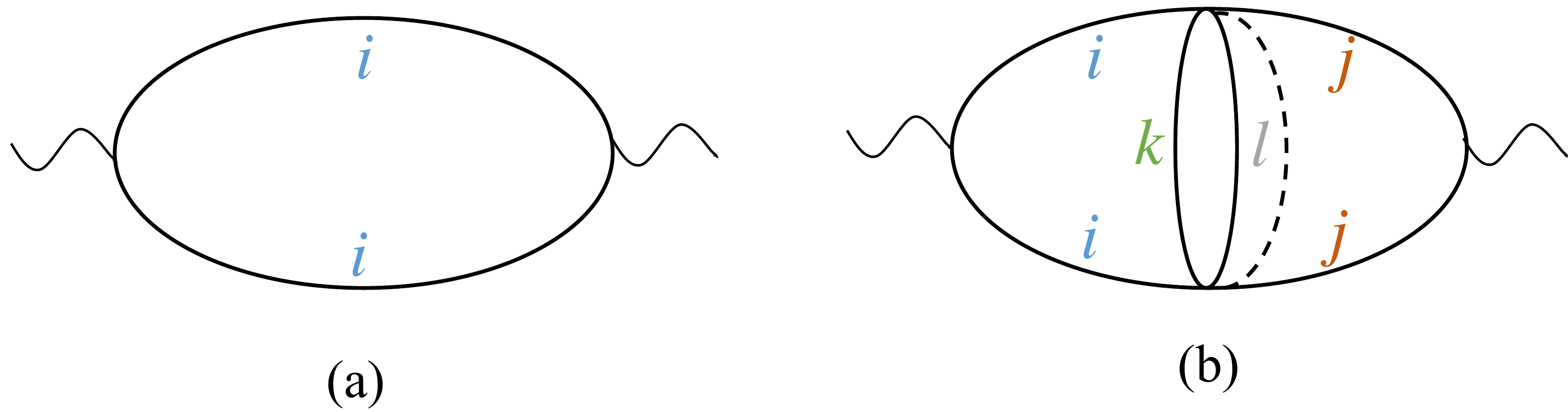}
\end{center}
\caption{Current-current correlation function for evaluating the conductivity in the one-band model. Wiggly line denotes the insertion of the current operator. The solid lines represent the fully dressed propagators. (a) Feynman diagram without vertex corrections. (b) The lowest order vertex correction diagram, which is subleading in the high temperature (LICM) regime. The dashed line represents a $U_c^2$ contraction, as before.}
\label{opt}
\end{figure}

Let us now discuss both the optical conductivity and the dc resistivity of the metallic phases introduced above. The real part of the optical conductivity is given by the Kubo formula 
\begin{equation}
\sigma_{xx}'(\omega) = \frac{\tn{Im}~\Pi_{J_x}^{\tn{ret}}(\omega)}{{\omega}},
\end{equation}
where $\Pi^{\tn{ret}}_{J_x}(\omega)$ is the retarded current-current correlation function for the current in $x$ direction. The total current operator is given by 
\beq
\J = \sum_i \J_i = \sum_\k \v^i_\k c_{\k i}^\dagger c_{\k i},
\eeq 
with $\J_i$ denotes the current from orbital $i$ and $\v^i_\k = \nabla_\k\ve^i_\k$. For the previously assumed identical dispersions for all the orbitals, the velocities are also the same. The leading diagrams which contribute to $\Pi^{\tn{ret}}_{J_x}(\omega)$ are shown in Fig.~\ref{opt}. In Fig. \ref{opt} (a), we show the leading graph without vertex corrections. 

In the high temperature ($T\gg\Omega_c^*$) regime, the vertex corrections (Fig. \ref{opt}b) are subleading. To see this, note that the electron velocity is odd in momentum, while all the Green's functions are momentum independent to lowest order in $W_c/U_c$. Then each of the loops over orbital $i$ and $j$ vanish individually. We therefore consider only the diagram in Fig.~\ref{opt}(a), and this results in
\begin{equation}
\sigma(\omega,T) = \sum_i \frac{1}{\omega}\int d\omega' \int_\k v_\k^2~A_\k(\omega')~A_\k(\omega+\omega')~[f(\omega') - f(\omega+\omega')],
\end{equation}
where $A_\k(\omega)$ is the electron spectral function and $f(...)$ represents the Fermi-Dirac distribution function. In the high temperature SYK-like regime ($T\gg\Omega_c^*$), the optical conductivity clearly satisfies $(\omega/T)$ scaling. At frequencies much higher than the temperature (i.e. by arranging $\Omega_c^*\ll T\ll\omega\ll U_c$), we find $\sigma(\omega)\propto Nv^2/(U_c~\omega)$. Focusing on the dc conductivity in this regime, we find (in units of $e^2/h$):
\begin{eqnarray}
\sigma_{dc} \propto \frac{N v^2}{U_c T},
\end{eqnarray}
where $v^2~(\sim t_c^2)$ represents an average over the Fermi-surface. As a result of the $\omega/T$ scaling, the crossover scale from the high-frequency to the dc limit is of order $T$. We note that in this incoherent regime, once the electronic correlation functions become locally critical, the previously established mechanism for incoherent transport in disordered SYK-like models \cite{Parcollet1,Balents} continues to be applicable as a result of the strong momentum dissipation in the lattice model.

\section{A simple view on the one-band results} 
\label{scaling}
In this section, we provide a simple alternate understanding of the physics of the one-band model that does not require a detailed analysis of the saddle-point equations in Eq.~(\ref{SP1}-\ref{SPc_b}). We carry out a simple scaling analysis for the extension of the usual SYK model (as defined above in Eq.~\ref{hc1}), as well as an extension of the model that involves higher than $2-$body interactions. The latter will be used later in Sec.~\ref{nfl} to obtain a non-Fermi liquid with a critical Fermi surface.

We begin by considering the limit where the hopping $t_c \ll U_c$. When $t_c = 0$ the different SYK islands are decoupled from each other. Further we know that within each island the electron has power law correlations in time with a scaling dimension $\Delta_c = \frac{1}{4}$. For small hopping $t_c$, we can study the relevance/irrelevance of the hopping term in the decoupled SYK theories. In the action, the hopping term becomes 
\begin{equation}
S_{\tn{hopping}} = -t_c \int d\tau \sum_{i, \langle \r \r' \rangle} c^\dagger_{\r i}(\tau) c_{\r'i}(\tau)  + c^\dagger_{\r'i}(\tau)c_{\r i}(\tau). 
\end{equation}
Clearly then under a scaling transformation $\tau \rightarrow \tau' = \frac{\tau}{s}$, $t_c \rightarrow t_c' = t_c ~s^{\frac{1}{2}}$ so that the hopping is relevant.  To study the system at a non-zero temperature $T$, we run the scaling until a scale $s_T = \frac{U_c}{T}$. The effective renormalized hopping at this scale is then $t_c(s_T) = t_c  \left( \frac{U_c}{T} \right)^{\frac{1}{2}}$. With decreasing temperature we will stay in the regime of  weak hopping until a temperature such that $t_c(s_T) \sim U_c$. This corresponds to a temperature scale $T_{\tn{coh}} \sim \frac{ t_c^2}{U_c}$ which matches exactly with the coherence scale identified in section \ref{spec}. 

For $T \gg T_{\tn{coh}}$ the physics will be that of weakly coupled SYK islands and we can calculate physical properties in perturbation theory in $t_c$.  For $T \ll T_{\tn{coh}}$ it is natural to expect that the coupling between the different islands leads to a Fermi liquid phase. 

 We can now understand the thermodynamics and transport through simple physical arguments. First we recall that for  the $(0+1)$-dimensional SYK model, the entropy is known \cite{Parcollet1,Maldacena_syk} to obey 
 \begin{equation}
 S(T) = N(S_0 + \gamma_0 T +...).
 \end{equation}
 The ground state entropy $S_0$ is nonzero in the limit $N \rightarrow\infty $, and then $T \rightarrow 0$. As argued in Sec.~\ref{thermo}, in the limit $t_c = 0$ this is obviously the entropy per site of the 
 lattice model.  When $t_c \neq 0$ and at sufficiently high temperature such that $T \gg T_{\tn{coh}}$, both the entropy and the compressibility only get small corrections when we perturb in $t_c$.  For $T \ll T_{\tn{coh}}$, however, the ground state entropy of the decoupled limit is relieved, and $\frac{S(T \rightarrow 0)}{M} \rightarrow 0$. ($M=N V$ is the total number of sites). In the low-$T$ Fermi liquid we expect $\frac{S(T \rightarrow 0)}{M} = \gamma T$. An estimate for $\gamma$ can be obtained by matching this entropy extrapolated to $T = T_{\tn{coh}}$ with the residual entropy of the high temperature phase. This gives 
 \begin{equation}
 \gamma \approx \frac{S_0}{T_{\tn{coh}}}~\sim \frac{U_c}{t_c^2}.
 \end{equation}
 
 In Fermi liquid theory the $\gamma$ coefficient directly gives the quasiparticle effective mass $m^* \sim  \frac{U_c}{t_c^2 a^2}$.  ($a$ is the lattice spacing.)
 Note that the ``bare"  mass determined from the hopping Hamiltonian is $m \sim \frac{1}{t_c a^2}$ .  Therefore the mass enhancement $\frac{m^*}{m} \sim \frac{U_c}{t_c} \gg 1$ in exact agreement with the solution of the self consistency equations in section \ref{spec}.  The behavior of the compressibility in both the high-$T$ and low-$T$ limits have already been described in section \ref{thermo}.

 Let us now turn to transport. In the high-$T$ regime in perturbation theory in $t_c$, the conductivity $\sigma_{dc}$ will be $\propto t_c^2$. In $d = 2$, $\sigma_{dc}$ is dimensionless in units of $\frac{e^2}{h}$. We thus expect that for $T \gg T_{\tn{coh}}$, $\sigma_{dc} \sim \frac{Ne^2}{h}\left( \frac{t_c(s_T)}{U_c}\right)^2$ where $t_c(s_T)$ is the effective renormalized hopping at a temperature $T$ introduced above. We therefore get 
 \begin{equation}
 \sigma_{dc} \sim  \frac{Ne^2}{h} \frac{t_c^2}{U_cT}~\sim  \frac{Ne^2}{h} \frac{T_{\tn{coh}}}{T}.
 \end{equation}
 This is again in exact agreement with the calculations in section \ref{trans1}. For $T \ll T_{\tn{coh}}$, if the Fermi surface is big enough to allow umklapp scattering of the low energy quasiparticles, we will get a resistivity 
 $\rho(T) = \frac{\tilde{A}}{N} T^2$. To estimate $\tilde{A}$, we require that when extrapolated to $T = T_{\tn{coh}}$ this matches the extrapolation of the high $T$ result down to $T_{\tn{coh}}$.  This  leads to $\tilde{A} \sim \frac{h}{e^2} \frac{1}{T_{\tn{coh}}^2}$. Note that in the low-$T$ Fermi liquid $\tilde{A} \sim \gamma^2$ thereby obeying the Kadowaki-Woods relationship~\cite{KW}. 
 
 The understanding above readily generalizes to the physics of coupled SYK models where the on-site interaction is composed of $q$ ($q \geq 4$ and even) fermion operators \cite{Gross17}; we studied a generalized two-band version of this model in section~\ref{nfl}. Specifically consider the model of just a single band of electrons with the Hamiltonian
 \beq
H_c = \sum_{\r,\r'} \sum_{\el} (-t^c_{\r,\r'} { - \mu_c \delta_{\r\r'} }) c_{\r \el}^\dagger c_{\r'\el} + \frac{\left(\frac{q}{2} !\right)}{N^{\frac{q-1}{2}}}\sum_{\{i_\ell\}}U^c_{i_1i_2...i_q} \bigg[c^\dagger_{\r,i_1}c^\dagger_{\r,i_2}...c^\dagger_{\r,i_{q/2}}c_{\r,i_{q/2+1}}...c_{\r,i_{q-1}}c_{\r,i_q} \bigg]. \nonumber\\
\label{hc}
\eeq
As before we take $U_{i_1i_2...i_q}$ and the hopping $t_c$ to be translationally invariant and $\overline{U_{i_1i_2...i_q}}= 0$, and $\overline{(U_{i_1i_2...i_q})^2}=U_c^2$. 
We focus on the small $t_c$ regime. For general $q$, the scaling dimension of the fermion is $\Delta(q) = 1/q$. It follows that a small $t_c$ is relevant at the decoupled fixed point and scales as 
\beq
t_c(s) = t_c~ s^{1- \frac{2}{q}}.
\eeq
Following the discussion above, we determine that the  physics will be that of weakly coupled islands until a coherence scale $T_{\tn{coh}} = t_c \left(t_c/U_c\right)^{\frac{2}{q-2}}$. 
In the high-$T$ regime, the entropy and compressibility have the same qualitative behavior as for $q = 4$. Importantly, there is a residual entropy $S_0$ (with a linear $T$ correction) and a finite non-zero compressibility. 
At $T \ll T_{\tn{coh}}$ we again expect a Fermi liquid. The residual entropy is relieved, and the low-$T$ heat capacity coefficient is $\gamma \sim \frac{S_0}{T_{\tn{coh}}}$. This can be converted into an estimate for the quasiparticle effective mass in the Fermi liquid. 

The electrical resistivity in the high-$T$ regime, estimated as above, is of the form
\beq
\rho_{dc} \sim \frac{h}{Ne^2} \left(\frac{U_c}{t_c}\right)^2 \left(\frac{T}{U_c}\right)^{\frac{2(q-2)}{q}}.
\eeq
Note that $\rho_{dc}$ increases faster than linearly with $T$, but slower than $T^2$. Thus the high-$T$ linear resistivity is not a generic property of coupled SYK models and  requires $q = 4$. As before (with umklapp scattering) the low-$T$ resistivity is $\rho(T) = \tilde{A} T^2$ with $\tilde{A} \sim \gamma^2$. 

\subsection{Explicit transport calculation at high-temperature} 
\label{perttrnsprt}
It is instructive to explicitly calculate the conductivity in the high-$T$ regime in perturbation theory in the hopping, taking special care with issues regarding disorder averaging. As the leading temperature dependence is $\propto t_c^2$, a second order perturbative calculation should give the exact answer for this leading term.  The imaginary frequency current-current correlator is readily related to the electron Green's function of each SYK island (details of such perturbative calculations are straightforward - see for example Appendix E of Ref.~\cite{TSDCP}):
\beq
\label{Pipert0}
\Pi_{J_x}(q = 0, i\omega_n) =  \frac{(et_c)^2}{\beta} \sum_{\omega_\nu} \sum_{ij} G_{ij} (\r, \r; i\omega_\nu) \left( G_{ji} (\r', \r'; i(\omega_n + \omega_\nu) - G_{ji}(\r', \r'; i\omega_\nu) \right)
\eeq
Note that we have not carried out the disorder averaging yet. $G_{ij} (\r, \r; i\omega_\nu)$ is the frequency dependent fermion Green's function within the SYK island at site $\r$ and $\r'$ is the site neighboring to $\r$ in the positive $x$ direction. 

We now wish to average this over disorder realizations. If the SYK interactions were independently random at different sites (like in the models studied in Ref. \onlinecite{Balents}), then obviously upon disorder averaging (indicated with an overline) the products $G_{ij} G_{ji}$ that appear above can be replaced by $\overline{G}_{ij} \overline{G}_{ji}$ for any $N$. In our translation invariant models, the SYK interactions are the same at every site. Thus strictly speaking we must instead take $\overline{G_{ij} G_{ji}}$.  Fortunately (as shown in Appendix \ref{app:selfaverage}) for $q \geq 4$ in the large-$N$ limit\footnote{The $q = 2$ case is special and will be discussed in detail in Appendix \ref{syk2}.}, the property 
\beq
\label{gfctr}
\overline{G_{ij} G_{ji}} = \overline{G}_{ij} \overline{G}_{ji}
\eeq
holds and we can continue to make this replacement in the products entering the correlation function. Further we know that when $N \rightarrow  \infty$, only $\overline{G}_{ii}$ is $O(1$)
and $\overline{G}_{ij}$ for $i \neq j$ is suppressed. \footnote{Note that there are $N$ ``diagonal" terms with $i = j$ while there are $O(N^2)$ off-diagonal terms where $i \neq j$. Thus it is necessary that the off-diagonal terms are suppressed by sufficiently high powers of $N$. For SYK$_q$ we show in Appendix \ref{app:selfaverage} that the off-diagonal contributions are of order $\sim N^{3-q}$ and hence can be ignored compared to the $O(N)$ diagonal contributions for $q > 4$. Clearly however they cannot be ignored at $q = 2$, and play an important role in obtaining the correct physics as we discuss in Appendix \ref{syk2}.} Therefore we will henceforth replace all Green's functions by their averages (and drop the overlines). Implicitly this has been done in all of the discussions in this paper. 
Analytically continuing Eq.~(\ref{Pipert0}) to real frequencies we get the familiar form for the real part of the conductivity
\beq
\label{sigmapert}
\sigma_{xx}'(\omega, T) = N\pi (et_c)^2 \int d\Omega A(\Omega) A(\omega + \Omega) \left(\frac{f(\Omega) - f(\omega + \Omega)}{\omega}\right) .
\eeq
Here $A(\omega)$ is the spectral function for the Green's function within a single SYK island. For SYK$_q$ (with $q \geq  4$) this satisfies $\omega/T$ scaling,
\beq
A(\omega, T) = \frac{1}{U_c} \left(\frac{U_c}{|\omega|}\right)^{1- \frac{2}{q}} F_q\left(\frac{\omega}{T} \right),
\eeq
with $F_q(...)$ a known universal function. It follows from Eq.~(\ref{sigmapert}) that the conductivity itself satisfies $\omega/T$ scaling. We get 
\beq
\sigma'_{xx}(\omega, T) = \frac{N e^2 t_c^2 }{U_c^2}  \left(\frac{U_c}{T}\right)^{2- \frac{4}{q}}{\cal S}_q \left(\frac{\omega}{T} \right), 
\eeq
with ${\cal S}_q(...)$ a universal function determined in terms of $F_q(...)$ by Eq.~(\ref{sigmapert}). In particular in the dc limit we reproduce the temperature dependence previously obtained for general $q \geq 4$. As a result of the $\omega/T$ scaling, it is easy to see that the frequency scale over which $\sigma'_{xx}(\omega)$ reaches its dc value is $\tau_{\tn{opt}}^{-1}\sim a T$ (in units of $k_B/\hbar$), with $a$ an $O(1)$ number. Moreover, the scaling function ${\cal{S}}_q(x)\sim 1/x^{2-4/q}$ at large $x$. Therefore, at frequencies much larger than the temperature, the conductivity has the form  $\sigma'_{xx}(\omega)\sim 1/\omega^{2-4/q}$.

\section{Two-Band Model --- Marginal Fermi Liquid} \label{sec:MFL}
In the previous section, we saw an example of a crossover from a Fermi-liquid to an incoherent metal, without any remnant of a Fermi-surface, in a one-band model. It is interesting to ask if a critical Fermi-surface~\cite{TS08} can emerge in the general class of translationally invariant models that are being considered here. Before proceeding further, it is useful to define precisely what we mean by a critical Fermi-surface. Within our definition, the criticality is associated with the gapless single-particle excitations of physical electrons over the entire Fermi surface, which remains sharply defined\footnote{In contrast sometimes in the literature the phrase `critical Fermi surface' is used to denote a Fermi surface of emergent fermions (not locally related to microscopic degrees of freedom) which themselves may be critical.}. However there are no Landau quasiparticles across the critical Fermi surface and the quasiparticle residue $Z$ is zero. We describe two classes of models in the next two sections that host such a critical Fermi surface. 

Let us begin with a model where we introduce an additional band of $f-$fermions with operators $f^{\dagger}_{\r,\el}$, $f_{\r,\el}$ ($\el=1,...,N$) and an associated conserved $U(1)$ charge density, $Q_f$ that may be tuned by a chemical potential $\mu_f$, which we set to zero. The modified Hamiltonian (with a $U(1)_c\times U(1)_f$ symmetry) is 
\beq
H = H_c + H_f + H_{cf},
\label{Htot}
\eeq
where $H_c$ is as described in Eq.~(\ref{hc}), and $H_f$ is defined in an identical fashion with translationally invariant hoppings $t^f_{\r\r'}$ and on-site interactions $U^f_{ijk\el}$. The form of the inter-band interaction is chosen to be
\beq
H_{cf} = \frac{1}{N^{3/2}}\sum_{\r}\sum_{ijk\el} V_{ijk\el} c^\dagger_{\r i}f^\dagger_{\r j} c_{\r k} f_{\r \el},
\label{hcf}
\eeq
where the coefficients, $V_{ijkl}$, are chosen to be identical at every site with $\overline{U^f_{ijkl}} = \overline{V_{ijkl}} =  0$, and where the distribution of the couplings satisfy $\overline{(U_{ijk\el}^{f})^2}=U_f^2,~\overline{(V_{ijk\el})^2}=U_{cf}^2$. We now assume that $t^f_{\r,\r'}\ll t^c_{\r,\r'}$, i.e. the bandwidth for the $f-$fermions is much smaller than the bandwidth for the $c-$fermions ($W_f\ll W_c$). The model described by (\ref{Htot}) therefore has some similarity to models for `heavy-Fermion' systems, with a specific form of interaction terms, and where the direct hybridization term, $H_{\tn{hyb}} = \sum_{\r, ij} M_{ij}~c_{\r i}^\dagger f_{\r j}$ has been set to zero.

\begin{figure}
\begin{center}
\includegraphics[width=1.0\columnwidth]{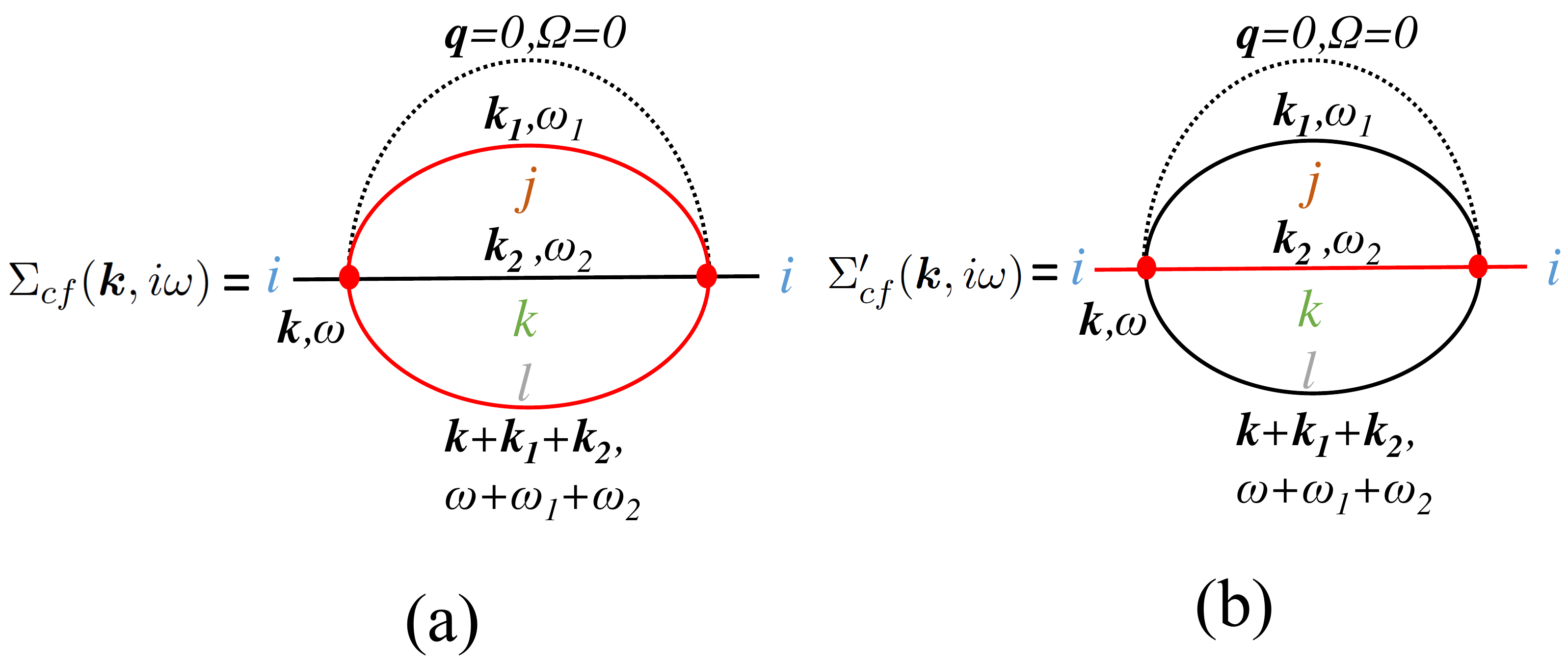}
\end{center}
\caption{The self-energy diagrams in the two-band model due to inter-band scattering for (a) $c-$fermions, $\Sigma_{cf}$, and, (b) $f-$fermions, $\Sigma'_{cf}$, with orbital index $i$. The solid black (red) lines represent fully dressed Green's functions, $G_c(\k,\omega)~(G_f(\k,\omega))$; see Eq.~(\ref{sceq_a}) and \ref{sceq_b}. The dashed lines correspond to $U_{cf}^2$ contractions respectively and carry no frequency/momentum.}
\label{se2}
\end{figure}

To leading order in $1/N$, the saddle point equations for the Hamiltonian defined in Eq.~(\ref{Htot}) are given by,
\begin{subequations}
\beq
G_c(\k,i\omega) &=& \frac{1}{i\omega - \ve_\k - \Sigma_c(\k,i\omega) - \Sigma_{cf}(\k,i\omega)},\label{sceq_a} \\
G_f(\k,i\omega) &=& \frac{1}{i\omega - \xi_\k - \Sigma_f(\k,i\omega) - \Sigma'_{cf}(\k,i\omega)} ,\label{sceq_b} \\
\Sigma_{f}(\k,i\omega) &=& { -} U_f^2 \int_{\k_1} \int_{\omega_1} G_f(\k_1,i\omega_1)~\Pi_f(\k+\k_1,i\omega+i\omega_1),\label{sceq_c} \\
\Sigma_{cf}(\k,i\omega) &=& { -} U_{cf}^2 \int_{\k_1} \int_{\omega_1} G_c(\k_1,i\omega_1)~\Pi_f(\k+\k_1,i\omega+i\omega_1),~ \label{sceq_d} \\
\Pi_f(\q,i\Omega) &=& \int_{\k} \int_{\omega} G_f(\k,i\omega)~G_f(\k+\q,i\omega+i\Omega),~\tn{and} \label{sceq_e} \\
\Sigma'_{cf}(\k,i\omega) &=& - U_{cf}^2 \int_{\k_1} \int_{\omega_1} G_f(\k_1,i\omega_1)~\Pi_c(\k+\k_1,i\omega+i\omega_1).\label{sceq_f}
\eeq
\end{subequations}
We have introduced $\xi_\k$ as the dispersion for the $f$ fermions and $\Sigma_c$, $\Pi_c$ are as defined earlier in Eqs.~(\ref{SPc_a}-\ref{SPc_c}). The watermelon-diagrams for the self-energies are shown in Fig.\ref{se2}.

Based on our analysis for the one-band model in section \ref{mod}, we see immediately that if $H_{cf}=0$, the two decoupled subsystems have a Fermi-liquid to LICM crossover at frequencies or temperatures of the order of $\Omega_c^*$ and $\Omega_f^*$ respectively ($\Omega_f^*\ll \Omega_c^*$). In the high-temperature regime, $T>\Omega_c^*\gg \Omega_f^*$, when both bands are in a LICM phase, adding $H_{cf}$ does not alter any of the features qualitatively and the resulting state is thus still described by a LICM phase. Similarly, at low-temperatures, $T<\Omega_f^*\ll \Omega_c^*$, when both species are in a Fermi-liquid phase, a finite $H_{cf}$ does not modify the qualitative aspects. There are two Fermi-surfaces for the $c$ and $f$ fermions consistent with their individual Luttinger count; adding a finite $H_{\tn{hyb}}$ hybridizes the two Fermi-surfaces, breaking the two independent $U(1)_c\times U(1)_f$ symmetries down to a single conserved $U(1)$ charge corresponding to the total fermion density. The low-energy description of the Fermi-liquid phase is similar to our considerations from the previous section. The key question that remains is what is the fate of the system in the intermediate regime $\Omega_f^*< T <\Omega_c^*$.

For the purpose of our subsequent discussion, we can set $U_c=0$, such that the $c$ band is uncorrelated and the scale $\Omega_c^*$ is pushed out to infinity (the bandwidth, $W_c$, is the physical UV scale).{\footnote{All of our results below remain qualitatively the same in the presence of a finite $U_c$. }}
In order to have a sharp meaning to the notion of a non-Fermi liquid with a critical Fermi surface, it is useful to also send the scale $\Omega_f^*$ to zero. This is conveniently done by setting $t^f = 0$ while keeping $U_f$ finite.  In this limit, we can pose sharp questions about the presence or absence of quasiparticles and Fermi surfaces in the $T \rightarrow 0$ limit.

\subsection{Fermion Green's Function}
\label{MFLgreen}
In the window of intermediate energies, where the $f$ fermions enter the LICM regime, while the $c$ fermions do not, one may find a Fermi surface formed by the lighter bands, with an anomalous single-particle lifetime due to scattering off the heavy band. As we show below, it is precisely in this regime that we obtain a {\it marginal Fermi liquid} regime with a critical Fermi surface of the $c$ fermions. In the next section, we will generalize the model to obtain a critical Fermi surface of the $c$ fermions with a singular frequency dependent self-energy with variable exponents.

In order to obtain the structure of the solution in the intermediate frequency regime, $\Omega_f^* < \omega <\Omega_c^*$, we begin by considering the effect of the inter-band interaction perturbatively. Later, we will check that the behavior we find is self-consistent. As emphasized earlier, our conclusions hold in this regime for a finite $U_c$, but we set $U_c=0$ for simplicity. We assume that the $f$ fermions are in the LICM regime, such that $G_f(\k,\tau)\sim i\mathrm{sgn}(\tau)/\sqrt{U_f |\tau|}$ for imaginary times $|\tau| \gg 1/U_f$, which is the familiar form in the SYK model~\cite{SY,Parcollet1,Parcollet2}. We ignore here the weak-momentum dependent correction to the $f$ Green's function in the LICM phase (as discussed in section \ref{spec}) by considering the limit of $W_f/U_f\rightarrow0$, with $U_f$ fixed (the crossover scale $\Omega_f^*$ also goes to zero in this limit). Then, $\Pi_f(\q,\omega)$ has the momentum independent form\footnote{Note that the spectral asymmetry in the $f$-Green's function cancels out in the product below.},
\beq
\Pi_f(\q,i\omega) = \int d\tau ~e^{i\omega \tau}~ G_f(\tau)~G_f(-\tau) \sim \frac{1}{U_f} \log\left( \frac{U_f}{|\omega|} \right).
\label{mflpif}
\eeq

Inserting $\Pi_f(\omega)$ into Eq.~(\ref{sceq_d}) we get for the $c$ self-energy in Fig.\ref{se2}(a) (see Appendix~\ref{MFLapp})
\beq
\Sigma_{cf}(i\omega) \sim  - \frac{\nu_0 U^2_{cf}}{2\pi^2 U_f}\, i \omega \log\left( \frac{U_f}{|\omega|} \right).
\label{SEmfl}
\eeq

The self-energy of the $c$ fermions then has a `marginal Fermi liquid' (MFL)~\cite{Varma} form. {It is important to note that the above result is valid at most up to scales at which the self-energy becomes of the order of the bandwidth, i.e. $\Sigma_{cf}\sim W_c$. This scale can be easily seen to be $\Omega_{cf}^*\sim U_f(W_c/U_{cf})^2$. 

In order to check that the form of $\Sigma_{cf}(i\omega)$ in (\ref{SEmfl}) is self consistent, we need to verify that it does not change qualitatively if it is evaluated using the full Green's functions. 
Moreover, we also need to evaluate $\Sigma'_{cf}(\omega)$ (the self-energy for the $f$ fermions due to coupling to the $c$ fermions; Fig. \ref{se2}b) using the renormalized $G_c$, and verify that its behavior is sub-leading to that of $\Sigma_f(\omega) \sim \sqrt{U_f \omega}$. We demonstrated that this solution is indeed self-consistent in Appendix~\ref{SCse}. 
In particular, focusing for simplicity on the case where $U_{cf} \ll W_c$, the contribution to the $f$ fermion self-energy due to the inter-species interaction consists of two contributions -- an analytic correction, which is given by
\beq
\Sigma_{cf,1}'(i\omega) \sim \nu_0\frac{U^2_{cf}}{\sqrt{U_fW_c}}i\omega,
\eeq
that renormalizes the bare `$i\omega$' term, and a singular (but subleading) correction (see Appendix \ref{MFLapp})
\beq
\Sigma'_{cf,2}(i\omega)\sim \frac{U_{cf}^2}{W_c^2\sqrt{U_f}}~ i|\omega|^{3/2}\tn{sgn}(\omega).
\eeq 
$\Sigma_{cf,1}'(\omega)$ is negligible compared to the bare $i\omega$ term if $U_f\gg W_c (U_{cf}/W_c)^4$.

In the limit $U_{cf} \ll W_c$, the MFL regime extends over a frequency (or temperature) window $\Omega_f^* \ll \omega ~(\tn{or}~ T) \ll \mathrm{min}(W_c, U_f)$. It is interesing to consider the case where $U_{cf} \sim U_f \gg W_c$; then, the MFL extends only up to temperatures of the order of $\Omega^*_{cf}$. We leave the behavior above $\Omega^*_{cf}$ for a future study. 

To conclude, we find a broad temperature regime above $\Omega_f^*$ where the $f$ fermions behave as a LICM, while the $c$ fermions follow MFL behavior, with a well-defined Fermi surface and marginally defined fermionic quasiparticles. 
As we discuss in section~\ref{nfl} below, a generalized version of the two-band model gives a critical Fermi surface where quasiparticles are not even marginally defined (a full-fledged non-Fermi liquid). 
In Sec.~\ref{nfl} we analyze the density response at the ``$2K_F$" wavevector and quantum oscillations in magnetization of such a critical Fermi surface.

Interestingly, in both the marginal Fermi liquid of the present section and in the non-Fermi liquid of Sec.~\ref{nfl}, the structure of the $c-$fermion self-energy is such that it is singular in the limit of $\omega\rightarrow0$ for all momenta, even far away from the Fermi-surface. This is a consequence of the fact that the fluctuations of the $f$ fermions are critical for all momenta. We analyze this structure more carefully in Appendix~\ref{SCse}.

\subsection{Thermodynamic Properties}
\label{thermomfl}

Let us begin by analyzing the specific heat in the MFL regime, where the entropy density has contributions from the critical Fermi-surface of the $c$ fermions as well as the $f$ fermions which are in an SYK-like regime. The total extrapolated zero temperature entropy is finite as a result of the finite entropy density from the SYK sector. However, this excess entropy is relieved as a result of the crossover to the Fermi-liquid below $\Omega_f^*$. In order to extract the contribution from the critical Fermi-surface, we can compute the free energy at a finite temperature using the standard Luttinger-Ward (LW) analysis (Appendix \ref{pif} and \ref{LWapp}). Let us consider the different contributions to the free energy, written as 
\beq
F = \tn{Tr}[\tn{log} ~G_c^{-1}] + \tn{Tr}[\Sigma_{cf}~G_c] + \tn{Tr}[\tn{log} ~G_f^{-1}] + \tn{Tr}[(\Sigma_f + \Sigma'_{cf})~G_f] - \Phi_{\mathrm{LW}}[G_c, G_f], 
\label{FLW}
\eeq
where $\Phi_{\mathrm{LW}}[G_c,G_f]$ is the Luttinger-Ward functional, which depends on the exact Green's functions of the $c$ and $f$ electrons. 
The first two terms, with the form of the MFL self energy in Eq.~(\ref{SEmfl}), gives rise to a low temperature singular logarithmic correction~\cite{MFLsp} to the linear in $T$ specific heat, i.e. it has a $T \ln(1/T)$ form. This feature is reminiscent of the results of Refs.~\cite{Varma,MFLsp}. However, we note that the self-energy alone does not fix the thermodynamic properties. In particular, the other contribution to the free energy arises from the LW term,

\begin{align}
\Phi_{\mathrm{LW}}[G_c,G_f] & =  \frac{U_{f}^2}{4} \sum_\r \int_0^\beta d\tau~ |G_f(\r,\tau) G_f(\r,-\tau)|^2 \nonumber \\ & + \frac{U_{cf}^2}{2} \sum_\r \int_0^\beta d\tau~ G_c(\r,\tau)G_c(\r,-\tau)G_f(\r,\tau)G_f(\r,-\tau). \label{LW}
\end{align}
[The derivation of (\ref{LW}) follows closely the derivation in the single-band case, outlined in Appendix~\ref{pif}.] Given the local character of the $f$ Green's function, we only need the local form of the $c$ fermion bubble above (which are the same as in a Fermi liquid). At low temperature, the first term in the LW functional above is proportional to $T$, and the second is proportional to $T^2$. Hence, the LW term does not lead to any singular modification of the results for the specific heat, $c_V = -T \partial^2 F/\partial T^2$.

 The MFL has a critical Fermi surface that satisfies Luttinger's theorem: $n_c = {\mathcal{A}_{\mathrm{FS}}}/{(2\pi)^d}$, where $\mathcal{A}_{\mathrm{FS}}$ is the area of the Fermi surface and $n_c$ is the density of $c$ fermions. This follows from a Luttinger-Ward analysis, applied to the $c-$fermion Green's function: $G_c = [i\omega - (\ve_\k - \mu_c) - \Sigma_{cf}(i \omega)]^{-1}$, and accounting for the fact that our model has two conserved $U(1)_c\times U(1)_f$ densities corresponding to the $c$ and $f$ fermions (see Appendix \ref{LWapp} for details of this analysis). The same analysis gives that the $c-$fermion compressibility $\chi_c \equiv \partial n_c/\partial \mu_c$ is finite and non-singular as a function of $U_{cf}$. In particular, for small $U_{cf}$ (and $U_c=0$) it is given by $\chi_c = \chi_0 + O[U_{cf}^2/(U_f W_c^2)]$, where $\chi_0$ is the non-interacting compressibility. The LW analysis for the conserved $f$ fermion density has been carried out in Ref.~\cite{Parcollet2}.

\subsection{Transport}

\label{transmfl}

Next, we consider the charge transport properties in the MFL regime. The arguments below will also apply to the discussion of transport for the non-Fermi liquid metals described in section~\ref{nfl}. We set $U_c=0$ and only consider the effects of inter-band scattering ($U_{cf}$) when the $f$ electrons are in an incoherent SYK-like regime. 

We are interested in the real part of the charge conductivity, which is given by 
\beq
\sigma'_{xx}(\Omega) = \frac{\mathrm{Im}~\Pi^{\mathrm{ret}}_{J_x}(\Omega)}{\Omega},
\eeq
where $\Pi^{\mathrm{ret}}_{J_x}(\Omega)$ is the retarded correlation function of the $x$ component of the current at a non-zero frequency. In particular, it is important to explore the role of vertex corrections of the current. In the 2-band model there are two independent global $U(1)$ symmetries associated with the separate conservation of $c$ and $f$ fermions. Correspondingly there are two independent conductivities associated with transport of the $c$ and $f$ fermions. Here we will be interested in the conductivity due to the $c$ fermions.  The conductivity due to the $f$ fermions will be essentially identical to the discussion in the one-band model and we will not elaborate further on it here.

\begin{figure}
\begin{center}
\includegraphics[width=0.6\columnwidth]{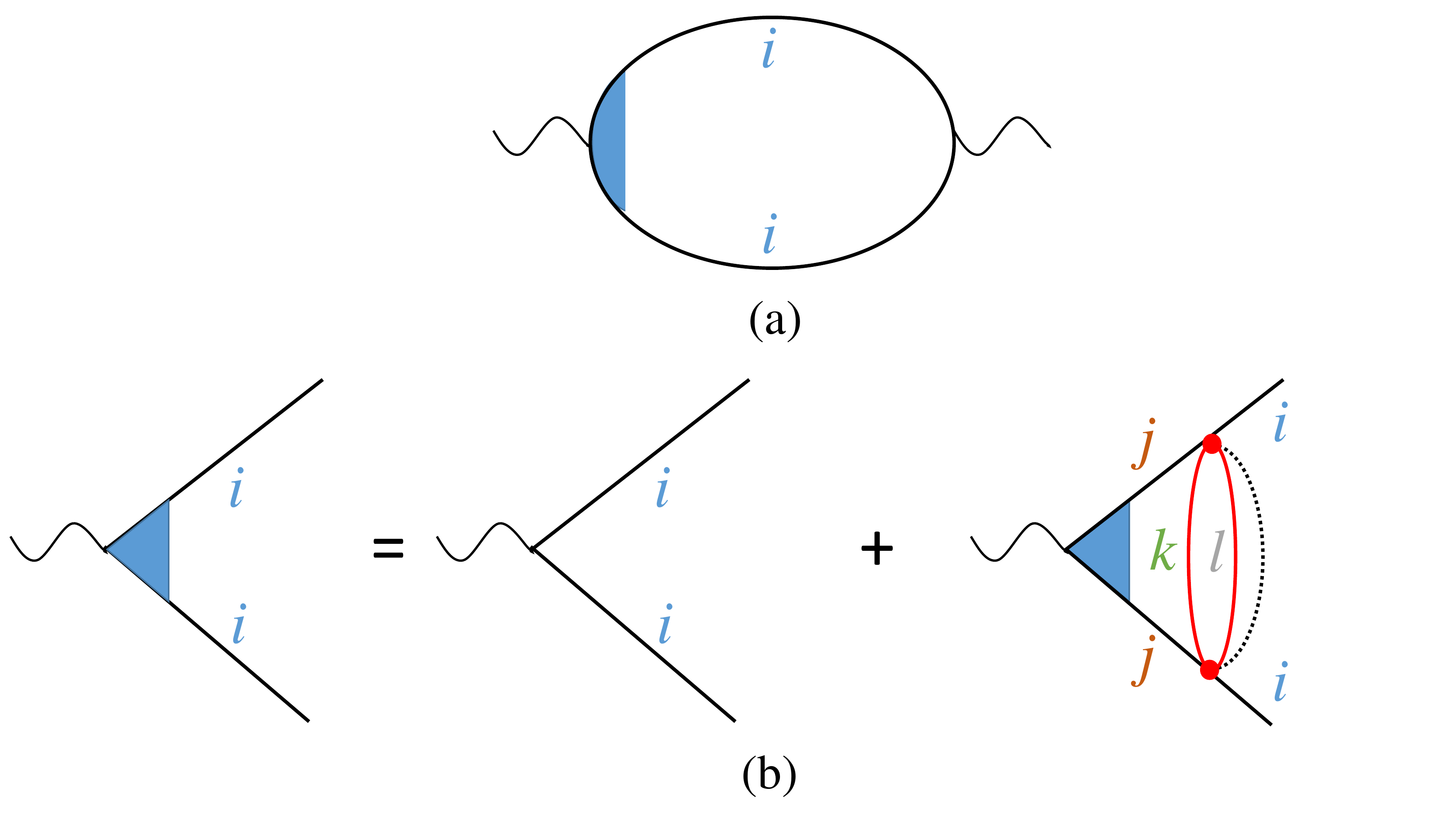}
\end{center}
\caption{(a) Diagram for the computation of $\sigma_{xx}(\Omega)$. Wiggly lines represent current operators and solid black (red) lines represent the full Green's functions, $G_c(\k,\omega)$ ($G_f(\k,\omega)$). (b) The self-consistent equation for the current vertex. }
\label{v2kf}
\end{figure}
To leading order in $1/N$, $\sigma_{xx}(\Omega)$ is given by the sum over the set of ladder diagrams shown in Fig.~\ref{v2kf}(a), where the self-consistent equation for the current vertex $\Gamma_{J_x}(\omega, \Omega)$, is described diagramatically in Fig.~\ref{v2kf}(b):
\begin{equation}
    \Gamma_{J_x}(\k,\omega, \Omega) = v^x_\k + U_{cf}^2 \int_\l\int_{\omega'} \Gamma_{J_x}(\l,\omega',\Omega)~ G_c(\l,\omega')~ G_c(\l,\omega'+\Omega)~ \Pi_f(\l-\k,\omega'-\omega),
    \label{Gammanfl}
\end{equation}
where $v^x_\k = \partial \varepsilon_\k / \partial k_x$ is the `velocity' along $x$ and we have assumed an identical dispersion for all the orbitals. It is important to recall that in a system that preserves inversion symmetry, the velocity (or equivalently, the current) vertex itself is odd with respect to the momentum label, $\k$, i.e. $v^x_{-\k} = -v^x_{\k}$ and $\Gamma_{J_x}(-\k,\omega,\Omega) = -\Gamma_{J_x}(\k,\omega,\Omega)$. At the same order in $1/N$, there is another contribution to the set of ladder diagrams and the current vertex{\footnote{This diagram is reminiscent of an `Aslamazov-Larkin' type contribution.}}, as shown in Fig. \ref{al}. However, this correction vanishes as a result of the local structure of the $f$ fermions, as explained below.

\begin{figure}
\begin{center}
\includegraphics[width=0.5\columnwidth]{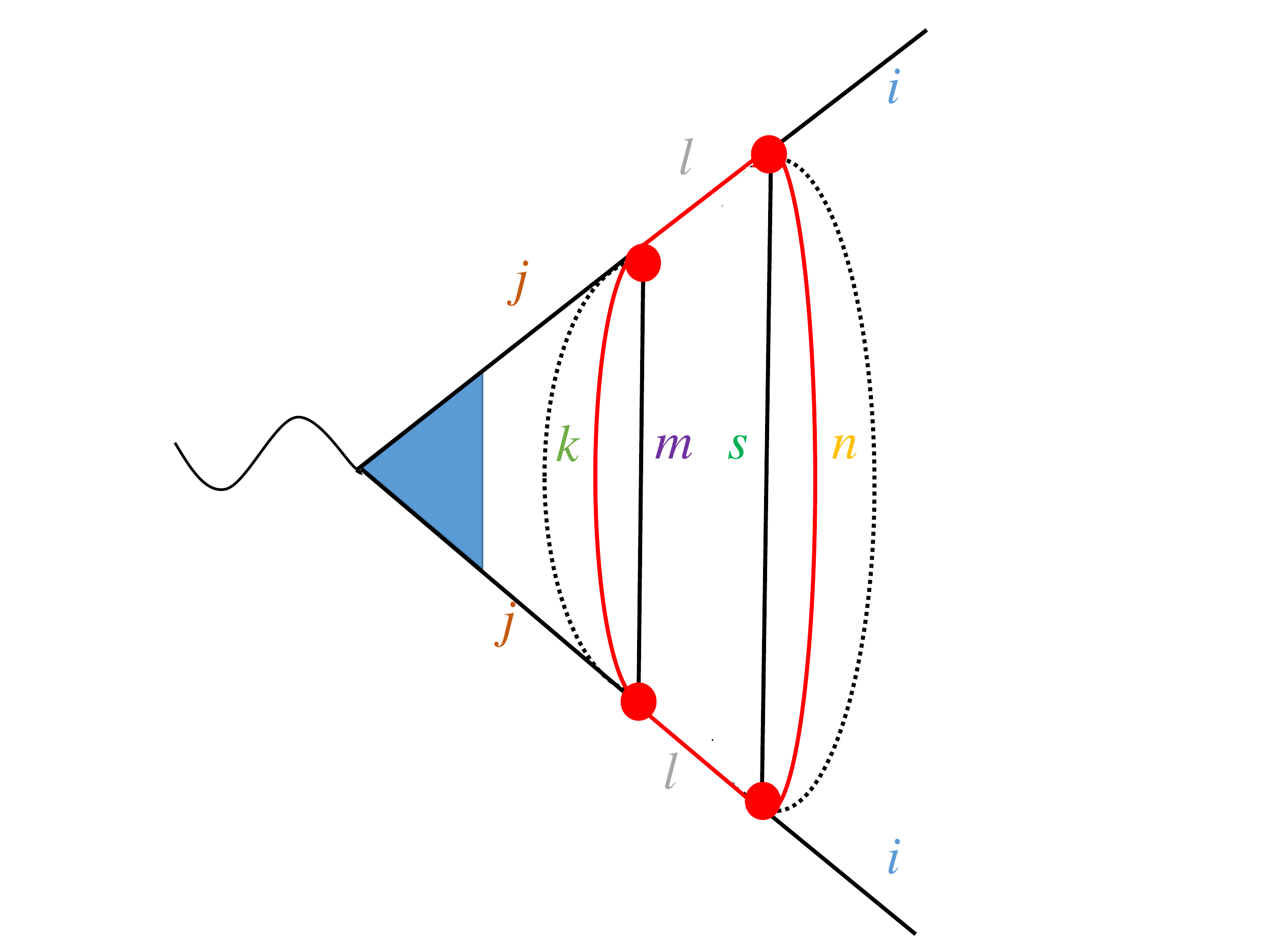}
\end{center}
\caption{Correction to the current vertex in Fig. \ref{v2kf}(b) at the same order in $1/N$.}
\label{al}
\end{figure}

In general, the above self-consistent equation is difficult to solve for the full current vertex. However, the ladder insertions in the above series of diagrams, $\Pi_f$, have a simple local structure that greatly simplifies the problem. At leading order in $\xi_\k/U_f$, as we have discussed above, $\Pi_f(\q,\omega)$ is independent of the momentum $\q$ and has the form shown in Eq.~(\ref{mflpif}) [see Eq.~(\ref{nflpif}) in section~\ref{nfl} for the generalization to the non-Fermi liquid case], which arises from the completely local form of the Green's function, $G_f$. If we ignore the momentum dependence of $\Pi_f$ in Eq.~(\ref{Gammanfl}) above, it is straightforward to see that the momentum integral in the second term vanishes, as the integrand is odd in $\l$. The correction in Fig. \ref{al} vanishes for the same reason. In this limit, we can therefore ignore the vertex corrections altogether, such that the conductivity is given only by the diagram in Fig. \ref{v2kf}(a) without any rungs, which reduces the expression for the conductivity to
\beq
\sigma'_{xx}(\Omega,T) = \frac{1}{\Omega}\int d\omega \int_\k (v^x_\k)^2~ A_\k(\omega)~A_\k(\Omega+\omega)~[f(\omega) - f(\Omega+\omega)],
\label{sigmaxx}
\eeq
where $A_\k(\omega)$ is the spectral function for the $c$ fermions and $f(...)$ represents the Fermi-Dirac distribution function. At frequencies much higher than the temperature ($\Omega\gg T$), this leads to:
\beq
\sigma'_{xx}(\Omega) \propto \frac{Nv^2~U_f}{U_{cf}^2} \frac{1}{\Omega~\ln(1/\Omega)^2},
\eeq
where $v^2$ is the average of $(v^x_\k)^2$ over the Fermi surface.

Let us now focus on the dc limit. 
We find that the scattering rate determined from the dc resistivity is determined by the single-particle scattering rate of the $c-$fermions. This result is not surprising, since in the regime that is being considered here where the $f-$fermions provide a momentum independent scattering channel, providing an effective `momentum sink' for the $c-$fermions{\footnote{ One might be tempted to associate the momentum relaxing scattering in the clean system above with `umklapp' scattering. However, we note that in the regime of interest here, there is no restriction on the respective $c$ or $f-$fermion densities. }}. Therefore, the resistivity (in units of $h/e^2$) is given by,
\beq
\rho_{dc}(T) \propto \frac{U_{cf}^2}{Nv^2 ~U_f} T.
\eeq
 We can now estimate the frequency scale at which the high-frequency form of the optical conductivity matches the low-frequency dc result. A simple analysis immediately reveals the crossover scale (in units of $k_B/\hbar$) to be 
 \beq
 \tau_{\tn{opt}}^{-1}\sim T/\ln^2(1/T).
 \label{toptmfl}
 \eeq 
 Note that the coefficient of the $T-$linear term in the scattering rate need not be $O(1)$ due to the $\ln^2(1/T)$ term in the denominator.

 In the regime $T \ll \Omega^*_{cf} = U_f (W_c/U_{cf})^2$ that we are considering here, the dc resistivity is always smaller than the Mott-Ioffe-Regel limit, $\rho_{dc} \ll h/(Ne^2)$. At higher temperatures, we expect the MFL behavior to break down. We shall not treat this regime here, leaving it to a future investigation.

\section{Two-Band Model --- Non Fermi Liquid}
\label{nfl}
In the previous section, we demonstrated an example of a metal with a critical Fermi-surface at which the electronic quasiparticles are only marginally defined. Is it possible to realize a more singular non-Fermi liquid with no well-defined quasiparticle excitations across a critical Fermi-surface? In this section we show that by generalizing the $f-$band Hamiltonian to the `SYK$_q$' form considered in section \ref{scaling} above, it is possible to obtain a non-Fermi liquid with a critical Fermi-surface and a more singular (and variable) self-energy. 

Let us reintroduce the $f-$electron operators $f_i,~f_i^\dagger$ with $i=1,..,N$ orbitals, as before. We generalize the interaction terms to have a $q-$fold term~\cite{Gross17} (with $q$ even; the models considered thus far correspond to $q=4$). The Hamiltonian is still given by $H =  H_c + \tilde{H}_f + H_{cf}$, where the modified Hamiltonian for the $f-$electrons is given by,
\beq
\tilde{H}_f = -\sum_{\r,\r'} \sum_{\{i_\ell\}} (t_{\r,\r'}^f - \mu_f\delta_{\r\r'}) f_{\r i_\ell}^\dagger f_{\r i_\ell} + \frac{(q/2)!}{N^{\frac{q-1}{2}}}\sum_{\{i_\ell\}}U^f_{i_1i_2...i_q} \bigg[f^\dagger_{\r,i_1}f^\dagger_{\r,i_2}...f^\dagger_{\r,i_{q/2}}f_{\r,i_{q/2+1}}...f_{\r,i_{q-1}}f_{\r,i_q} \bigg].\nonumber\\
\eeq 
The hopping matrix-elements $t^f_{\r,\r'}$ are translationally invariant and diagonal in orbital-space. The on-site inter-orbital interactions, $U^f_{i_1...i_q}$, are assumed to be random 
with $\overline{U_{i_1i_2...i_q}}= 0$, $\overline{(U_{i_1i_2...i_q})^2}=U_f^2$ and taken to be identical on every site. The model is therefore a translationally invariant generalization of the SYK$_q$ model \cite{Gross17} with uniform hoppings. Moreover, since we have already discussed the special case of $q=4$ in the previous section, we shall only consider the case of $q>4$ from now onwards.

\subsection{Fermion Green's Function}
As before, we are interested in the regime above a crossover-scale, $\Omega_{f}^*(q)$, where the $f$ band realizes an incoherent metallic state without any remnant of a Fermi-surface. This crossover scale for the $q-$fold interactions is given by $\Omega_{f}^*(q) = W_f (W_f/U_f)^{2/(q-2)}$, which reduces to the standard expression for $q=4$. In this regime, the scaling dimension of the $f-$operators is $\Delta(q) = 1/q$, such that the Greens function has the form $G_f(\tau) \sim \tn{sgn}(\tau)/(U_f~|\tau|)^{2\Delta(q)}$, or equivalently, $G_f(i\omega)\sim i\tn{sgn}(\omega)/(U_f^{2\Delta(q)} |\omega|^{1-2\Delta(q)})$. 

Let us now address the self-energy of the $c$ fermions as a result of the quartic inter-band scattering in $H_{cf}$. The bubble, $\Pi_f(\q,\omega)$, has a momentum-independent form,
\beq
\Pi_f(\q,i\omega) &=& \int \frac{d\Omega}{2\pi}~G_f(i\omega+i\Omega)~G_f(i\Omega)\nonumber\\
 &\sim& \frac{1}{U_f^{4\Delta(q)}}\frac{1}{|\omega|^{1-4\Delta(q)}}.
 \label{nflpif}
\eeq
Solving for the $c$ fermion self-energy self-consistently (see appendix \ref{SCse}), we obtain
\beq
\Sigma_{cf}(i\omega) \sim \frac{\nu_0 U_{cf}^2}{U_f^{4\Delta(q)}} ~i |\omega|^{4\Delta(q)}\tn{sgn}(\omega),
\label{SEnfl}
\eeq
which has a strong non-Fermi liquid form with an exponent $4\Delta(q)<1$ for $q>4$. This behavior is valid to scales up to which the self-energy becomes of the order of bandwidth, which immediately gives $\Omega_{cf}^*(q) \sim U_f (W_c/U_{cf})^{2/4\Delta(q)}$. Once again, for simplicity we restrict our attention to the case where $U_{cf}\ll W_c$ (which implies $\Omega_{cf}^*(q)\gg U_f$). Just like in the case of the MFL, a natural question to ask is if the feedback of the $c$ fermions on the $f$ fermions as a result of the inter-band scattering modifies the SYK form of their self-energy. There will be an analytic correction that renormalizes the bare `$i\omega$' term, with a coefficient $(W_c/U_f)^{2\Delta(q)} (U_{cf}/W_c)^2$. However this correction can be made small compared to the bare $i\omega$ term if $U_f\gg W_c(U_{cf}/W_c)^{1/\Delta(q)}$. In addition, an explicit computation of the singular (but subleading) correction to the $f$ self-energy as a result of the inter-species interaction leads to
\beq
\Sigma_{cf}'(i\omega) \sim \frac{\nu_0 U_{cf}^2}{W_c U_f^{2\Delta(q)}} i|\omega|^{1+2\Delta(q)} \tn{sgn}(\omega),
\eeq
which, as before, is subleading to $\Sigma_f(i\omega)$ at frequency (or temperature) scales small compared to $\Omega_{cf}^*(q)$.

We therefore conclude that in the intermediate regime between the scales $\Omega_f^*(q)$  and $\tn{min}(U_f,W_c)$, the $f$ fermions have a local SYK$_q$ form of correlations, while the c fermions have a NFL character with a well-defined Fermi surface but no sharply defined Landau quasiparticles.

\subsection{Thermodynamic Properties}
\label{thermonfl}
We turn to discuss the thermodynamic properties of the intermediate non-Fermi liquid regime. The free energy for general $q$ can be computed similarly to the $q=4$ case, using a Luttinger-Ward formulation [see Eq.~(\ref{FLW})]. The entropy is then obtained through $S=-\partial F/\partial T$. 
This gives three contributions to the entropy density ${\cal{S}}(T) = S(T)/(2NV)$: 
\beq
{\cal{S}}(T) &=&  {\cal{S}}_c(T) + {\cal{S}}_f(T) + {\cal{S}}_{int}(T),\\
{\cal{S}}_f(T) &=& {\cal{S}}_{0,q} + \gamma_q ~T,\\
{\cal{S}}_c(T) &\sim&   T^{1/z} \sim  T^{4\Delta(q)},\\
%{\cal{S}}_c(T) &\sim&  \int_{\tn{FS}} T^{1/z} \sim \int_{\tn{FS}} T^{4\Delta(q)},\\
{\cal{S}}_{int}(T) &\sim& T^{1 + 4 \Delta(q)}.  
\eeq
Here, ${\cal{S}}_f(T)$ is the entropy of a single SYK$_q$ model (where ${\cal{S}}_{0,q}$ and $\gamma_q$ have been computed in Ref.~\cite{SS17}), ${\cal{S}}_c(T)$ comes from the first and second terms in Eq.~(\ref{FLW}), and ${\cal{S}}_{int}(T)$ originates from the inter-species interaction term $\delta \Phi_{\mathrm{LW}}\propto U_{cf}^2 \int d\tau G_f^2 G_c^2$ in the LW functional. The extrapolated zero temperature entropy ${\cal{S}}(T\rightarrow0)={\cal{S}}_{0,q}$ is finite in the above regime. However, as described earlier, there is a crossover to a Fermi-liquid regime below the scale $\Omega_f^*$, where the excess entropy is relieved. In the non-Fermi liquid regime, the specific heat scales as $c_V = T\partial {\cal{S}}/\partial T \sim T^{4\Delta(q)}$.

The compressibility associated with the conserved densities for both species of fermions follows from our discussion in section~\ref{thermomfl}. In particular, the $c-$fermions continue to satisfy Luttinger's theorem, such that their density is given by the area inside the critical Fermi-surface (see Appendix \ref{LWapp}). Moreover, the compressibility for the $c-$fermions, that are scattering off the incoherent $f$ fermions, is a non-singular function of $U_{cf}$.
For $U_c=0$, the compressibility is given by that of non-interacting $c$-fermions, up to a correction of the order of $U_{cf}^2$.}

\subsection{Transport}
\label{transnfl}

The transport properties of the non-Fermi liquid regime considered here follow from a straightforward generalization of our results in section \ref{transmfl}. As a result of the completely local form of the Green's function, $G_f$, at temperatures above the crossover scale ($\Omega_f^*$), we can continue to ignore the vertex corrections to the current vertex in Fig. \ref{v2kf}(a). The optical conductivity for the $c$ fermions is then given by Eq.~(\ref{sigmaxx}). It is clear from the form of the spectral function in the non-Fermi liquid regime that the optical conductivity satisfies $\Omega/T$ scaling. At frequencies much higher than the temperature ($\omega\gg T$),
\beq
\sigma'_{xx}(\Omega) \propto \frac{Nv^2~U_f^{4\Delta(q)}}{U_{cf}^2} \frac{1}{\Omega^{4\Delta(q)}},
\eeq
which is determined by the single-particle scattering rate of the $c$ fermions. Following the discussion in section \ref{transmfl}, the dc resistivity is given by,
\beq
\rho_{dc}(T) \propto \frac{U_{cf}^2}{Nv^2 ~U_f^{4\Delta(q)}}~ T^{4\Delta(q)},
\label{rhonfl}
\eeq
which reduces to the marginal Fermi liquid form for $q=4$. However note that just as we discussed in the context of the MFL in Sec.\ref{transmfl} above, in the regime $T\ll\Omega_{cf}^*(q)$ that we consider here, the dc resistivity is always smaller than the Mott-Ioffe-Regel limit. As a result of the $\Omega/T$ scaling that holds in the same window of temperature and frequency scales, the conductivity can be expressed as,
\beq
\sigma_{xx}'(\Omega,T) \propto \frac{1}{\Omega^{4\Delta(q)}}~H\bigg(\frac{\Omega}{T}\bigg),
\eeq
where $H(...)$ is a universal scaling function. 

The dc resistivity in Eq.~(\ref{rhonfl}) displays a strong departure from the ``Planckian'' form, since $\rho_{dc}\sim T^{4\Delta(q)}$ with $4\Delta(1)<1$. However the scattering rate associated with the temperature dependent frequency scale that determines the crossover from the high frequency to the dc behavior, $1/\tau_{\mathrm{opt}}$, saturates the Planckian bound: $1/\tau_{\mathrm{opt}} \sim a k_B T/\hbar$ with $a=O(1)$. 

\subsection{$2K_F$ Singularities}
\label{2kfnfl}

The sharp structure associated with a Fermi-surface in momentum space in a conventional Fermi-liquid leads to a singular $2K_F$ response. A natural question to ask here is if the non-Fermi liquid states considered in this paper have a $2K_F$ response that is different from other known examples of (non-)Fermi liquids \cite{AIM,mross}? In this section, we analyze the modification to the singularity for the non-Fermi liquid regime discussed above and find that it is fixed by the scaling dimension of the fermions, $\Delta(q)$. We assume that the temperature, while higher than the crossover scale $\Omega_f^*$, continues to be much smaller than the Fermi-energy for $c$ Fermions such that effects of thermal smearing can be ignored. 

We are interested in studying the response of the critical Fermi-surface to an external source field that couples to the fermion density (for any orbital $i$) at a $2K_F$ momentum. Naively, we expect there to be a suppression of the $2K_F$ response as a result of the smearing of the Landau quasiparticle due to scattering off the `local', incoherent $f-$electrons. Moreover, we have here a situation where the vertex corrections from the interactions, which are usually important in determining the $2K_F$ response, are only weakly momentum dependent (since $\Pi_f(\q,\omega)$ has a weak dependence on $\q$). 

Let us then consider an external source field that couples to the $2K_F$ fermion density, i.e. a particle at $K_F$ and a hole at $-K_F$,  through the following term in the action (it will be sufficient to consider a pair of antipodal patches for this purpose),
\beq
\delta H = u\sum_i \int d^2\x~d\tau \bigg[c^\dagger_{i,L} c_{i,R} + \tn{H.c.}\bigg],
\label{delH}
\eeq 
where $R$ and $L$ correspond to the two antipodal patches at $\pm K_F$. The $2K_F$ operator is defined as $\rho_{2K_F}(\x,\tau) = \sum_i c^\dagger_{i,L}(\x,\tau) c_{i,R}(\x,\tau)$ and we are interested in the long range behavior of the correlation function $C_{2K_F}(\x,\tau) = \langle \rho_{2K_F}^* (\x,\tau)~\rho_{2K_F}({\vec{0}},0)\rangle$. 

We can obtain the singular structure for the correlation function by scaling, when the low energy physics is scale invariant under the following scaling transformation ($z$ is the dynamical exponent),
\beq
\omega' &=& \omega~b^{z/2},\\
p_x' &=& p_x~b,\\
p_y' &=& p_y~b^{1/2}.
\eeq
Let us suppose that $u' = u~b^{\phi}$ ($\phi=1$ under the above rescaling, but we allow for a general $\phi$ to account for the possible singular renormalization from corrections to be considered below). Then the $2K_F$ operator satisfies,
\beq
\rho'_{2K_F}(\x',\tau') = b^\alpha~ \rho_{2K_F}(\x,\tau),
\eeq
where 
\beq
\alpha = \frac{z+3}{2} - \phi.
\eeq
The fourier transform then satisfies the scaling,
\beq
C_{2K_F}(\p,\omega) = b^{(3+z)/2 - 2\alpha} ~C'_{2K_F}(\p',\omega'),
\eeq
where $\p$ in this case represents the deviation of the full momentum away from $2K_F \hat{x}$, i.e. $p_x~(p_y)$ is the direction perpendicular (parallel) to the Fermi-surface. Then we can immediately write the scaling form,
\beq
C_{2K_F}(\p,\omega) = \frac{1}{\omega^{1 + \frac{3 - 4\alpha}{z}}} ~Y\bigg(\frac{\omega}{|p_y|^z},\frac{p_x}{p_y^2} \bigg).
\eeq 
For a conventional Fermi liquid, $\phi=1$ and $z=2$, which leads to the famous $\sqrt{\omega}$ singularity in the $2K_F$ correlations. For the non-Fermi liquid considered above, {if we ignore vertex corrections (which will be shown to be negligible below)}, $\phi=1$ and $z=1/(2\Delta(q))$, which leads to the singular dependence
\beq
C_{2K_F}(\p,\omega) \sim \omega^{1-2\Delta(q)},
\eeq 
which is what we would naively obtain by computing the density-density response $\sim\int_\k \int_\Omega~G_c(\p+\k,\omega+\Omega)~G_c(\k,\Omega)$ with the above self-energy in Eq.~(\ref{SEnfl}).

\begin{figure}
\begin{center}
\includegraphics[width=0.4\columnwidth]{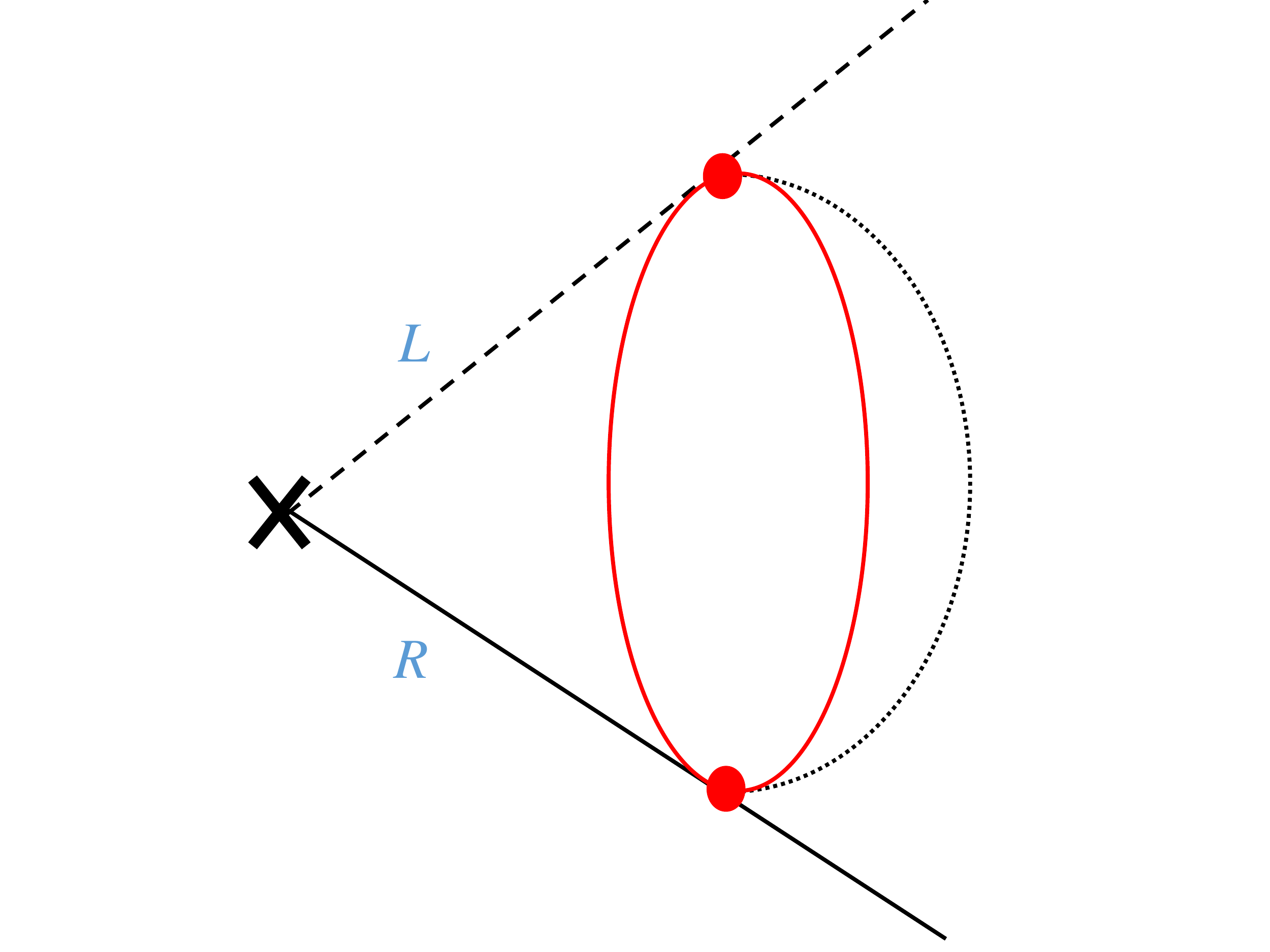}
\end{center}
\caption{One-loop vertex correction to the $2k_F$ operator, $\delta H$ in Eq.~(\ref{delH}) (denoted by cross). The solid and dashed lines correspond to the $c$ Fermion Green's functions for the two antipodal patches $R$ and $L$. Red lines denote $f$ Green's functions and red dots represent the $V_{ijkl}$ vertex.}
\label{2kf}
\end{figure}

Let us now compute the one-loop vertex correction (Fig. \ref{2kf}), which may {\it apriori} change the singular structure. For simplicity, we set all the external momenta and frequencies to zero. The expression for the diagram is then given by,
\beq
\delta u \sim U_{cf}^2 \int_{\k,\Omega}~\frac{\Pi_f(\k,\Omega)}{i\tn{sgn}(\Omega)~|\Omega|^{4\Delta(q)} - \ve_{\k}^+} \frac{1}{i\tn{sgn}(\Omega)~|\Omega|^{4\Delta(q)} - \ve_{\k}^-}, \nonumber\\
\eeq
where $\ve_\k^\pm = \pm v k_x + k_y^2$ and $v$ denote the dispersions and Fermi-velocities near the R/L patches; we have set the curvature to unity. Then,
\beq
\delta u \sim \frac{U_{cf}^2}{v~U_f^{4\Delta(q)}} \int_{k_y,\Omega} \frac{|\Omega|^{4\Delta(q)}}{|\Omega|^{8\Delta(q)} + k_y^4}~ \frac{1}{|\Omega|^{1-4\Delta(q)}}.
\eeq 
The above $k_y$ integral is convergent and leads to,
\beq
\delta u \sim \frac{U_{cf}^2}{v~U_f^{4\Delta(q)}} \int_{\Omega} |\Omega|^{2\Delta(q)-1} \sim |\omega|^{2\Delta(q)}.
\eeq
We may now include the effect of this vertex on the density-density response, in order to compute the correction $\delta C_{2K_F}(\p,\omega)\sim\int_\k\int_\Omega \delta u ~G_c(\p+\k,\omega+\Omega)~G_c(\k,\Omega)\sim \omega$. This is clearly less singular than the result obtained from scaling (or equivalently, the bare correlation function without vertex corrections). This is simple to understand within our model, where the completely local form of the fluctuations associated with the incoherent $f-$fermions leads to scattering at all momenta for the $c-$fermions, and in particular no additional singularities arise as a result of any special scattering across the anti-podal patches. Note, however, that the vertex correction will be important for density correlations near $q \approx 0$, and indeed are needed to obtain the finite non-zero compressibility that we argued characterizes these states.

\subsection{Quantum Oscillations}
\label{qo}

A hallmark of a Fermi liquid with well-defined quasiparticle excitations across the Fermi-surface is the observation of quantum oscillations as a function of an inverse external magnetic field ($B$) in a number of physical observables that depend on the density of states (e.g. magnetization).  In all of the translationally invariant models of non-Fermi liquids considered in this paper, there is a sharply defined Fermi-surface of electrons at $\k=\k_F$ in momentum space, but the electronic quasiparticles are destroyed as a result of coupling to the locally critical degrees of freedom. Here we address the question of whether the non-Fermi liquids considered display quantum oscillations periodic in $1/B$ \cite{YBK95} and if they are different in character from oscillations in Fermi-liquids. The fate of quantum oscillations in the marginal Fermi liquid has been addressed before \cite{wasserman,pelzer_qo,shekhter}.

It is useful to treat the problem of oscillations in three-dimensions, where all of our previously obtained results for the self-consistent solutions to the saddle point equations continue to be true. We focus on the example of the non-Fermi liquid (with $q>4$). We note that strictly speaking, we should work at fixed density and account for  the oscillations of the chemical potential as a function of the magnetic field. However, our calculations below will be done at a fixed chemical potential rather than a fixed density. In three dimensions, this is a justified approximation as an expansion in leading powers of the ratio of the magnetic field to the cross-sectional area of the Fermi-surface. At a fixed density the chemical potential has an oscillatory correction to its value at zero field whose amplitude vanishes linearly in field \cite{shoenberg}. We do not include the effect of such chemical potential oscillations, that are subleading in powers of the ratio described above.

Let us now study the effect of a uniform magnetic field, $B$, along the $z-$direction through its orbital coupling to the $c$ fermions (we assume that there is no orbital coupling to the $U(1)$ charge associated with the $f$ fermions, which is explicitly true when we set $W_f\rightarrow0$). We analyze the structure of the saddle-point equations in the presence of the magnetic field in Appendix \ref{appqo:saddle}. A  key property of the solution is that even at $B \neq 0$ both the $c$ and $f$- self energies are completely local in space. In the NFL regime, the $f$ fermions continue to be described in terms of the $(0+1)$ dimensional SYK model and the self-energy for the $c$ fermions as a result of the coupling to the $f$ fermions can be written as,
\beq
\Sigma_{cf}(i\omega) = \Sigma_{cf}(i\omega,B=0) + \tilde\Sigma_{cf}(i\omega,B\neq0).
\eeq
We study the effect of the first term above (independent of $B$) on all of the oscillatory phenomena; the effects arising from the explicit dependence of $\tilde\Sigma_{cf}$ on $B$ are of higher order in $(\omega_c/\mu_c)$, where $\omega_c=eB/m^*$ is the cyclotron frequency{\footnote{ For simplicity, we assume a spherical Fermi-surface for the $c$ fermions with $\ve_\k = \k^2/(2m^*) - \mu_c$.}}. The magnetic field leads to a singular modification of the kinetic energy of the $c$ fermions into Landau ``bands" in three dimensions that disperse along the direction of the field{\footnote{Unlike Landau levels (LL) in two dimensions.}}.  

The Green's function for the $c$ fermions in the LL basis is given by,
\begin{subequations}
\beq
\label{LLG}
G_c(n,p_z,i\omega_m) &=& \frac{1}{i\omega_m - \epsilon_n(p_z) + \mu_c -\Sigma_{cf}(i\omega_m)},~\tn{where}\\
\epsilon_n(p_z) &=&  \bigg(n+\frac{1}{2}\bigg) \omega_c + \frac{p_z^2}{2m^*}.
\label{enp}
\eeq
\end{subequations}
We are interested in the oscillatory contribution to two quantities: (i) the spectral density of states, and, (ii) the (orbital) magnetization. The oscillatory component of the spectral density of states in the limit of $\omega\rightarrow0$ at a finite $T$ is of the form,
\beq
N_{\tn{osc}}(\omega\rightarrow0,T) = \frac{N(0)}{2\pi} \sum_{k=1}^{\infty} \frac{(-1)^k}{(2k)^{1/2}} ~ \sin\bigg[\frac{2\pi k \mu_c}{\omega_c} - \frac{\pi}{4}\bigg] ~e^{-\frac{2\pi k}{\omega_c} \frac{N(0)U_{cf}^2T^{4\Delta(q)}}{U_f^{4\Delta(q)}}}\sqrt{\frac{\omega_c}{\mu_c}}.
\label{dosqo}
\eeq
($N(0)$ is the density of states of the non-interacting problem in the absence of $B$). Interestingly, we find that the density of states at zero energy has oscillations in $1/B$ even in the absence of quasiparticle excitations, with the period set by the standard cross-sectional area of the critical fermi-surface. The damping of amplitude of the oscillations is determined by the imaginary part of the self-energy, which has an unconventional form compared to the standard fermi-liquids. The details appear in Appendix \ref{app:qodos}. 

Let us now focus on the oscillatory component of the orbital magnetization, $M_{\tn{osc}}$, which is a thermodynamic quantity. It is possible to write down the oscillatory component of the free energy and compute the magnetization by taking appropriate derivatives. Instead, we compute the magnetization in a different manner here by noticing that the dependence on $B$ enters only through the kinetic energy of the $c$ fermions (as already described above). The  magnetization density defined per unit area then is given by,
\beq
M(B) = -\frac{1}{NA} \bigg\langle \frac{\partial H_c}{\partial B}\bigg\rangle,
\eeq
where only the kinetic part of $H_c$ in the presence of magnetic field enters the above expression: $H_c = -\sum_{\r\r'} h_{\r\r'} c^\dagger_{\r} c_{\r'} + \tn{H.c.}$, where after Peierls' substitution  $h_{\r\r'} = t^c_{\r\r'}~e^{iA_{\r\r'}}$ ($A_{\r\r'}\equiv$vector-potential corresponding to uniform $B$ along $z-$direction). In the LL basis, the magnetization is then given by,
\beq
M(B) = \sum_{n,\alpha,p_z} \langle c_{n\alpha,p_z}^\dagger c_{n\alpha,p_z}\rangle \sum_{\r\r'} \phi_{n\alpha}^*(\r) \phi_{n\alpha}(\r') \frac{\partial h_{\r\r'}}{\partial B},
\eeq
where  $n$ labels the LL index, $\alpha$ denotes all of the degenerate states within each LL and $\phi_{n\alpha}(\r)$ is the LL wave function. The latter sum over $\r,\r'$ can be carried out to yield,
\beq
M(B) = \sum_{n,\alpha,p_z} \langle c_{n\alpha,p_z}^\dagger c_{n\alpha,p_z}\rangle \frac{\partial \epsilon_n(p_z)}{\partial B}.
\eeq
In the above equation $\epsilon_n(p_z)$ is as denoted in Eq.~(\ref{enp}). Equivalently, this can be obtained directly by writing the Hamiltonian in the LL basis as,
\beq
H_c = \sum_{n,\alpha,p_z} \epsilon_n(p_z) c_{n\alpha,p_z}^\dagger c_{n\alpha,p_z}.
\eeq
We then have,
\beq
M(B) = \frac{1}{2\pi\beta}\sum_{\omega_m}\sum_n\int_{-\infty}^\infty \frac{dp_z}{2\pi} \frac{(n+1/2)\frac{B}{m^*}}{i\omega_m - (n+1/2)\omega_c + \mu_c - \frac{p_z^2}{2m^*} - \Sigma_{cf}(i\omega_m)}.
\label{mag}
\eeq
Using the Poisson resummation formula, the oscillatory component of the magnetization is then,
\beq
M_{\tn{osc}}(B) = \frac{1}{2\pi\beta}\sum_{\omega_m}\int_{-\infty}^\infty \frac{dp_z}{2\pi} \sum_{k=-\infty}^\infty \int_0^\infty dn \frac{(n+1/2)\frac{B}{m^*}~e^{2\pi i kn}}{i\omega_m - (n+1/2)\omega_c + \mu_c - \frac{p_z^2}{2m^*} - \Sigma_{cf}(i\omega_m)}.
\eeq
After some standard manipulations, details of which appear in Appendix \ref{appqo:mag}, we obtain,
\beq
M_{\tn{osc}}(B) \approx \frac{N(0)}{4\pi} \sum_{k=-\infty}^\infty (-1)^k e^{i\pi/4} \frac{1}{(2k^3)^{1/2}}\bigg(\frac{\mu_c}{m^*}\bigg)\sqrt{\frac{\omega_c}{\mu_c}} e^{2\pi ik\mu_c/\omega_c}{\cal{A}}\bigg(\frac{2\pi k}{\omega_c}\bigg).
\label{mosc}
\eeq
Here, ${\cal{A}}(...)$ is a purely real amplitude for the oscillation of the $k^{\tn{th}}-$harmonic [an explicit expression for the amplitude appears in Eq.~ (\ref{alambda})]. We find that the period of oscillations is determined by the cross-sectional area of the Fermi-surface and remains unaffected by the form of the self-energy. The amplitude, on the other hand, is affected by the non-Fermi liquid form of the self-energy and has a non Lifshitz-Kosevich form {\footnote{Non Lifshitz-Kosevich forms for the amplitude of magnetization oscillations have been obtained in earlier holographic calculations \cite{HartnollQO}.  }}. 

The universal scaling structure for the temperature dependence of the amplitude of the oscillations can be determined to be as follows (see Appendix \ref{appqo:mag})
\beq
{\cal{A}}(\lambda_k) = \bigg(\frac{2\pi k}{\omega_c}\bigg)^{1-\frac{1}{4\Delta(q)}} R(2\pi k x),
\label{ampscal}
\eeq
where $R(x)$ is a scaling function of $x=(U_{cf}^2T^{4\Delta(q)}/W_cU_f^{4\Delta(q)}\omega_c)$ that decays exponentially at large $x$. The scale for damping of the amplitude for any given harmonic at any finite temperature is then given by $T^*\sim \omega_c^{1/4\Delta(q)}$.

\section{General Constraints on Local Criticality}
\label{LC}

Both the incoherent metal regime in the single band model (Sec.~\ref{mod}) and the marginal/non Fermi liquid regime in the two-band model (Sec.~\ref{sec:MFL},\ref{nfl}) display ``local quantum critical'' behavior. 
By that, we mean that the temporal correlation functions decay as power laws (up to a correlation time $\xi_\tau \sim 1/T$), whereas the spatial correlations decay exponentially over a temperature-independent length-scale of a few lattice constants.\footnote{Note that this is different from the scenario where the system has long range spatial correlations in addition to the power law correlations in time, but only the frequency dependent correlations have anomalous dimensions~\cite{Si2} - a situation also referred to as ``local quantum criticality.'' This behavior has been invoked in the context of heavy Fermion quantum criticality~\cite{Si1}. Here, we refer only to the situation where the spatial correlations are strictly local.} In both of our models, the local quantum critical regime is unstable at sufficiently low temperatures: below a certain ``coherence temperature,'' a crossover to a different, more conventional behavior occurs. This is consistent with the fact that in both models, the entropy in the local quantum critical regime extrapolates to a non-zero value in the limit $T\rightarrow 0$, violating the third law of thermodynamics. Instead, the one- and two-band models cross over to a Fermi liquid regime below the energy scales $\Omega_c^*$ and $\Omega_f^*$, respectively, and relieve the excess entropy.

This raises the question whether, in generic lattice models with a finite number $N$ of degrees of freedom per unit cell, local quantum critical behavior can be stable down to $T=0$ (either as a quantum phase or at a quantum critical point). 
On scaling grounds, it has been argued that local quantum criticality must be accompanied by a finite entropy density in the limit $T\rightarrow 0$~\cite{Kristan2011}, although some caveats have been pointed out~\cite{Hartnoll2012}.\footnote{For a hyperscaling-violating theory in $d$ spatial dimensions, the entropy density scales as $S\sim T^{(d-\theta)/z}$, where $\theta$ is the hyperscaling violation exponent. Naively, $z\rightarrow \infty$ implies a finite ground state entropy density.  Ref.~\cite{Hartnoll2012} pointed out that this can be avoided if $\theta \rightarrow -\infty$.}
Here, we argue that in any translationally invariant lattice model with finite $N$, local quantum criticality is not possible down to arbitrarily low temperature. 
More generally, systems with a weaker form of quantum criticality, where the correlation length diverges sub-polynomially in $1/T$ (as in Refs.~\cite{aji}),
must have an entropy that scales as a power of the linear dimension $L$ in the limit $T\rightarrow 0$. We expect this large residual entropy to lead to an instability at sufficiently low temperature, resulting in a lower entropy state.

 First, consider a translationally invariant system where the correlation time $\xi_\tau \sim 1/T$, while the correlation length $\xi$ is independent of $T$. 
In a finite cluster of linear size $L = \alpha \xi$, % at a finite, low temperature $T$,
the temporal correlation functions of local operators approach their values in the thermodynamic limit for sufficiently large $\alpha$. Hence, the temporal correlations decay as a power law up to times of the order of $\xi_\tau$. Since the system is finite, $\xi_\tau$ cannot exceed the inverse of the mean level spacing near the ground state, $\xi_\tau \leq 1/\delta(L)$; i.e., $\delta(L) \leq T$. 
Therefore, in a generic system with a finite $N$, the local quantum critical behavior cannot persist to arbitrarily low $T$, otherwise $\delta(L)\rightarrow 0$. 

Next, we consider systems with a weaker version of local quantum criticality in which the correlation time, $\xi_\tau$, grows faster than polynomially as a function of the correlation length, $\xi$. The dynamical critical exponent, $z$ (defined via $\xi_\tau \sim \xi^z$) is still infinite. 
Repeating the argument above for a finite cluster of linear size $L = \alpha \xi$ at $T=0$,  

we get that $\xi_\tau$ cannot exceed the inverse of the level spacing near the ground state, $\xi_\tau \leq 1/\delta(L)$.
Hence, $\delta(L)$ must decrease faster than polynomially in $L$. In contrast, the level spacing near the ground state in generic many-body systems with local interactions is expected to depend polynomially on the system size \cite{RMT}. 

The anomalously small level spacing near the ground state has consequences for the entropy in the limit $T\rightarrow 0$. In the microcanonical ensemble, the low-temperature entropy scales as $S(T\rightarrow 0) \sim \log[\Delta E /\delta(L)]$, where $\Delta E$ is a sub-extensive energy shell. As a concrete example, suppose that $\xi \sim \log(\xi_\tau)$, as proposed in Refs.~\cite{chakravarty,aji} for certain quantum critical points. In this case, following the considerations above, $\delta(L) \leq e^{-L / \alpha}$.  
Therefore, we find that $S(T\rightarrow 0) \sim L/\alpha$. Even though such behavior does not violate the third law of thermodynamics in spatial dimension $d>1$, we do not expect it to hold down to $T=0$. The high density of low energy states generically leads to an instability that lifts the near-degeneracy of the ground state. Similarly, if the correlation time scales as $\xi_\tau \sim [\log(\xi)]^\gamma$, we get that $S(T\rightarrow 0) \sim L^{1/\gamma}$. 

We note some interesting exceptions to this rule. The disorder-averaged correlations of disordered systems at infinite randomness fixed points~\cite{DSF} are known to display $z=\infty$ behavior. This behavior comes from rare regions where the correlation time is much longer than the typical one. However, we do not expect such rare region effects in generic translationally invariant systems. Another exception is found in certain three-dimensional topologically ordered states, called ``fracton states''~\cite{haah,fracton1,fracton2}, that have $S(T\rightarrow 0) \sim L$ without any fine tuning. However, this property probably does not lead to quantum critical behavior of local correlation functions, since local operators have vanishingly small matrix elements between the topologically distinct near-degenerate states that are responsible for the low-temperature entropy. 

\section{Discussion}
\label{sec:discussion}
In this work, we have defined a class of translationally invariant models that can be solved in the large $N$ limit. Even though the ground states of these models are conventional (although strongly renormalized) Fermi liquids, they exhibit a crossover at an intermediate energy scale - which can be parametrically smaller than the microscopic coupling constants - into a non-Fermi liquid regime. This regime is characterized by local quantum critical scaling of certain correlation functions - i.e., the correlation time diverges as $\xi_\tau \sim 1/T$, while the correlation length is nearly temperature-independent. 

Interestingly, many of the properties of the non-Fermi liquid regimes are reminiscent of those seen in different quantum materials. In the one-band model of Sec.~\ref{mod}, the resistivity grows linearly with temperature, and does not saturate at the Mott-Ioffe-Regel limit. The two-band version of the model (Sec.~\ref{sec:MFL},\ref{nfl}) exhibits a regime where the light band has a critical Fermi surface - either a marginal Fermi liquid or a non-Fermi liquid, depending on the precise nature of the interactions between the heavy and light bands. The  resistivity grows as $\rho \propto T$ in the MFL and as $\rho\propto T^{4/q}$ with an exponent $q>4$ in the NFL.

In this section, we will put these results in the context of previous work, and discuss their possible implications either to more generic models (in particular, ones that do not involve the limit of a large number of degrees of freedom per unit cell), as well as to strongly correlated materials.

\subsection{Relation to other work} 

Several models composed of lattices of coupled SYK dots have been studied recently~\cite{Gu17,SS17, Yao, Balents, mcgreevy,Zhang17,shenoy}. 
Of these, the one-band model we introduce here is closest to the model solved by Song, Jian, and Balents~\cite{Balents}, who studied a lattice of SYK dots coupled by single-particle hopping. 
The main difference between this work and the present one is that the model studied here is translationally invariant, whereas the model of Ref.~\cite{Balents} is strongly disordered - both the interactions and the hopping matrix elements vary from site to site. The translational invariance allows us to address the properties of the Fermi surface in the low-temperature Fermi liquid regime. In the strong coupling limit, we find a strongly renormalized Fermi-liquid with a momentum independent self-energy. Interestingly, however, the properties of the high-temperature ($T\gg W_c^2/U_c$) LICM phase are similar in the two models. This is a consequence of the fact that, in this regime, the correlations become short-range in space; hence, the presence of translational invariance does not modify the properties of the system in a fundamental way. For example, even with translational symmetry, there is no remnant of a Fermi surface, and the resistivity is linear in temperature in both cases. In our model, the resistivity scales as $T^2$ in the low temperature regime ($T\ll W_c^2/U_c$), as expected in a Fermi liquid; in contrast, in the model studied in Ref.~\cite{Balents}, we expect the resistivity to saturate to a temperature-independent constant, due to the presence of strong disorder. 

Earlier work~\cite{Parcollet1} considered a model of localized moments with long-ranged, random in sign interactions, coupled to a band of itinerant electrons. Even this model is very different from the two-band model considered here - in particular, our model is translationally invariant, and has only local interactions - the properties of the intermediate temperature ``marginal Fermi liquid'' regime realized in both models are similar. Hence, our model demonstrates that this regime - as well as the non-Fermi liquid regime discussed in Sec.~\ref{nfl} - can be realized even in the clean limit and can host a critical Fermi-surface of electrons. 

Finally, our results are -- not surprisingly\footnote{It is the possibility of a simple holographic description of the $0+1$-D  SYK model that has partly contributed to the tremendous recent interest in this model. Our two-band model is roughly similar in spirit to the ``semi-holographic" theory in Ref.~\cite{SH11}, although of course the details are very different.} --  similar to those found in strongly coupled theories that can be solved using holographic dualities~\cite{SH11,Liu1,Liu2,tong}. These models give locally quantum critical behavior associated with a non-vanishing entropy in the limit $T\rightarrow 0$. Upon coupling the locally quantum critical degrees of freedom to itinerant fermions, marginal Fermi liquid and non-Fermi liquid states can result (see also Ref.~\cite{mcgreevy}). As in the case of lattices of SYK dots, these models involve taking the limit of a large number of local degrees of freedom. Moreover, as in our model, the locally quantum critical regime is unstable at low energies to the formation of either long-range ordered states or a heavy Fermi liquid.

\subsection{Bounds on transport}

It is interesting to discuss our results in the context of possible ``universal bounds'' on transport coefficients. It has been proposed ~\cite{QPT,Zaanen04} that the relaxation time (or ``dephasing time''~\cite{QPT}) is bounded by the Planckian time, $1/\tau_P \le a k_B T/\hbar$, where $a$ is an unknown constant of order unity. Following this idea, a number of bounds on transport coefficients have been proposed~\cite{son,Hartnoll15,Blake16,Hartnoll17}. An interesting conjectured bound on the heat and charge diffusion constants in Ref. \cite{Blake16} involved the many-body `chaotic' properties of the system (see Appendix \ref{chaos}). However explicit calculations \cite{Lucas1,Lucas2} have demonstrated violation of such bounds in different settings (at present there is no known counterexample to the bound proposed in Ref. \cite{Hartnoll17}). Empirically, the transport lifetime of many metals where the resistivity is linear in $T$ has been found to be not far from $\hbar/(k_B T)$~\cite{Bruin13}.

As we already described in the introduction, there is no unique choice of a transport scattering rate that may have an associated universal bound. In order to compare with the procedure adopted in Ref.~\cite{Bruin13}, where the scattering rate was extracted by fitting the transport data to a Drude-like form, let us focus on the case of the MFL and NFL states discussed in sections~\ref{sec:MFL},\ref{nfl} above. For the model in section~\ref{sec:MFL}, we may extract the renormalized mass $m^*/m \sim (\nu_0 U_{cf}^2/U_f) \ln(1/T_{\tn{coh}})$ from the low temperature FL regime (as measured in quantum oscillations) below $T_{\tn{coh}}
\sim \Omega_f^*$. Using $\sigma=ne^2 \tau_{\tn{dc}}/m^*$ to define $\tau_{\tn{dc}}$ in the MFL regime at high temperatures leads to $1/\tau_{\tn{dc}} \sim T/\ln(1/T_{\tn{coh}}) \ll T$, which satisfies a Planckian bound for the particular choice of the dc scattering rate. Note, however, that the resistivity in the MFL is $\rho \propto T$ with no logarithmic corrections, i.e., it is not simply proportional to $1/\tau_{\tn{dc}}$. A similar procedure adapted to the NFL regime of the two band model with $q>4$ in section~\ref{nfl} leads to a lifetime with a strongly non-Planckian form,
\beq
\frac{1}{\tau_{\tn{dc}}} \sim \frac{T^{4\Delta(q)}}{T_{\tn{coh}}^{4\Delta(q) - 1}}.
\eeq
However it is still true that $1/\tau_{\tn{dc}} < T$ (since in the NFL regime, $T > T_{\tn{coh}}$).

It is interesting to point out that in all the cases studied here, the ``optical scattering rate,'' $1/\tau_{\tn{opt}}$, defined as the frequency scale at which the high frequency optical conductivity approaches its dc value, satisfies $1/\tau_{\tn{opt}} < a k_B T/\hbar$ with $a=O(1)$. In the incoherent regime of the one-band model and in the two-band non-Fermi liquid, $1/\tau_{\tn{opt}} \sim T$; in the two-band MFL, $1/\tau_{\tn{opt}} \sim T/\ln^2(1/T)$ [see Eq. (\ref{toptmfl})]. Thus, $1/\tau_{\mathrm{opt}}$ satisfies a Planckian-type bound, but the temperature dependence of the dc resistivity does not necessarily follow that of $1/\tau_{\mathrm{opt}}$.

\subsection{Implications for generic models}

Clearly, the models (\ref{hc1},\ref{Htot}) are fine-tuned in many ways. In particular, the number of local degrees of freedom, $N$, is taken to be large, and the interactions ${U_{ijk \el}}$ are taken to be independent, random variables whose average is precisely zero. It is thus important to ask which of the properties of the solution are peculiar to these models, and which are expected to hold more generically, even in less fine-tuned models with a finite number of degrees of freedom per unit cell. 

Here, we will discuss possible implications of our results for generic models (with a finite number of degrees of freedom per unit cell). In particular, we describe how local quantum critical behavior may arise in an intermediate temperature window in systems where the coherence scale (e.g., the effective Fermi energy or the Bose condensation temperature) is much smaller than the microscopic scale. We then formulate a conjecture for an effective ``coarse grained'' description of non-Fermi liquid states in generic models, inspired by the models constructed in this work, based on notions of many-body quantum chaos.

\subsubsection{Local quantum criticality in generic models}

The local quantum critical behavior found in some regimes of our models is unlikely to be stable in generic models down to zero temperature. A diverging correlation time without a corresponding diverging correlation length clearly requires an infinite number of local degrees of freedom. Moreover, as we have argued in Sec.~\ref{LC}, even a correlation length that diverges sub-polynomially with the correlation time implies a divergent (although not necessarily macroscopic) entropy in the $T\rightarrow 0$ limit. Hence an instability is likely to occur at a sufficiently low temperature. 

Nevertheless, we speculate that local quantum critical behavior (with a correlation time that scales as $\hbar/T$ and a nearly temperature-independent correlation length) can appear generically in strongly correlated metals, over a finite but broad temperature window. To see how such behavior can arise, consider a model with a metallic (Fermi-liquid) ground state. At low temperature, the single-particle lifetime scales as $1/\tau \sim T^2/\Omega^*$, where $\Omega^*$ is a non-universal ``coherence scale'' that depends on the strength and form of the inter-particle interactions. If the interactions are sufficiently strong, $\Omega^*$ may be much smaller than the microscopic coupling constants of the model, such as the hopping or the interaction strength. We expect that $\Omega^* \sim E_F^* \sim v_F^* k_F$, where $E_F^*$, $v_F^*$ are the renormalized Fermi energy and Fermi velocity, respectively\footnote{This is certainly not universally the case; for example, in the vicinity of a metallic quantum critical point with a ${\mathbf Q}=0$ order parameter, $\Omega^*$ and $v_F^* k_F$ are parametrically different. Here, we are assuming that there is a single energy scale $\Omega^*$, which is small not because of the proximity to a quantum critical point, but due to strong microscopic interactions.}. This is indeed the case in the one-band model of Sec.~\ref{mod}. In the Fermi liquid regime, temporal correlations decay exponentially over a timescale $\xi_\tau \sim 1/T$.  Spatial correlations decay over the thermal length, $\xi_T \sim v_F^*/T$. Thus, crudely extrapolating to $T \sim \Omega^*$, where the Fermi liquid behavior starts to break down, we get that the correlation length at the crossover temperature becomes $\xi_T \sim v_F^*/\Omega^* \sim \lambda_F$, implying that the correlation length reaches the microscopic length scale set by the Fermi wavelength. (In a typical metal, this is of the same order of magnitude as the lattice spacing.) On the other hand, the correlation time at this temperature is of the order of $1/T$. If the renormalized Fermi energy is much smaller than the microscopic energy scales (set by the interaction strength and the hopping), then at $T \sim \Omega^*$, the correlations extend over a time which is much longer than the inverse of the ``bare'' Fermi energy.\footnote{A classic example of a Fermi system with a low coherence temperature is the normal state of $^3$He; the renormalized Fermi energy is significantly smaller than the bare one, due to the strong inter-particle interactions. The temperature window above the renormalized Fermi energy, where Fermi liquid behavior breaks down but the system is still quantum mechanical, has been termed a ``semi-quantum liquid''~\cite{Andreev1979}.}

What happens at temperatures higher than $\Omega^*$? The spatial correlations already decay over a microscopic length scale at $T\sim \Omega^*$, so it is natural to assume that the correlation length is not strongly temperature dependent in this regime. We argued above that the correlation time at $T\sim \Omega^*$ is $\xi_\tau \sim 1/T$. Further, we assume that the ``scrambling rate'' (discussed in Appendix~\ref{chaos}) at this temperature is close to saturating the bound~\cite{Maldacena2016}, $\lambda_L \sim T$. Therefore, one can guess that the bound remains nearly saturated at $T>\Omega^*$. 
The natural appearance of a Planckian time scale implies that the correlation time $\xi_\tau$ remains of the order of $1/T$ even above $\Omega^*$.
Hence, if the temperature window between the renormalized Fermi energy and the bare one can be made very large, then we expect this window to exhibit some form of ``local quantum criticality.''\footnote{Interestingly, the scenario discussed here is similar to the behavior found in the ``spin-incoherent Luttinger liquid'' regime~\cite{sill}; above the spin coherence temperature, the single-particle Green's function decays in space over a length scale set by the inter-particle spacing, but displays power-law behavior in time, up to $\tau \sim 1/T$. However, in this case, the spatial correlations of other operators - such as the density - still decay as a power law. In contrast, we are assuming that not only the single particle correlation functions become short ranged at $T\sim\Omega^*$, but {\it all} correlation functions do.}

\newpage

\subsubsection{Towards a ``coarse-grained" description of non-Fermi liquid behavior in correlated materials}
\label{conj}
As outlined in the introduction, there is a zoo of materials that display non-Fermi liquid behavior, in terms of their single-particle properties and transport, over a broad range of temperatures. However, even amongst all of these materials there is a varying degree to which the non-Fermi liquid behavior persists down to the lowest temperatures. A general observation across the wide variety of systems displaying non-Fermi liquid properties are as follows:
\begin{enumerate}
\item In many correlated metals, the dc resistivity is often linear in temperature{\footnote{ Other power-laws have also been reported, e.g. in some families of the ruthenates \cite{allen}.}}, i.e. $\rho_{\tn{dc}}\sim T$, and persists over a broad intermediate range of temperatures with a temperature independent slope. Moreover, it shows no sign of saturation and exceeds the Mott-Ioffe-Regal limit. 
\item In a number of materials where the above is true, there is a low {\it coherence scale} below which there is a departure from the non-Fermi liquid behavior and a crossover to more conventional Fermi liquid type behavior (and possibly to other ordered phases). Moreover, the extrapolated zero temperature entropy from the finite temperature non-Fermi liquid regime is finite and has been reported in certain members of the  ruthenates family \cite{allen} and in the cobaltates \cite{bruhwiler}. This excess entropy is relieved below the coherence scale associated with the low temperature Fermi liquid.

There are a number of outliers to the above description, most prominent amongst them being the optimally doped cuprates and certain quantum critical heavy-Fermion materials, where the non-Fermi liquid behavior observed at intermediate temperatures persists down to the lowest temperatures without any changes or characteristic crossovers. Similarly, the extrapolated zero temperature entropy in the non-Fermi liquid regime is zero (see e.g. Ref. \cite{loram} for the cuprates).

\item The intermediate scale behavior is remarkably similar in a wide variety of these systems, in spite of the microscopic details being totally distinct. This is particularly surprising, since it appears that there is an emergent universal behavior and the details of the microscopic physics are somehow not important. However, e.g. the coefficient of the $T-$linear transport scattering rate can generically be different and dependent on the underlying details. 
\end{enumerate}

The above experimental observations pose an interesting theoretical challenge. In particular the apparent universality of the phenomena suggests that the explanation does not rely too much on the precise microscopic details of any single system but instead is generic to strong correlations between the electrons at the lattice scale.
The theoretical models studied by us in this paper are consistent with a number of these empirical observations. Is it then possible to draw some general lessons from this exercise in order to bridge the gap between a realistic description of materials and the solvable models considered by us? 

Below we will consider the possibility of  a coarse-grained description over scales much longer than any microscopic scale in the problem with a few key assumptions, that allows us to reproduce the features described above. We propose  one possible route that allows us to give such a coarse-grained description of non-Fermi liquid metals in a general setting below. This will allow us to  place the specific models studied in this paper in context within a conceptual framework that applies to generic strongly correlated materials.

The apparent universality of intermediate scale non-Fermi liquid physics in diverse correlated systems naturally leads to the possibility that  there is a universal coarse-grained description.  After all if the macroscopic behavior is universal it makes sense that the universality has set in  at some finite length/time scale  large compared to microscopic scales. 
This length/time scale will itself be non-universally related to the microscopic scales but the subsequent behavior at even longer scales will be universal. There will thus be a universal coarse grained description (much like in hydrodynamics or other theories of universal macroscopic phenomena). We will use the notion of `many-body' quantum chaos to formulate our conjectures below (see Appendix \ref{chaos} for a brief exposition to the subject).

\begin{itemize}
\item {\bf Conjecture 1 ~ (C1)---} {\it For systems that display non-Fermi liquid behavior over a wide range of temperatures above a low crossover scale $(\Omega^*)$, there is an intermediate emergent lengthscale $\ell$, with $a\ll \ell\ll L$ ($a\equiv$lattice spacing and $L\equiv$system size), such that a sub-system defined within a region of size $\ell$ is maximally chaotic. The entire system may or may not be maximally chaotic globally (on scales $\sim L$).}

\item {\bf Conjecture 2 ~ (C2)---} {\it For a patch of size $\ell$ the assumption of maximal chaos severely restricts the structure of general $n$-point correlators, i.e, it restricts them to a set of universality classes of possibilities}. 
\end{itemize} 

Let us state the first conjecture a bit more sharply. 
Consider the squared commutator for generic local operators, $W$ and $V$, 
\beq
\C(t,\r) = \langle [V(\r,t),~W(\vec{0},0)]^2 \rangle_\beta \sim \epsilon ~e^{\lambda_L t}, 
\eeq
where $\epsilon$ depends on $\r$. The statement of Conjecture C1 is that for ``normal" non-fermi liquid systems, there is a length scale $\ell \gg a$ ($a$ = microscopic length scale) such that for $ a  \ll |\r| \ll \ell$, and  for times $\ell/v_B \gg   t \gg |\r|/v_B$  the Lyapunov exponent  $\lambda_L = 2\pi T$ thereby saturating the chaos bound.  These time scales are long enough for two local operators at $\x, \x'$ within a patch to mix but short enough that information has not moved between patches. On the other hand, for $|\r|\gg\ell$, the system need not be maximally chaotic with $\lambda_L\leq 2\pi T$. 

Conjecture C2 simply says that the physics of a maximally chaotic bubble is restricted to some universality classes. 

A coarse grained description of the system would then consist of ``islands" of typical size $\ell$ that are maximally chaotic, which are coupled to each other by generic hopping and interaction terms. 

Why might we expect these conjectures to be true? Let us start with C1. We have already noted that it is natural that there exists a long length scale $\ell$ at which universality first emerges in a ``normal" NFL system. Sufficiently complex and generic strong local interactions may make it  natural that the dynamics is maximally chaotic at these length scales (with no guarantee of course that maximal chaos persists out to macroscopic scales).  We regard this as roughly analogous to the assumption of molecular chaos in the kinetic theory of gases.   

As for Conjecture C2, given the existence of a bound on the Lyapunov exponent it is again natural that systems that saturate the bound are very special and have universal properties. 

Further inspiration for these conjectures comes from current ideas on strongly coupled continuum quantum field theories and their relationship to quantum black holes.  Consider a UV  field theory with a conserved global $U(1)$ symmetry that is sufficiently strongly coupled that it has  a classical gravity dual.  We assume the theory is at a non-zero density of the global $U(1)$ charge. This UV theory will flow under the RG to some IR behavior that in general will describe different physics. As the temperature is decreased there will be a change from a  regime controlled by the strongly coupled UV theory to whatever IR theory emerges under the RG flow. 
In the high temperature regime, in the gravity description of the UV theory we should include a charged black hole.  It is well known that this black hole has a residual zero temperature entropy.  Thus the corresponding high temperature behavior of the boundary quantum field theory is IR-incomplete and has an extrapolated ground state entropy. Now, it is believed that black holes are the ``fastest scramblers" \cite{sekino}, {\it i.e} they saturate the chaos bound. Thus in the high-$T$ regime the quantum field theory we are considering will satisfy the chaos bound. However there is no guarantee that this will continue to be the case as the temperature is decreased. The restricted behavior  of maximally chaotic systems can then be plausibly related to the different universality classes of systems captured  holographically by charged black holes. 

This situation mimics the situation we envisage for generic, complex, strongly coupled lattice models. Of course the presence of the lattice (and the concomitant finite number of degrees of freedom/unit cell) requires that maximal chaos can only develop on some length scale much bigger than $a$. 

We of course leave for the future  explorations of these conjectures and their development into a useful coarse grained description of non-fermi liquids.  Here they provide a conceptual context within which we can place the solvable models studied in this paper. Each SYK island is a specific example of a maximally chaotic system. Thus we can view our models as a toy example of a macroscopic system made out of coupling maximally chaotic bubbles.
We note however that a future development of a universal coarse-grained NFL description will need to be more refined than simply modeling each bubble by an SYK island \footnote{It may be tempting to contemplate that the large number  - $O[(\ell/a)^d]$ - of degrees of freedom within each bubble provides the large-$N$ necessary to obtain the solvable SYK island as a model for the bubble. However we believe this is incorrect and it is too naive to simply expect exactly SYK-like physics to emerge at the scale $\ell$. Note that though the charged SYK model  strictly speaking only has $U(1)$ global internal symmetry it has a statistical $U(N)$ internal symmetry after the averaging over realizations. This will not be the case in generic models. One manifestation of the $U(N)$ symmetry is the emergence -  in the low-$T$ Fermi liquid phase - of $N$ degenerate Fermi surfaces of electrons. We certainly do not expect this to happen in generic models even if they develop intermediate scale maximal chaos. }. The refinement will need to include spatial locality within each bubble. Further it will need to include microscopic lattice symmetries as effective internal symmetries at scale $\ell$. Finally  it will have to incorporate the right microscopic Luttinger/Lieb-Schultz-Mattis constraints  involving both charge conservation and the microscopic lattice symmetries. Despite these deficiencies  we find it encouraging that the toy model of coupled SYK islands leads to behavior reminiscent of experiments.

\section{Acknowledgements} 
We thank E. Altman, R. Gopakumar, D. Jafferis, A. Nahum, S. Sachdev, and B. Swingle for discussions. DC is supported by a postdoctoral fellowship from the Gordon and Betty Moore Foundation, under the EPiQS initiative, Grant GBMF-4303 at MIT. DC acknowledges the hospitality of the Aspen Center for Physics, which is supported by NSF grant PHY-1607611. DC, EB, and TS acknowledge the hospitality of KITP at UCSB, where this work was initiated, which is supported by NSF grant PHY-1125915. TS is supported by a US Department
of Energy grant DE-SC0008739, and in part by a
Simons Investigator award from the Simons Foundation.

\newpage

{\it Note added:} As this paper was being completed for submission, we became aware of a related work \cite{SSmagneto} that studies {\it disordered} higher dimensional generalizations of the SYK model. Our point of strongest overlap is in the discussion of the two-band model where both constructions find a marginal Fermi liquid (in addition, we also obtain a critical Fermi surface in this example). In Ref.~\cite{SSmagneto}, the authors analyze the magnetotransport properties of such a disordered metallic regime; we study the fate of such a critical Fermi surface under the effect of a magnetic field, that gives rise to quantum oscillations even in the absence of quasiparticles. 

\appendix

\section{Model for translationally invariant random matrix (SYK$_2$) with uniform hoppings}
\label{syk2}
It is instructive to consider a version of the single-band model, Eq.~(\ref{hc}), where the intra-site Hamiltonian is quadratic in the fermion operators. The Hamiltonian is given by
\begin{equation}
    H_{c,2} = -\sum_{\r,\r'} \sum_{\el} t^c_{\r,\r'} c_{\r \el}^\dagger c_{\r'\el} + \frac{1}{N^{1/2}}\sum_{\r}\sum_{ij} J^c_{ij} c^\dagger_{\r i}c_{\r j}.
    \label{hc2}
\end{equation}
Here, $J^c_{ij}$ are random, site-independent coupling constants that satisfy $\overline{J^c_{ij}}=0$, $\overline{(J^c_{ij})^2}=J^2$. This can be viewed as a generalization of the Hamiltonian (\ref{hc}) where every site is an SYK$_2$ model. The self-consistent equation for the fermion self-energy is written as
\begin{equation}
    \Sigma(\k,i \omega) = J^2 G(\k,i \omega) = \frac{J^2}{i\omega - \varepsilon_\k - \Sigma(\k,i \omega)}.
    \label{eq:sig2}
\end{equation}
Solving this equation, and requiring that the self-energy vanishes for $J=0$, we get
\begin{equation}
    \Sigma(\k,i \omega) = \frac{i\omega - \varepsilon_\k}{2} + \mathrm{sgn}(\varepsilon_\k) \sqrt{\left(\frac{i\omega - \varepsilon_\k}{2}\right)^2 - J^2}.
    \label{eq:sig22}
\end{equation}
Analytically continuing to real times, the spectral function is given by
\begin{align}
A(\k,\omega) & = -\frac{1}{\pi} \mathrm{Im} G(\k,\omega + i 0^+) = \begin{cases}
\frac{1}{\pi J^{2}}\sqrt{J^{2}-\left(\frac{\omega-\varepsilon_{k}}{2}\right)^{2}} & ,\,\,\,\left|\omega-\varepsilon_{k}\right|\le2J,\\
0 & ,\,\,\,\left|\omega-\varepsilon_{k}\right|>2J.
\end{cases}
\end{align}
$A(\k,\omega)$ has the familiar Wigner semi-circle law centered at $\varepsilon_\k$. Note that, unlike the case of quartic interactions, the Green's function does not display Fermi liquid-like behavior, even at the lowest frequencies; e.g., there is no quasi-particle pole at the Fermi surface. This is a result of the averaging over realizations of $J^c_{ij}$. For any given realization, Eq.~(\ref{hc2}) describes free fermions with no disorder, and the spectral function at every momentum is a set of $N$ $\delta-$functions. The spectral function becomes continuous after averaging, since the band energies of every realization are different. This is unlike the interacting case, where we expect no $\delta-$function singularity in the spectral function except at the Fermi surface, even within a single realization. 

\begin{figure}
\begin{center}
\includegraphics[width=0.6\columnwidth]{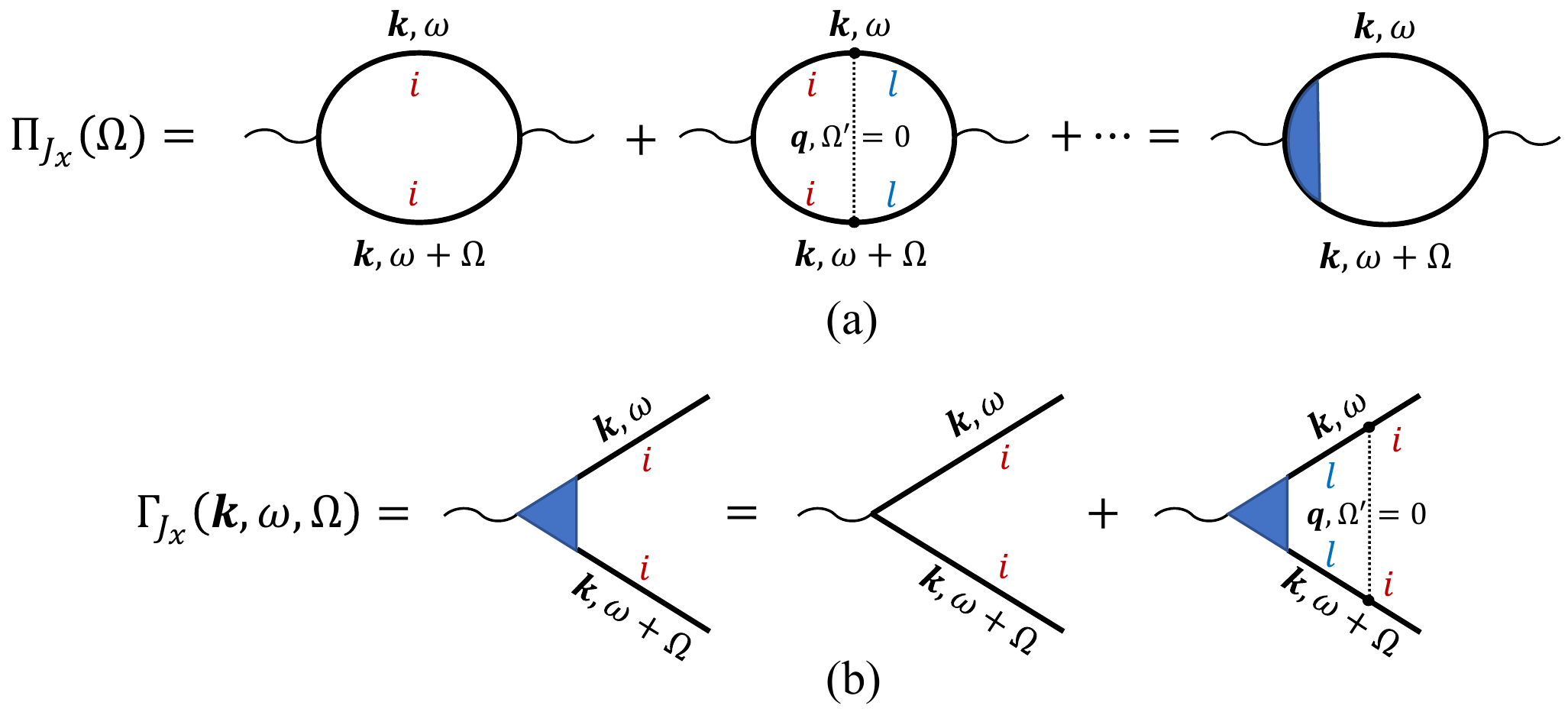}
\end{center}
\caption{Diagrams for the computation of $\sigma_{xx}(\Omega)$ in the $q=2$ model of Eq.~(\ref{hc2}).}
\label{fig:q2}
\end{figure}

\begin{figure}
\begin{center}
\includegraphics[width=0.35\columnwidth]{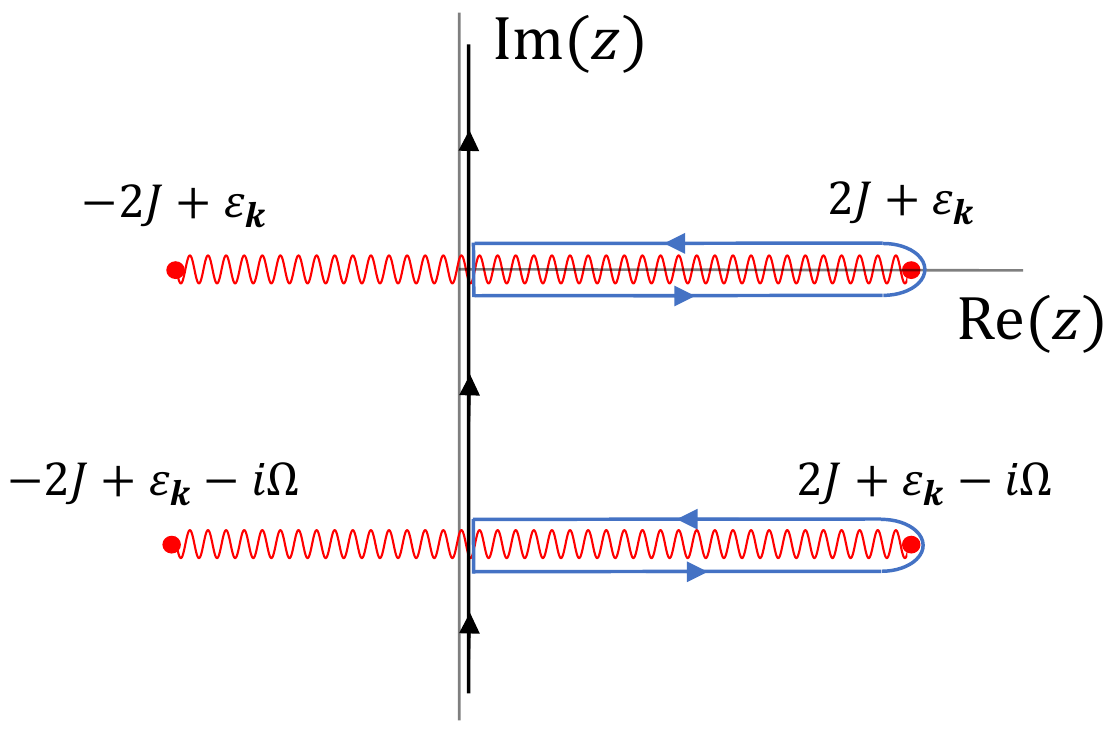}
\end{center}
\caption{Contour for evaluating the integral in Eq.~(\ref{Pi2}). The original integral is along the contour $z = i\omega$. The contour can be deformed into the two closed contours shown in blue above, going around the branch cuts of the integrand.}
\label{fig:branch}
\end{figure}

Since every realization of Eq.~(\ref{hc2}) is a free electron model with translational invariance, we expect the real part of the frequency-dependent conductivity to contain a Drude-like contribution, $\sigma'_{xx,\mathrm{Drude}}(\omega) = D\delta(\omega)$, where $D$ is a temperature-dependent Drude weight. Moreover, since the current operator is diagonal in orbital space, $\sigma'_{xx}(\omega) = \sigma'_{xx,\mathrm{Drude}}(\omega)$ for {\it all} realizations of $J^c_{ij}$; i.e., there is no ``regular'' background in the optical conductivity. Let us demonstrate this for the zero-temperature case. The real part of the conductivity at non-zero frequency is given by $\sigma'_{xx}(\Omega) = \mathrm{Im} \Pi^{\mathrm{ret}}_{J_x}(\Omega)/\Omega$, where $\Pi^{\mathrm{ret}}_{J_x}(\omega)$ is the retarded correlation function of the $x$ component of the current. To leading order in $1/N$, $\sigma_{xx}(\Omega)$ is given by the sum over the set of ladder diagrams shown in Fig.~\ref{fig:q2}(a). To compute this sum, it is useful to first solve the self-consistent equation for the current vertex $\Gamma_{J_x}(i \omega, \Omega)$, shown in Fig.~\ref{fig:q2}(b):
\begin{equation}
    \Gamma_{J_x}(\k,i \omega, i \Omega) = v^x_\k + J^2 \Gamma_{J_x}(\k,i \omega,i \Omega) G(\k,i \omega) G(\k,i \omega + i\Omega),
\end{equation}
where $v^x_\k = \partial \varepsilon_\k / \partial k_x$ is the band dispersion along $x$. Solving this equation and inserting $\Gamma_{J_x}$  into the bubble in the last equality in Fig.~\ref{fig:q2}(a), we get that that the Matsubara frequency $\Pi_{J_x}(i \Omega)$ is given by
\begin{equation}\label{Pi2}
    \Pi_{J_x}(i \Omega) = \int_{\k,\omega} \frac{(v^x_\k)^2}{1 - J^2 G(\k,i \omega) G(\k,i \omega+i \Omega)} G(\k,i \omega) G(\k,i \omega + i\Omega). 
\end{equation}
Evaluating the $\omega$ integral gives that $\Pi^{\mathrm{ret}}_{J_x}(i \Omega) = 0$ for any $\Omega \ne 0$. 
To show this, we analytically continue the integrand to the complex plane, $i\omega \rightarrow z$. A little bit of algebra shows that the denominator of the integrand in Eq.~(\ref{Pi2}) does not vanish for any $z$ unless $J=0$ or $\Omega=0$. [This is shown using the explicit form of $G(\k,z)$ from Eqs.~(\ref{eq:sig2},\ref{eq:sig22}).] Therefore, the only singularities in the integrand are the branch cuts of $G(\k,z) G(\k,z + i\Omega)$, shown in Fig.~\ref{fig:branch}. Since the integrand decays as $|z|^{-2}$ at $|z|\rightarrow\infty$, we can deform the integration contour into a pair of contours that enclose the parts of the branch cuts to the right of the imaginary $z$ axis (shown in blue in Fig.~\ref{fig:branch}). The integral then becomes
\begin{align}
\Pi_{J_x}(i \Omega) & = \int_\k \int_0^{2J + \varepsilon_\k} \frac{d\omega'}{2\pi} (v^x_\k)^2 \bigg[ \frac{1}{G^{-1}(\k,\omega'-i0^{+}) G^{-1}(\k,\omega'+ i\Omega) - J^2}
- \frac{1}{G^{-1}(\k,\omega'+ i0^{+}) G^{-1}(\k,\omega'+ i\Omega) - J^2}
\nonumber \\ 
& + \frac{1}{G^{-1}(\k,\omega'-i0^{+}) G^{-1}(\k,\omega' - i\Omega) - J^2}
- \frac{1}{G^{-1}(\k,\omega'+ i0^{+}) G^{-1}(\k,\omega' - i\Omega) - J^2}
\bigg]. 
\end{align}
The integrand is purely imaginary, as can be seen by noting that the fourth term in the square brackets is the complex conjugate of the first, and the third term is the complex conjugate of the second. [This follows from the fact that $G(\k,z^*) = G(\k,z)^*$.] On the other hand, the original integral in Eq.~(\ref{Pi2}) is a real function of $\Omega$, as can be seen performing a change of variables, $\omega \rightarrow - \Omega - {\omega}$. Therefore, the integral in Eq.~(\ref{Pi2}) vanishes.

The discussion above shows that $\Pi_{J_x} (i \Omega) = 0$ for any $\Omega\ne 0$; analytically continuing to real frequency, we get that $\Pi^{\mathrm{ret}}_{J_x}(\Omega)=0$ for $\Omega\ne 0$. This is a direct consequence of the fact that, for any realization of our model, the current is an exactly conserved quantity. Therefore, $\sigma(\Omega\ne 0) = 0$. According to the conductivity f-sum rule,
\begin{equation}
    \int \frac{d\Omega}{\pi} \, \sigma_{xx} (\Omega) = e^2 \sum_\el \int_\k \frac{\partial^2 \varepsilon_\k}{\partial k_x ^ 2} \langle c^\dagger_{\k,\ell} c_{\k,\ell}\rangle \equiv D,
\end{equation}
where $D$ is the Drude weight. We conclude that $\sigma_{xx} = D\delta(\Omega)$, as expected.

The calculation above was at $T = 0$. However we expect that the zero frequency delta function in the conductivity actually holds at all temperatures. To see this explicitly consider the calculation of the conductivity at high temperature  from the standpoint of the  perturbation theory in the hopping described in Sec. \ref{perttrnsprt}. In contrast to the SYK$_q$ models with $ q \geq 4$, at $q = 2$, we cannot  replace $\overline{G_{ij} G_{ji}}$  by  $\overline{G}_{ij} \overline{G}_{ji}$: 
\begin{equation}
\overline{G_{ij} G_{ji}} \neq \overline{G}_{ij} \overline{G}_{ji},~~~~q = 2.
\end{equation}
This can be checked by explicit calculation of both sides. It is readily seen that the correct averaging $\overline{G_{ij} G_{ji}}$ leads to the expected $\delta(\omega)$ peak in $\sigma_{xx}$ in the high temperature limit.

\section{Self averaging of the correlation functions at large $N$}
\label{app:selfaverage}

\begin{figure}
\begin{center}
\includegraphics[width=0.6\columnwidth]{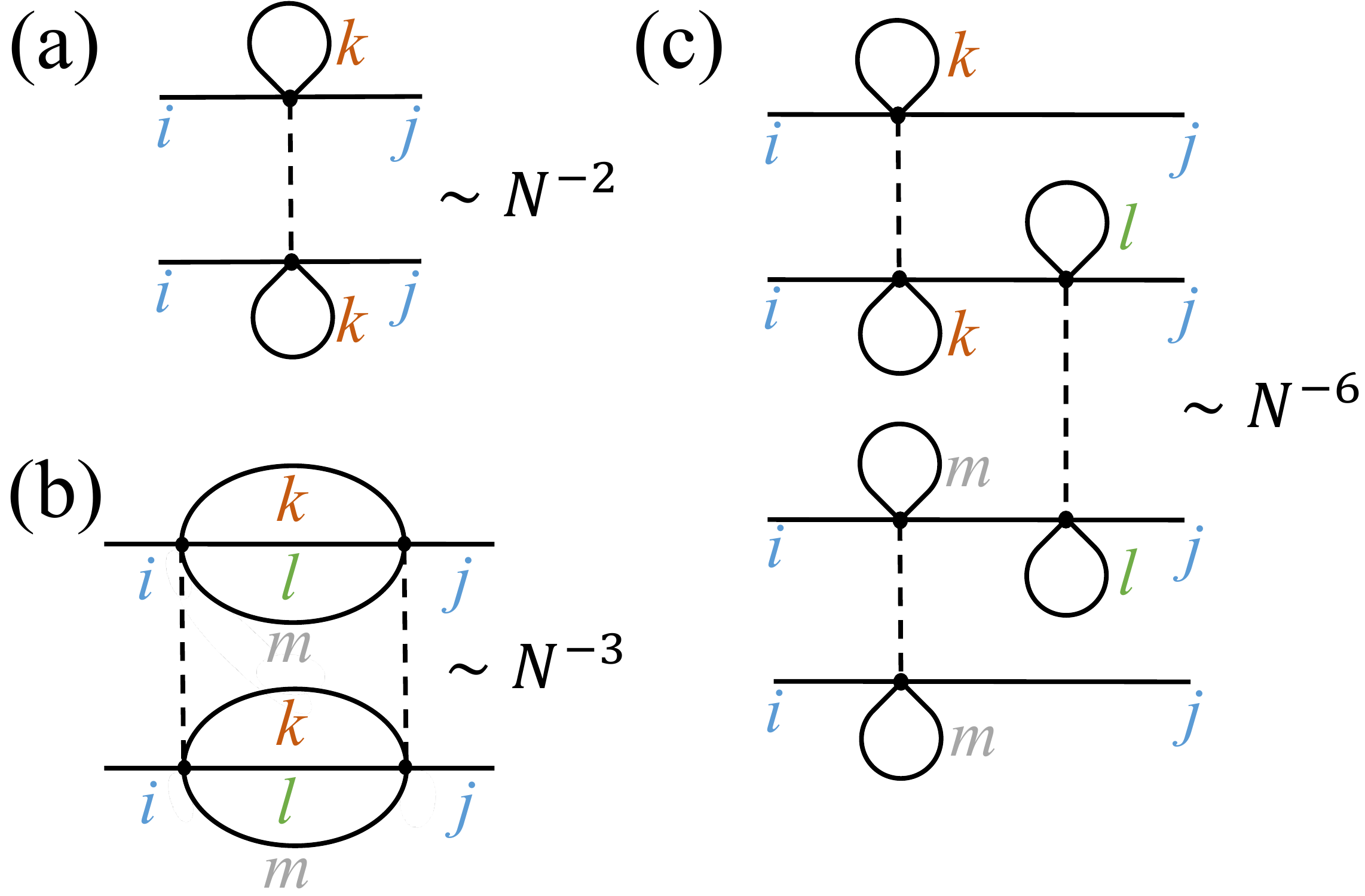}
\end{center}
\caption{Low order contributions to the second and fourth cumulants $C_2$, $C_4$ of the $(i,j)$ element of the single-particle Green's function. (a) $O(N^{-2})$ contribution (lowest order) to $C_2$; (b) $O(N^{-3})$ contribution to $C_2$; (c) Lowest order contribution to $C_4$, of order $N^{-6}$.}
\label{fig:stddev}
\end{figure}

In this Appendix, we show that in the large $N$ limit, the correlation functions of a single realization of the model (\ref{hc}) are essentially the same as the averaged correlation functions over realizations of $U_{ij k \el}$. Consider, for example, the orbital-diagonal single-particle Green's function $G_{c,ii}(\k,i\omega)$. We define $\delta G_{c,ij}(\k,i\omega) = G_{c,ij}(\k,i\omega) - \overline{G_{c,ij}(\k,i\omega)}$ as the deviation of the Green's function of a single realization from the mean, where the overline denotes averaging over realizations of the interaction $U_{ijk\el}$. The variance of $G_{c,ij}$ is given by
\beq
C_2(\k,i\omega) = \overline{[\delta G_{c,ij}(\k,i\omega)]^2} = \overline{\left[G_{c,ij}(\k,i\omega)\right]^2} - \left[\overline{G_{c,ij}(\k,i\omega)}\right]^2.
\label{eq:C2}
\eeq
This quantity can be represented as a sum of all the diagrams with two Green's functions connected by at least one interaction line. [The disconnected terms are subtracted off by the last term in Eq.~(\ref{eq:C2}).] We examine some of the leading order diagrams that contribute to $C_2$ in Fig.~\ref{fig:stddev}(a,b). The lowest-order contribution, Fig.~\ref{fig:stddev}(a), scales as $1/N^2$. Therefore, we conclude that the standard deviation of the Green's function is much smaller than the average in the large $N$ limit. Similar considerations hold for any correlation function. 

Moreover, we can estimate higher cumulants of the Green's function. Consider, for example, the fourth order cumulant, $C_4 = \overline{\left[ \delta G_{c,ij}(\k,i\omega)\right]^4} - 3\left[\overline{\left[ \delta G_{c,ij}(\k,i\omega)\right]^2}\right]^2$. The leading order diagram for $C_4$ is shown in Fig.~\ref{fig:stddev}(b). As can be seen from the figure, $C_4 \sim 1/N^6$. Similarly, one can show that the $n$th cumulant of $\delta G_{c,ij}$, $C_{n} \sim 1/N^{2n-2}$. Hence, all the higher cumulants $C_n$ decrease rapidly with $n$, and we expect the distribution of the Green's function to become approximately Gaussian in the large $N$ limit. 

\section{Path integral formulation}
\label{pif}
Here we briefly describe the path integral formulation of our translationally invariant models. This gives an alternate view on the self-consistency equations as a saddle point approximation (which becomes exact in the $N \rightarrow \infty$ limit) to the path integral. It leads immediately to the Luttinger-Ward functional used in many places in the paper. 
Our discussion will closely follow the treatment described in detail in previous work on SYK models \cite{SY,SS15,Maldacena2016}. We will mainly emphasize the minor differences arising from the translation invariant form of the SYK interactions in our models. We begin with the one-band model. The partition function is given by the imaginary time path integral 
\begin{eqnarray}
Z & = &  \int {\cal D}c ~ e^{- S_o[c] -  S_U[c]} \\
S_o[c] & = & \int d\tau \sum_{\r, i} c^\dagger_{\r i} \left(\frac{\partial }{\partial \tau}  - \mu \right) c_{\r i}  + 
t^c \sum_{i, \langle \r \r' \rangle} \bigg(c^\dagger_{\r i}(\tau) c_{\r' i}(\tau)  + c^\dagger_{\r' i}(\tau)c_{\r i}(\tau)\bigg) \\
S_U & = & \int d\tau \frac{1}{(2N)^{\frac{3}{2} }}\sum_r \sum_{ijk\el} U^c_{ijk\el} c^\dagger_{\r i}c^\dagger_{\r j} c_{\r k} c_{\r \el}.
\end{eqnarray}
To deal with the SYK interactions we should average over their probability distribution. Strictly speaking this should be done using replicas. However as is well known from previous SYK studies (and as we demonstrated in Appendix \ref{app:selfaverage} due to the self-averaging property of our version of the model), the physical propeties of interest can be extracted by averaging a single replica, i.e. by directly averaging the partition function. We therefore just study $\overline Z$ (and drop the overline henceforth). After disorder averaging we find 
\begin{eqnarray}
Z & = & \int {\cal D}c ~e^{- S_0[ c] - S_{int}[c]} \\
S_{int}[c] & = &  - \frac{U_c^2}{4N^3} \int d\tau d\tau' \sum_{\r,\r'} \bigg|\sum_i c^\dagger_{\r i}(\tau) c_{\r' i} (\tau')\bigg|^4.
\end{eqnarray}
Most importantly, the independence of $U^c_{ijkl}$ on $\r$ leads to a sum over all pairs of lattice sites $\r, \r'$ in $S_{int}$. If we had instead chosen $U^c$ to be independent random variables at different sites, $S_{int}$ would have only involved on-site interactions. 

Now define the function $G(\r', \tau'; \r, \tau)$ through
\beq
G(\r', \tau'; \r, \tau) = \frac{1}{N} \sum_i c^\dagger_{\r i}(\tau) c_{\r'i} (\tau')
\eeq
Inserting the following identity 
\beq
1  = \int {\cal D}G~  \delta\bigg( G(\r', \tau'; \r, \tau) - \frac{1}{N} \sum_i c^\dagger_{\r i}(\tau) c_{\r'i} (\tau')\bigg)
\eeq
into the path integral, we rewrite the delta function as 
\beq
 \delta\bigg( G(\r', \tau'; \r, \tau) - \frac{1}{N} \sum_i c^\dagger_{\r i}(\tau) c_{\r'i} (\tau')\bigg) = \int {\cal D} \Sigma ~e^{ \int d\tau d \tau' \Sigma(\r, \tau; \r', \tau') (N G(\r', \tau'; \r, \tau) - \sum_i c^\dagger_{\r i}(\tau) c_{\r'i} (\tau'))}.\nonumber\\
 \eeq
 The disordered averaged interaction can be expressed directly in terms of $G$ as 
 \beq
 S_{int} = -\frac{N U_c^2}{4} \sum_{\r, \r'}  \int d\tau d\tau' |G(\r', \tau'; \r, \tau)|^4
 \eeq
 After these manipulations the fermion integral is quadratic. Performing it we get 
 \begin{eqnarray} 
 Z & = & \int {\cal D}G {\cal D}\Sigma ~ e^{-  NS[\Sigma, G]} \\
S[\Sigma, G] & = &  \tn{Tr} \ln (\partial_\tau - \mu + \epsilon_c  + \Sigma )  +  \int d\tau d \tau' \sum_{\r, \r'} \Sigma(\r, \tau; \r', \tau') G(\r', \tau'; \r, \tau) \nonumber \\
& & - \frac{ U_c^2}{4} \sum_{\r, \r'}  d\tau d\tau' |G(\r', \tau'; \r, \tau)|^4 
\end{eqnarray}
The first term is written schematically with $\epsilon_c$ being the bare dispersion of the electrons. The action now has an overall factor of $N$ multiplying it. Thus in the large-$N$ limit the $\Sigma, G$ integrals can be done in saddle point. It is readily seen that the  saddle point equations are precisely the self-consistency equations described in the main text. 
$S[\Sigma, G]$ then directly gives us the Luttinger-Ward functional for this model. 

The two  band model can be treated in an identical manner. It is readily seen that it leads to the appropriate self-consistency equations and Luttinger-Ward functional.

\section{Green's function for the one-band model}
\label{M1A}
In this appendix we present calculations for the Green's function for the $c$ fermions in the one-band model.
\subsection{Polarization function}
We anticipate a Fermi-liquid to locally incoherent critical metal crossover as a function of temperature. Hence, including a self-energy of the form in Eq.~(\ref{Sigma_c}) in the Fermi-liquid regime, we assume the following form for the Green's function, and search for a self-consistent solution to the set of saddle-point equations [Eq.~(\ref{SPc_a}-\ref{SPc_c})],
\begin{equation}
G_c(\k,i\omega) \sim \left\{
\begin{array}{cc}
\frac{1}{i Z^{-1}\omega - \tilde\varepsilon_{\k}}, & \omega \ll \Omega^*_c,\\
\frac{i\mathrm{sgn}(\omega)}{\sqrt{U_c |\omega|}}, & \Omega^*_c \ll \omega \ll U_c.
\end{array}
\right.
\end{equation}

We separate the polarization function into two components, 
\begin{equation}
\Pi_c(\q,i\Omega) = \Pi^1_c(\q,i\Omega)  + \Pi^2_c(\q,i\Omega),
\end{equation}
where $\Pi^1_c(\q,i\Omega)$ includes contributions of electrons with energies below $\Omega_c^* (= W_c^2/U_c)$, and $\Pi^2_c(\q,i\Omega)$ takes into account electrons with energies above $\Omega_c^*$. 

The contribution from the low lying electron-hole excitations is calculated in the usual way~\cite{AltlandandSimons} and is given by
\begin{eqnarray}
\Pi^1_c(\q,i\Omega) &=& -\int_\k\int_{-\Omega_c^*}^{\Omega_c^*}\frac{d\omega}{2\pi} \frac{1}{iZ^{-1}\omega-\tilde\ve_\k}\frac{1}{iZ^{-1}(\omega+\Omega)-\tilde\ve_{\k+\q}},\\
&\approx&-\int_\k\int_{-\infty}^{\infty}\frac{d\omega}{2\pi} \frac{1}{iZ^{-1}\omega-\tilde\ve_\k}\frac{1}{iZ^{-1}(\omega+\Omega)-\tilde\ve_{\k+\q}},\nonumber\\
&\approx&-Z^2\int_\k\left[n_F(Z\tilde\ve_{\k+\q})-n_F(Z\tilde\ve_{\k})\right]\frac{1}{i\Omega-Z(\tilde\ve_{\k+\q}-\tilde\ve_{\k})}\nonumber.
\end{eqnarray}

In the second line above, we have made an approximation by extending the domain of frequency integration. The frequencies between $\Omega_c^*$ and $U_c$ lead to a correction of the order of $Z^2/\Omega_c^*$ for small frequencies $\Omega\ll (\Omega_c^*)^2/U_c$; at strong coupling (where $Z\sim 1/\nu_0U_c$) this results in a numerical modification of Eq.~\ref{eq:Piform} below, and we therefore neglect it.

For $|\q|\ll k_F$, and at low temperatures $T\ll \Omega_c^*$, we can approximate
\begin{eqnarray}
n_F(Z\tilde\ve_{\k+\q})-n_F(Z\tilde\ve_\k)&\approx& -\delta(\tilde\ve_\k)~\tv q\cos\theta\nonumber,\\
\tilde\ve_{\k+\q}-\tilde\ve_{\k} &\approx&~ \tv q\cos\theta,
\end{eqnarray}
where $\theta$ is the angle between $\q$ and $\k=\k_F$, and we therefore get
\begin{eqnarray}\label{eq:Piform}
\Pi^1_c(\q,i\Omega) &=& Z\nu_0\int\frac{d\theta}{2\pi}\left[1-\frac{i\Omega}{i\Omega-Z\tv q\cos(\theta)}\right],\\
&=&Z\nu_0\left[1-\frac{|\Omega|}{\sqrt{\Omega^2+(Z\tv q)^2}}\right]\nonumber,
\end{eqnarray}
where $\tv$ is the renormalized Fermi-velocity and $\nu_0$ is the density of states at the Fermi-energy.

The contribution to the polarization function from high-energy electron-hole excitations has a completely local $\q-$independent form and is given by
\begin{eqnarray}
\Pi^2_c(\q,i\Omega) &=& -\frac{1}{U_c}\int_{\Omega_c^*}^{U_c}\frac{d\omega}{2\pi}\frac{1}{\sqrt{\omega}}\frac{1}{\sqrt{\omega+\Omega}},\\
&\approx&-\frac{1}{U_c}\log\left(\frac{U_c}{\tn{max}\{|\Omega| ,\Omega^*_c\}}\right).\nonumber
\end{eqnarray}

\subsection{Electronic self-energy}
Having calculated the polarization function, it is straightforward to calculate the electron self-energy $\Sigma_c(\k,i\omega)$ in the Fermi-liquid regime. Once again, we separate the self-energy into two components, arising from $\Pi^1_c(\q,i\Omega)$ and $\Pi^2_c(\q,i\Omega)$:
\begin{eqnarray}
\Sigma_c(\k,i\omega) &=& \Sigma_c^1(\k,i\omega) + \Sigma_c^2(\k,i\omega).
\end{eqnarray}
We are concerned with the self energy at and near the Fermi surface, where Eq.~(\ref{Sigma_c}) holds. We begin with the contribution of the low-lying excitations:
\begin{eqnarray}
\Sigma_c^1(\k,\omega) &=& -Z^2U_c^2\nu_0\int_\q\int\frac{d\Omega}{2\pi}\frac{1}{i(\omega+\Omega)-Z\tilde\ve_{\k+\q}}\left[1-\frac{|\Omega|}{\sqrt{\Omega^2+(Z\tv q)^2}}\right].
\end{eqnarray}
Let us first evaluate the self-energy at $\k=\k_F$ and finite $\omega$. The important contribution comes from small momentum scattering, $|\q|\ll k_F$; for such wavevectors, we can approximate $\tilde\ve_{\k_F+\q}\approx \tv q\cos(\theta)$, with $\theta$ being the angle between $\q$ and $\k_F$. We thus get
\begin{eqnarray}
\Sigma_c^1(k_F,\omega) &=& -Z^2U_c^2\nu_0\int_\q\int\frac{d\Omega}{2\pi}\frac{1}{i(\omega+\Omega)-Z\tv q\cos(\theta)}\left[1-\frac{|\Omega|}{\sqrt{\Omega^2+(Z\tv q)^2}}\right]\nonumber\\
&=& i Z^2U_c^2\nu_0\int \frac{q~dq}{2\pi}\int\frac{d\Omega}{2\pi}\frac{\sign(\omega+\Omega)}{\sqrt{(\omega+\Omega)^2+(Z\tv q)^2}}\left[1-\frac{|\Omega|}{\sqrt{\Omega^2+(Z\tv q)^2}}\right].
\end{eqnarray}

The integrand vanishes for $\Omega\gg Z\tv q$; we thus consider only the limit where $\Omega,\omega\ll Z\tv q$:
\begin{eqnarray}
\Sigma_c^1(k_F,\omega) 
&=& i U_c^2\frac{Z\nu_0}{\tv}\int \frac{dq}{2\pi}\int\frac{d\Omega}{2\pi}\sign(\omega+\Omega)\left[1-\frac{|\Omega|}{Z\tv q}\right]\\
&=& i U_c^2\frac{Z\nu_0}{\tv}\left[\frac{k_F\omega}{2\pi^2}-\frac{1}{Z\tv}\int \frac{dq}{4\pi^2}\frac{1}{q} \omega^2\sign(\omega)\right]\nonumber
\end{eqnarray}
The $\omega^2$ term displays the well-known logarithmic divergence in the imaginary part which occurs in the calculation of the self-energy of two-dimensional Fermi liquids; this can be fixed by recalling that in the $q$ integral, the expression is valid only for $q$ much larger than $\omega/Z\tv$. Introducing the appropriate IR cutoff leads to,
\begin{eqnarray}
\Sigma_c^1(k_F,\omega) 
= \nu_0^2 U_c^2 \left[ i Z \omega + i \alpha \nu_0 |\omega|^2 \ln(Z\tv k_F/|\omega|) \mathrm{sign}(\omega) \right],\nonumber
\end{eqnarray}
with $\alpha$ is a number of order unity.

We now focus on the contribution from the high-energy excitations that arise from the component $\Pi^2_c$:
\begin{eqnarray}
\Sigma_c^2(k_F,\omega) &\sim& -ZU_c\int_\q\int\frac{d\Omega}{2\pi}\frac{1}{i(\omega+\Omega)-Z\tilde\ve_{\k_F+\q}}\log\left(\frac{U_c}{W_c}\right)\\
&=& i ZU_c\int \frac{q~dq}{2\pi}\int\frac{d\Omega}{2\pi}\frac{\sign(\omega+\Omega)}{\sqrt{(\omega+\Omega)^2+(Z\tv q)^2}}\log\left(\frac{U_c}{W_c}\right)\nonumber\\
&\sim&i U_c\nu_0\log\left(\frac{U_c}{W_c}\right)\omega\nonumber,
\end{eqnarray}
which at strong coupling, where $Z\sim 1/\nu_0U_c$, is comparable to $\Sigma_c^1(k_F,\omega)$, up to logarithmic factors.

Let us now evaluate the self-energy at $\omega=0$ and finite (but small) $k = |\k-\k_F|$. The contribution from the low lying excitations is,
\beq
\Sigma_c^1(\k,0) &=& -Z^2U_c^2\nu_0\int_\q\int\frac{d\Omega}{2\pi}\frac{1}{i\Omega-Z\tilde\ve_{\k+\q}}\left[1-\frac{|\Omega|}{Z\tv |\q|}\right].
\eeq
We are interested in the contribution from the second term, which gives the leading $k$ dependence. In particular, the real part is given by,
\beq
\tn{Re}[\Sigma_c^1(\k,0)] &=& -Z^2U_c^2\nu_0 \int_\q \int\frac{d\Omega}{2\pi} \frac{\tilde\ve_{\k+\q}}{\Omega^2 + Z^2\tilde\ve_{\k+\q}^2}\frac{|\Omega|}{\tv |\q|}\\
&=& -Z^2U_c^2\nu_0 \int_\q \frac{\tilde\ve_{\k+\q}}{\tv |\q|}\log\bigg|\frac{\tv|\q|}{\tilde\ve_{\k+\q}} \bigg|,
\eeq
where in the $\Omega$ integral, we have assumed that $\Omega<Z\tv |\q|$. 
Consider now the external momentum $\k = (k_x,0)$. The typical internal momenta $q_x\lesssim q_y^2/k_F$ with $q_y<k_F$. Restricting ourselves to momenta close to the Fermi-surface, 
\beq
\tn{Re}[\Sigma_c^1(\k,0)] = -Z^2U_c^2\nu_0 \int_0^{k_F} dq_y~\int_0^{q_y^2/k_F} dq_x~\frac{k_x+q_x}{q_y} \log\bigg|\frac{q_y}{k_x+q_x} \bigg|.
\eeq
The term linear in $k_x$ has a contribution from two terms above,
\beq
\tn{Re}[\Sigma_c^1(\k,0)] &=& -Z^2U_c^2\nu_0 \int_0^{k_F} dq_y~\int_0^{q_y^2/k_F} dq_x~\bigg[ \frac{k_x}{q_y} \log\bigg|\frac{q_y}{q_x}\bigg| - \frac{q_x}{q_y}\frac{k_x}{q_x}\bigg],\\
&=& -\frac{Z^2U_c^2\nu_0}{k_F} \int_0^{k_F} dq_y~k_x~(q_y - q_y\log q_y - q_y) \approx - \zeta Z^2U_c^2\nu_0 k_F k_x,
\eeq
where $\zeta$ is an $O(1)$ number that depends on the fermi-surface geometry.

\section{Momentum dependence in LICM phase}
\label{licm}
In this appendix, we examine the weak, analytic momentum-dependence of the Green's function in the high-temperature local incoherent metal. We begin with an ansatz for the 
self-energy of the form,
\beq
\Sigma_c(\k,i\omega) = \Sigma_c^{(0)}(i\omega) + \Sigma^{(1)}_c(\k,i \omega),
\label{eq:sigma_exp}
\eeq 
where $\Sigma_c^{(0)}(i\omega) = i\tn{sgn}(\omega)\sqrt{U_c|\omega|}$ is the single-site SYK self-energy, and $\Sigma^{(1)}_c(\k,i \omega)$ is the correction due to the electronic dispersion $\varepsilon_\k$. We view Eq.~(\ref{eq:sigma_exp}) as an expansion in powers of the bare bandwidth $W_c$. The Green's function is given by
\beq
G_c(\k,i \omega) &=& \left[i\omega - \varepsilon_\k - \Sigma_c(\k,i\omega)\right]^{-1} \nonumber \\
&\approx& - \frac{1}{\Sigma^{(0)}_c(i\omega)} + \frac{ \varepsilon_\k + \Sigma_c^{(1)}(\k,i\omega) }{[ \Sigma_c^{(0)}(i\omega)]^2} + \dots \nonumber \\
&\equiv& G^{(0)}_c(i \omega) + G^{(1)}_c(\k, i \omega) + \dots.
\label{eq:G_c_exp}
\eeq
In the second line, we have neglected the $i\omega$ term, assuming that we are at frequencies and temperatures such that $|\Sigma^{(0)}(i\omega)| \gg |\omega|$, and expanded to leading order in $W_c$. The third line defines $G^{(0)}_c(i \omega)$ and $G^{(1)}_c(i \omega)$ as an expansion of $G_c(i\omega)$ in powers of $W_c$.

The self-energy satisfies the self-consistent equation~(\ref{selfenergyc}). Expanding this equation to leading order in $W_c$, we get:
\begin{align}
\Sigma^{(1)}_c(\r, \tau)  = &  - 2 U_c^2 G_c^{(0)}(\tau) G_c^{(0)}(-\tau) \int \frac{d^d \k}{(2\pi)^d} G^{(1)}(\k, \tau) \nonumber \\
& - U_c^2 [G_c^{(0)}(\tau)]^2  \int \frac{d^d \k}{(2\pi)^d} G^{(1)}(\k, -\tau).
\label{eq:sigma_1}
\end{align}
Eq.~(\ref{eq:sigma_1}) shows that the leading-order correction to the self-energy is independent of position (or momentum), since the right-hand side does not depend on $\r$. Therefore, the leading-order momentum dependence of $G_c$ is of the form $\varepsilon_\k / [\Sigma^{(0)}_c(i\omega)]^2$ [that appears in the second line of Eq.~(\ref{eq:G_c_exp})]. This leads to the expression in Eq.~(\ref{limits}) for $G_c$ in the LICM regime ($\omega \gg W_c^2/U_c$).

\section{Green's function for the two-band model}
\label{MFLapp}
In this appendix we present calculations for the Green's function of the $c$ and $f$ fermions in the two-band model. In particular, it follows from appendix \ref{M1A} that in the low-temperature regime  $T<\Omega_f^*\ll\Omega_c^*$, both species are in a Fermi liquid phase. Similarly, in the high-temperature regime $T>\Omega_c^*\gg\Omega_f^*$, it follows that both species are in a LICM phase. The remainder of this appendix is devoted to deriving the form of the Green's functions in the intermediate regime where one obtains a marginal Fermi liquid. We analyze the saddle point equations in a fully self-consistent fashion for the MFL and NFL models considered above in appendix \ref{SCse}, and find results that are consistent with ones obtained below.  

\subsection{Self energy of the $c-$electrons}
In the regime where the $f-$electrons form a LICM (i.e. $T>\Omega_f^*$), we use the following momentum independent form of the polarization function,
\begin{eqnarray}
\Pi_f(\q,\omega) \sim -\frac{1}{U_f} \log\left( \frac{U_f}{|\omega|} \right).
\end{eqnarray}
Using this to calculate the self-energy of the $c$-electrons leads to,
\begin{eqnarray}
\Sigma_{cf}(\k,\omega) = \frac{ZU_{cf}^2}{U_f} \int_\q\int\frac{d\Omega}{2\pi} \frac{1}{i(\omega+\Omega)-Z\ve_{\k_F+\q}}\log\left(\frac{U_f}{|\Omega|}\right).
\end{eqnarray}

We again approximate $\ve_{\k_F+\q}\approx v_Fq\cos(\theta)$, with $\theta$ the angle between $\q$ and $\k$, and perform the integral over $\theta$ to obtain
\begin{eqnarray}
\Sigma_{cf}(\k,\omega) = -i\frac{ZU_{cf}^2}{2\pi U_f} \int qdq\int\frac{d\Omega}{2\pi} \frac{\sign(\omega+\Omega)}{\sqrt{(\omega+\Omega)^2+(Zv_Fq)^2}}\log\left(\frac{U_f}{|\Omega|}\right).
\end{eqnarray}
The most singular contribution comes from the region where $\Omega\ll Zv_Fq$, and thus
\begin{eqnarray}
\Sigma_{cf}(\k,\omega) &\approx& -i\frac{U_{cf}^2}{2\pi U_f}	 \frac{k_F}{v_F} \int\frac{d\Omega}{2\pi} \sign(\omega+\Omega)\log\left(\frac{U_f}{|\Omega|}\right)\\
&=& -i\frac{U_{cf}^2\nu_0}{(2\pi)^2U_f}\left[\log\left(\frac{U_f}{|\omega|}\right)-1\right]\omega\nonumber
\end{eqnarray}
for small $\omega<U_f$, this becomes the marginal Fermi liquid form,
\begin{eqnarray}
\Sigma_{cf}(\k,\omega) &\sim& -\frac{U_{cf}^2\nu_0}{U_f}i\omega\log\left(\frac{U_f}{|\omega|}\right).
\end{eqnarray}

\subsection{Feedback of $c-$fermions on the $f-$fermions}
\label{polc}

In this section, we analyze to what extent the $c-$fermions modify the self-energy of the $f-$fermions, and in particular if the $f$ fermions retain their local SYK-like character (as a result of self-interactions, $U_f$) even when they scatter off the $c$ fermions (as a result of the inter-band scattering, $U_{cf}$). 

The self-energy of interest is given by,
\beq
\Sigma_{cf}'(\k,i\omega) = -U^2_{cf} \int_{\k_1}\int_\Omega G_f(\k+\k_1,i\omega+i\Omega) ~\Pi_c(\k_1,i\Omega).
\eeq

It is clear from the momentum-independent form of $G_f$ that the self-energy will be independent of momentum, {\it i.e.} will also retain a local character. Let us first carry out the analysis in frequency domain; we will show later that the analysis in time domain follows more simply. In the MFL, the polarization function for the $c$ fermions retains a Fermi-liquid-like form, with a low frequency and wave-vector polarization function of the form
\begin{equation}
\Pi_c(\k_1,i\Omega) = \nu_0\left(1-\frac{|\Omega|}{\sqrt{\tilde{v}_F^2k_1^2+\Omega^2}}\right).
\end{equation}

It is clear from the above expression that the leading frequency dependence of the self-energy will arise from the constant term. Therefore, we have
\begin{eqnarray}
\Sigma_{cf}'(i\omega) = -i\nu_0\frac{U^2_{cf}}{\sqrt{U_f}}\int_\Omega \frac{\sign(|\omega+\Omega|)}{\sqrt{|\omega+\Omega|}} + ...,
\end{eqnarray}
where $...$ denote higher order corrections that we will comment on below. At low frequencies, this results in
\begin{eqnarray}
\Sigma_{cf}'(i\omega) &=& -i\nu_0\frac{U^2_{cf}}{\sqrt{U_f}}\int_{-W_c-\omega}^{W_c-\omega}\frac{\sign(\Omega)}{\sqrt{|\Omega|}}d\Omega
=-2i\nu_0\frac{U^2_{cf}}{\sqrt{U_f}}\left[\sqrt{W_c-\omega}-\sqrt{W_c+\omega}\right]\nonumber\\
&\sim&2\nu_0\frac{U^2_{cf}}{\sqrt{U_fW_c}}i\omega\nonumber.
\end{eqnarray}

Thus, we find that there is an analytic ({\it i.e.} non-singular) correction to the self-energy of the $f$ fermions as a result of coupling to the density of $c$ fermions. Strictly speaking, the above correction renormalizes the bare `$i\omega$' term, {\it i.e.} at this scale the appropriate self-energy for the $f$ fermions (without including the `SYK' piece) is
\beq
\Sigma_{cf}'(i\omega) \sim \bigg(1 + \frac{2\nu_0 U_{cf}^2}{\sqrt{U_fW_c}}\bigg) i\omega.
\eeq
The correction is small as long as the following condition is met:
\beq
\bigg(\frac{U_{cf}}{W_c} \bigg)^2 \sqrt{\frac{W_c}{U_f}} \ll 1.
\eeq

It is worth analyzing the singular correction to $\Sigma_{cf}'$, even though it is subleading to the analytic piece computed above. In order to compute this correction, it is simpler to study the self-energy in the time domain. Using the local character of the self-energy, we can rewrite it in time domain as,
\beq
\Sigma'_{cf}(\tau) = -U_{cf}^2 G_f(\tau) ~\Pi_c(\tau),
\eeq
where now we only need the local form of the polarization bubble for the $c-$fermions. Even in the MFL, this is the same as in a Fermi liquid, and decays as $1/\tau^2$. This leads to 
\beq
\Sigma'_{cf}(i\omega)\sim \frac{U_{cf}^2}{W_c^2\sqrt{U_f}}~ i|\omega|^{3/2}\tn{sgn}(\omega).
\eeq

\section{Self-consistent solutions for two-band model}
\label{SCse}
For the two-band model, it is useful to analyze the saddle-point equations for the $c$ fermions self-consistently, unlike what was done in Section \ref{MFLgreen}. This will also shed some light on one of the key differences between the non-Fermi liquid metals and critical Fermi-surfaces being considered here and the more conventional `quantum-critical' non-Fermi liquids. As before, we can set $U_c=0$ and focus on the effects of the inter-band scatterings only in the presence of a finite $U_f$ that drives the $f$ Fermions in a locally critical regime. The rest of this analysis is applicable to both the MFL as well as the NFL models in sections \ref{sec:MFL} and \ref{nfl}, respectively.

It is then reasonable to assume that the self-energy of the $c$ fermions has the following scaling form:
\beq
\Sigma_{cf}(\omega,\k) = \omega^\phi~ \tn{sign}(\omega)~H\bigg(\frac{\omega}{k_\perp} \bigg),
\eeq
where $k_\perp = |\k-\k_F|$ measures the deviation from the Fermi-surface, $\phi$ is an exponent to be determined and $H(x)$ is a scaling function with the following property,
\beq
H(x\rightarrow\infty) = \tn{const}.
\eeq
The above property of $H(x)$ simply encodes the feature that in the limit $\k\rightarrow\k_F$, there is a singular frequency dependent self-energy (if $\phi\leq1$) at and near the Fermi-surface. 

On the other hand, we may ask what is the limit of $H(x\rightarrow0)$, when the momenta are taken far away from the Fermi-surface at small enough frequencies? In examples of quantum-critical non-Fermi liquids, the singular structure is restricted to the vicinity of the Fermi-surface, and hence $H(x\rightarrow0)\sim 1/x^\phi$. However, if the singular structure persists everywhere in momentum-space (as is usually the case in locally critical systems), there is no reason for the above to be true and the limit of $H(x\rightarrow0)$ can be a constant. 

In explicit terms, the above saddle-point equation for $\Sigma_{cf}$ becomes
\beq
\Sigma_{cf}(\omega,\k) = -U_{cf}^2 \int_{\q}\int\frac{d\Omega}{2\pi}\frac{1}{i(\omega+\Omega) - \ve_{\k+\q} - (\omega + \Omega)^\phi~ \tn{sign}(\omega + \Omega)~H[(\omega+\Omega)/(\k+\q)_\perp]}~\Pi_f(\q,\Omega).\nonumber\\
\eeq
Using the fact that $\Pi_f(\q,\Omega)$ is nearly independent of $\q$, we can shift $\q\rightarrow(\q-\k)$ such that the $\k$ dependence drops out completely {\footnote{This step can't be carried out in theories of Fermi-surfaces coupled to critical bosons, unless one restricts to the near vicinity of the Fermi-surface.}} the above simplifies to,
\beq
\Sigma_{cf}(\omega,\k) \approx -\nu_0 U_{cf}^2 \int d\ve \int\frac{d\Omega}{2\pi}\frac{1}{i(\omega+\Omega) - \ve - (\omega + \Omega)^\phi~ \tn{sign}(\omega + \Omega)~H[(\omega+\Omega)/\ve]}~\Pi_f(\Omega).\nonumber\\
\label{SCsigcf}
\eeq
It is then straightforward to see that taking $H(x)$ to be a constant and integrating over $\ve$ leads to a self-consistent solution for $\Sigma_{cf}$. The exponent, $\phi=4\Delta(q)$, is then fixed by the $\Omega$ dependence of $\Pi_f(\Omega)$ (determined by $\Delta(q)=1/q$). The key difference from the quantum-critical metals is that this singular frequency dependence persists everywhere in momentum-space, and arises from the locally critical `bath' that is coupled to the $c$ fermions.

\section{Luttinger-Ward analysis}
\label{LWapp}
In this section, we use the Luttinger-Ward (LW) functional to analyze the fate of Luttinger's theorem for the non-Fermi liquids with a critical Fermi-surface considered above. This analysis will also be useful for determining thermodynamic properties, such as the compressibility. As we already mentioned in sections \ref{thermomfl} and \ref{thermonfl}, the LW analysis for the conserved $f$ fermion density has been carried out in Ref. \cite{Parcollet2} for the $q=4$ case and for general $q$ in Ref. \cite{SS17}. 

The LW analysis for the conserved $c$ fermion density proceeds as follows. The conserved $c$ electron density is,
\beq
n_c = \int_\k \int \frac{d\omega}{2\pi} ~ G_c(\k,i\omega)~e^{i\omega0^+},
\eeq
where $G_c$ is given by,
\beq
 G_c(\k,i\omega) = \frac{1}{i\omega - (\ve_\k - \mu_c) - \Sigma_{cf}(i\omega,\k)}.
 \label{Gcgen}
\eeq

Following \cite{AGD}, we may write the above as,
\beq
n_c = i\int_\k \int \frac{d\omega}{2\pi}\bigg[\frac{\partial}{\partial\omega}\ln G_c(\k,i\omega) - G_c(\k,i\omega)\frac{\partial}{\partial\omega}\Sigma_{cf}(\k,i\omega) \bigg]~e^{i\omega0^+}.
\label{LWnc}
\eeq
Consider the first term on the right-hand side above (the second term will be shown to be zero momentarily),
\beq
n_c = i\int_\k \int \frac{d\omega}{2\pi}\bigg[\frac{\partial}{\partial\omega}\ln G_c(\k,i\omega)\bigg]~e^{i\omega0^+} &=& i\int_\k \int_{-\infty}^0 \frac{dz}{2\pi} \frac{\partial}{\partial z} \ln \frac{G_c(\k,z+i0^+)}{G_c(\k,z+i0^-)} \nonumber\\
&=& \frac{i}{2\pi}\int_\k  \ln \frac{G_c(\k,i0^+)}{G_c(\k,i0^-)}.  
\eeq
Let $\varphi_\k(z)$ denote the phase of the function $G_c^R(\k,z)$. The density can then be expressed in terms of the difference between $\varphi_\k(0^+)$ and $\varphi_\k(0^-)$,
\beq
n_c = -\frac{1}{\pi} \int\frac{d^2\k}{(2\pi)^2} ~[\varphi_\k(0^+) - \varphi_\k(0^-)],
\label{ncp}
\eeq
where the values of these phases are determined by the sign of Re $G_c(\k,0)$. We have,  $\varphi(0)=0$ if Re $G_c^R(\k,0)>0$ and $\varphi(0)=\pi$ if Re $G_c(\k,0)<0$, i.e. the phase changes at the $\k-$space location where $G_c^{-1}(\k=\k_F,\omega=0)=0$. This leads to the statement of Luttinger's theorem:
\beq
n_c = \int\frac{d^2\k}{(2\pi)^2} \Theta(\mu_c-\ve_\k) = \int_{|\k|\leq k_F}\frac{d^2\k}{(2\pi)^2}, 
\label{lutt}
\eeq
where $\Theta(x)$ is the heavyside-theta function. The above is the familiar form of Luttinger's theorem, relating the conserved $U(1)_c$ density to the area of the (critical) Fermi-surface.

Let us now revisit the second term on the right-hand side of Eq.~(\ref{LWnc}). To show that it evaluates to zero, we assume there exists a Luttinger-Ward functional, $\Phi[G_c(\k,i\omega)]$, which has the following two properties:
\beq
\rm{I}&:&\Sigma_{cf}(\k,i\omega) = \frac{\delta \Phi[G_c(\k,i\omega)]}{\delta G_c(\k,i\omega)},\\
\rm{II}&:& \Phi[G_c(\k,i\omega+i\epsilon)] = \Phi[G_c(\k,i\omega)].
\eeq
For the model being considered here, the LW functional has a simple form and is given by,
\beq
\Phi[G_c] = U_{cf}^2 \int d\tau~\prod_{i}\int_{\k_i}~ G_c(\k_1,\tau)~G_c(\k_2,-\tau)~G_f(\k_3,\tau)~G_f(\k_4,-\tau)~\delta\bigg(\sum_i\k_i\bigg).
\eeq
It is therefore clear that $\delta\Phi[G_c] = \int_\omega \int_\k \Sigma_{cf}(\k,i\omega)~\delta G_c(\k,i\omega)$, for $\Sigma_{cf}$ defined earlier from the saddle-point equations. Using property $\rm{II}$ of the functional $\Phi[G_c]$, it then follows that if the frequencies running along the $G_c$ lines are shifted by a tiny amount $\omega_0$,
\beq
\frac{\delta \Phi}{\delta \omega_0} = \int_\omega \int_\k \Sigma_{cf}(i\omega)~\frac{\partial G_c(\k,i\omega)}{\partial \omega} = 0,
\eeq
which in turn leads to the vanishing of the second term in Eq.~(\ref{LWnc}).

\section{Two band model in magnetic field}
In this appendix, we provide additional details for the computations of quantum oscillations in the spectral density of states and magnetization in the two band models of non-Fermi liquids in Section \ref{qo}.  
\subsection{Saddle point equations}
\label{appqo:saddle}
Let us begin by examining the structure of the saddle point equations in real space for SYK$_4$ models, which immediately sheds light on the modifications required in the presence of a magnetic field. The calculation is similar when the $f$ fermions have a SYK$_q$ form of interactions.

The self-energy for the $c$ fermions is given by (we set $U_c=0$) 
\beq
\Sigma_{cf}(\r,\tau;\r',\tau') &=& -U_{cf}^2~ G_c(\r,\tau;\r',\tau')~G_f(\r,\tau;\r',\tau')~G_f(\r,\tau';\r',\tau),\\
\Sigma_{f}(\r,\tau;\r',\tau') &=& -U_{f}^2~ [G_f(\r,\tau;\r',\tau')]^2~G_f(\r,\tau';\r',\tau),\\
\Sigma'_{cf}(\r,\tau;\r',\tau') &=& -U_{cf}^2~ G_f(\r,\tau;\r',\tau')~G_c(\r,\tau;\r',\tau')~G_c(\r,\tau';\r',\tau).
\eeq
The above set of equations can be solved self-consistently by a completely local form of the self-energy (and Green's function) for the $f$ fermions with $\Sigma_f(\r,\tau;\r',\tau') = \Sigma_f(\tau-\tau') \delta_{\r\r'}$ and $G_f(\r,\tau;\r',\tau') = G_f(\tau-\tau') \delta_{\r\r'}$; as discussed before  $\Sigma_f(\tau-\tau')$ then has the usual SYK-like form. 

The self-energy for the $c$ fermions then also has a local character
\beq
\Sigma_{cf}(\r;\tau-\tau') &=& -U_{cf}^2~ G_c(\r,\r;\tau-\tau')~G_f(\tau-\tau')~G_f(\tau'-\tau).
\eeq
As a result, in the presence of a magnetic field, when we express the equations in the LL basis, the self-energy for the $c$ fermions has a $B-$independent piece (i.e. which does not depend explicitly on the LL index) and has the usual marginal Fermi-liquid character (for $q=4$) and non-Fermi liquid character (for $q>4$) as discussed earlier. The oscillations therefore arise from the effect of the magnetic field on the kinetic energy of the $c$ fermions (through the formation of Landau bands). 

The above simplification arises from the absence of a kinetic energy term for the $f$ fermions. Even as in the presence of coupling to the $c$ fermions, the local structure of the $f$ fermion Green's function survives.

\subsection{Density of states oscillations}
\label{app:qodos}
We are only interested in the oscillatory component of the spectral density of states and so let us begin by considering,
\beq
A(i\omega_m) = \frac{B}{2\pi}\int_{-\infty}^\infty\frac{dp_z}{2\pi}\sum_n ~G_c(n,p_z,i\omega_m),
\eeq
where the Green's function is of the form shown in Eq.~(\ref{LLG}). Upon using the Poisson summation formula this yields,
\beq
A_{\tn{osc}}(i\omega_m) &=& \frac{B}{2\pi}\int_{-\infty}^\infty\frac{dp_z}{2\pi} \sum_{k=-\infty}^{\infty} \int_0^\infty dn ~\frac{e^{2\pi i kn}}{i\omega_m - (n+1/2)\omega_c + \mu_c - \frac{p_z^2}{2m^*} - \Sigma_{cf}(i\omega_m)},\\
&=& \frac{m^*}{2\pi} \int_{-\infty}^\infty\frac{dp_z}{2\pi} \sum_{k=-\infty}^{\infty} (-1)^k ~ e^{2\pi i k\mu_c/\omega_c}\int_{-\mu_c+\frac{\omega_c}{2}}^\infty dx ~\frac{e^{2\pi i kx/\omega_c}}{i\omega_m-x-\frac{p_z^2}{2m^*}-\Sigma_{cf}(i\omega_m)}.
\eeq
Analytically continuing to real frequencies, $i\omega_m\rightarrow\omega+i0^+$, we find that the $x$ integral has a pole at $x=\omega-\frac{p_z^2}{2m^*} - \Sigma_{cf}^R(\omega) - i\Sigma_{cf}^I(\omega)$. Extending the lower limit of the $x$ integral to $-\infty$ (as explained below), we note that we get a finite result only when $k>0$ {\footnote{For $k>0$, we close the contour in the upper half plane, which encloses the pole ($\Sigma_{cf}''(\omega)<0$), while for $k<0$ we have to close the contour in the lower half plane.}}. The above quantity then becomes,
\beq
A_{\tn{osc}}(\omega) &=& -\frac{m^*}{2\pi} \sum_{k=1}^{\infty} (-1)^k ~ e^{2\pi i k(\mu_c + \omega - \Sigma_{cf}^R(\omega))/\omega_c} ~e^{-2\pi k|\Sigma_{cf}^I(\omega)|/\omega_c} \int_{-\infty}^\infty\frac{dp_z}{2\pi} e^{-\frac{\pi i k}{m^*\omega_c}p_z^2},\\
&=& -\frac{N(0)}{2} \sum_{k=1}^{\infty} \frac{(-1)^k}{(2k)^{1/2}} ~ e^{2\pi i k(\mu_c + \omega - \Sigma_{cf}^R(\omega))/\omega_c} ~e^{-2\pi k|\Sigma_{cf}^I(\omega)|/\omega_c} e^{-i\pi/4}\sqrt{\frac{\omega_c}{\mu_c}},
\eeq
where we have expressed the answer in terms of the density of states, $N(0)$, of the non-interacting problem at $B=0$.

The spectral density of states, $N(\omega)$, is defined as,
\beq
N_{\tn{osc}}(\omega) &=& -\frac{1}{\pi}\tn{Im} ~A_{\tn{osc}}(\omega),\\
&=& \frac{N(0)}{2\pi} \sum_{k=1}^{\infty} \frac{(-1)^k}{(2k)^{1/2}} ~ \sin\bigg[\frac{2\pi k}{\omega_c}(\mu_c + \omega - \Sigma_{cf}^R(\omega)) - \frac{\pi}{4}\bigg] ~e^{-2\pi k|\Sigma_{cf}^I(\omega)|/\omega_c} \sqrt{\frac{\omega_c}{\mu_c}}.
\eeq
In the limit of $\omega\rightarrow0$ at a finite $T$ the above reduces to the quoted form in Eq.~(\ref{dosqo}).

\subsection{Magnetization oscillations}
\label{appqo:mag}

The oscillatory piece of the orbital magnetization is expressed in Eq.~(\ref{mag}). It can be re-expressed as,
\beq
M_{\tn{osc}}(B) = \frac{1}{2\pi\beta\omega_c}\sum_{\omega_m}\int_{-\infty}^\infty \frac{dp_z}{2\pi} \sum_{k=-\infty}^\infty (-1)^k \int_{-\mu_c+\omega_c/2}^\infty d\ve \frac{(\ve+\mu_c)~e^{2\pi i k(\ve+\mu_c)/\omega_c}}{i\omega_m - \ve - \frac{p_z^2}{2m^*} - \Sigma_{cf}(i\omega_m)}.
\eeq
We can now re-express the summation over the fermionic Matsubara frequencies $\omega_m=(2m+1)\pi T$ as an integral,
\beq
M_{\tn{osc}}(B) = \frac{1}{2\pi\omega_c}\int_{-\infty}^\infty \frac{dp_z}{2\pi} \sum_{k=-\infty}^\infty (-1)^k \int_{-\mu_c}^\infty d\ve (\ve+\mu_c)~e^{2\pi i k(\ve+\mu_c)/\omega_c}\int_{-\infty}^\infty \frac{d\Omega}{\pi} f(\Omega) \tn{Im}G_R\bigg(\ve+\frac{p_z^2}{2m^*},\Omega\bigg),\nonumber\\
\eeq
where we have dropped $\omega_c/2$ compared to $\mu_c$ in the lower limit of the $\ve$ integral above and $f(...)$ is the Fermi-Dirac distribution function. We use the notation $G_R(\ve,\Omega)=(\Omega-\ve-\Sigma_{cf}^R(\Omega)-i\Sigma_{cf}^I(\Omega))^{-1}$ and $A(\ve,\Omega)=-\tn{Im} G_R(\ve,\Omega)/\pi$. Let us now define
\beq
n\bigg(\ve+\frac{p_z^2}{2m^*}\bigg) = \int_{-\infty}^\infty d\Omega f(\Omega) A\bigg(\ve+\frac{p_z^2}{2m^*},\Omega\bigg), 
\eeq
which describes the mean occupation of $c$ fermions in the single-particle states with energy levels $\ve+p_z^2/2m^*$. The magnetization then becomes ($\lambda_k=2\pi k/\omega_c$),
\beq
M_{\tn{osc}}(B) = \frac{1}{2\pi\omega_c}\int_{-\infty}^\infty \frac{dp_z}{2\pi} \sum_{k=-\infty}^\infty \frac{(-1)^k}{i} \frac{\partial}{\partial\lambda_k}\int_{-\mu_c}^\infty d\ve ~e^{i\lambda_k (\ve+\mu_c)}~ n\bigg(\ve+\frac{p_z^2}{2m^*}\bigg).
\eeq
It is useful to express the integrand for the $\ve$ integral in terms of a derivative over $n(\ve)$. We express
\beq
\int_{-\mu_c}^\infty d\ve ~e^{i\lambda_k(\ve+\mu_c)}~ n\bigg(\ve+\frac{p_z^2}{2m^*}\bigg) = -\frac{1}{i\lambda_k}\int_{-\mu_c-p_z^2/2m^*}^\infty d\ve~ e^{i\lambda_k(\ve+\mu_c-p_z^2/2m^*)} \frac{dn(\ve)}{d\ve},
\eeq
where we have dropped a term $\sim n(-\mu_c-p_z^2/2m^*)$ that does not contribute to the oscillatory piece. It is clear that $n'(\ve)$ is peaked near $\ve\approx0$ and we therefore extend the lower limit of the $\ve$ integral to $-\infty$. The magnetization then is,
\beq
M_{\tn{osc}}(B) \approx -\frac{1}{2\pi\omega_c}\int_{-\infty}^\infty \frac{dp_z}{2\pi} \sum_{k=-\infty}^\infty \frac{(-1)^k}{i} \frac{\partial}{\partial\lambda_k} \bigg[\frac{1}{i\lambda_k}\int_{-\infty}^\infty d\ve~ e^{i\lambda_k(\ve+\mu_c-p_z^2/2m^*)} \frac{dn(\ve)}{d\ve}\bigg].
\eeq
We can now carry out the integral over $p_z$,
\beq
M_{\tn{osc}}(B) &\approx& \frac{1}{(2\pi)^2\omega_c} \sum_{k=-\infty}^\infty (-1)^k e^{-i\pi/4}\frac{\partial}{\partial \lambda_k} \bigg[\frac{1}{\lambda_k}  \sqrt{\frac{2\pi m^*}{\lambda_k}}\int_{-\infty}^\infty d\ve~ e^{i\lambda_k(\ve+\mu_c)} \frac{dn(\ve)}{d\ve}\bigg],\\
&=& \frac{1}{(2\pi)^2\omega_c} \sum_{k=-\infty}^\infty (-1)^k e^{-i\pi/4}\frac{\partial}{\partial \lambda_k} \bigg[\frac{1}{\lambda_k}  \sqrt{\frac{2\pi m^*}{\lambda_k}} e^{i\lambda_k\mu_c}{\cal{A}}(\lambda_k)\bigg].
\eeq
The amplitude ${\cal{A}}(\lambda_k)$ is defined in Eq. \ref{alambdadef} below. From the explicit expression for $n(\ve)$, it is easy to see that $n'(\ve)$ is even and then ${\cal{A}}(\lambda_k)$ is purely real (i.e. represents an amplitude and does not contribute to the oscillatory phase). In the limit of small fields, the $\lambda_k$ derivative is dominated by the oscillating term and so we get,
\beq
M_{\tn{osc}}(B) \approx \frac{1}{(2\pi)^2} \sum_{k=-\infty}^\infty (-1)^k e^{i\pi/4} \sqrt{\frac{2\pi m^*}{\lambda_k^3}}\bigg(\frac{\mu_c}{\omega_c}\bigg) e^{i\lambda_k\mu_c}{\cal{A}}(\lambda_k),
\eeq
which is the quoted form of the result in Eq.~(\ref{mosc}).

We can simplify the expression for the amplitude, which is expressed as an integral over two variables as follows:
\beq
{\cal{A}}(\lambda_k) = \int_{-\infty}^\infty d\ve~ e^{i\lambda_k\ve} \frac{dn(\ve)}{d\ve} &=& -\frac{1}{\pi}\tn{Im}\int_{-\infty}^\infty d\ve \cos(\lambda_k\ve) \frac{d}{d\ve} \int_{-\infty}^\infty d\Omega \frac{f(\Omega)}{\Omega-\ve-\Sigma_{cf}^R-i\Sigma_{cf}^I}.
\label{alambdadef}
\eeq
Let us now carry out the integral over $\ve$, by making the simplification
\beq
{\cal{A}}(\lambda_k)&=& -\frac{1}{\pi}\tn{Im}\int_{-\infty}^\infty d\Omega \frac{d}{d\Sigma_{cf}^R}\int_{-\infty}^\infty d\ve \cos(\lambda_k\ve) \frac{f(\Omega)}{\Omega-\ve-\Sigma_{cf}^R-i\Sigma_{cf}^I},\\
&=& \int_{-\infty}^\infty d\Omega ~f(\Omega)~e^{-|\lambda_k\Sigma^I_{cf}|} \frac{d}{d\Sigma^R_{cf}} \cos(\lambda_k(\Omega - \Sigma_{cf}^R))  =  \lambda_k \int_{-\infty}^\infty d\Omega ~f(\Omega)~e^{-|\lambda_k\Sigma^I_{cf}|} \sin(\lambda_k(\Omega - \Sigma_{cf}^R)). \nonumber\\
\label{alambda}
\eeq
%The above integral can be evaluated numerically. 
We can extract the universal scaling structure of the above integral from the following simple arguments. For the NFL model, the spectral function
\beq
A(\ve,\Omega) &=& \frac{1}{|\Omega|^{4\Delta(q)}}~ S\bigg(\frac{\Omega}{\ve^{1/4\Delta(q)}} \bigg),\\
\frac{dn(\ve)}{d\ve} &=& |\ve|^{\frac{1}{4\Delta(q)}-2}~ \tilde{S}\bigg(\frac{\ve}{T^{4\Delta(q)}} \bigg),
\eeq
where $S(...)$ and $\tilde{S}(...)$ are universal scaling functions. The amplitude then has a scaling form, as in Eq.~(\ref{ampscal}).

\section{Many-body quantum chaos}
\label{chaos}
This appendix serves as a self-contained  resource for some of the key aspects of many-body quantum chaos. We use these ideas to formulate  our conjectures for a universal description of non-Fermi liquid metals in Section \ref{conj} above. It has been proposed in recent years that the spread of information (or information {\it scrambling} \cite{HaydenPreskill}) can be diagnosed by studying special correlation functions which involve squared (anti-)commutators of local operators \cite{sekino,ShenkerStanford2014,kitaev_talk}. Such correlators were considered decades ago in a different context \cite{LO69} and have been employed more recently in a variety of different settings. They have been shown to diagnose quantum chaos in
black hole physics \cite{ShenkerStanford2014,Shenker2014,kitaev_talk,kitaevsuh}, which are supposed to be the {\it fastest} scramblers in nature \cite{MSS15}.

The squared anti-commutators for local Fermionic operators can be  defined as,
\beq
\C(t,\r) = \frac{1}{N^2}\sum_{i,j}\tn{Tr}\bigg[\rho~ \{c_{\x,i}(t),c_{\x',j}^\dagger(0) \} \{c_{\x,i}(t),c_{\x',j}^\dagger(0) \}^\dagger \bigg],
\eeq
where $\rho = e^{-\beta H}$ is the density matrix at a temperature $T=\beta^{-1}$ and $c_{\x,i}(t) = e^{iHt} c_{\x,i} e^{-iHt}$; $\r=\x-\x'$. For spatially well separated operators, these (anti-)commutators start out small and then grow at late times. For generic non-integrable systems and in systems with a large number of local degrees of freedom, the growth is expected to be of the form, $\C(t,\r) \sim \epsilon ~e^{\lambda_L t}$, where $\epsilon$ in general depends on $t,~\r$ and on the number of degrees of freedom in the system and the growth rate is denoted the `Lyapunov exponent' ($\lambda_L$). There is a fundamental limit on how large $\lambda_L(\leq 2\pi k_BT/\hbar)$ can be and black-holes are known to saturate the bound. An interesting feature of the $(0+1)-$dimensional SYK model in the large $N$ limit \cite{kitaev_talk,Polchinski16,Maldacena_syk,kitaevsuh}, as well as its higher-dimensional generalizations that preserve the SYK form of the interactions \cite{Gu17,SS17}, is that they are also maximally chaotic with $\lambda_L = 2\pi k_BT/\hbar$. Such correlation functions have been computed recently for a variety of field-theoretic problems (a number of which rely on some form of large $N$ expansion) \cite{aleiner16,DCon,SScfs,Stanford2016,EBelph} and none of these models show signs of being maximally chaotic.  

For operators that are separated spatially by an amount $\r$ as above, the spatial structure of $\C(t,\r)$ also contains valuable information about the spreading of chaos and entanglement in the system. In holographic calculations \cite{RSS15,StanfordStringy,ShenkerStanford2014} and in calculations involving a chain of coupled disordered SYK islands \cite{Gu17}, the exponential growth in time is accompanied by a spatial structure of the form $\C(t,\r) \sim e^{\lambda_L(t-|\r|/v_B)}$, leading to a ballistic growth of chaos, where $v_B$ is known as the `butterfly-velocity'. This form is reminiscent of a Lieb-Robinson type bound \cite{LiebRobinson} and $v_B$ has been argued \cite{BSLR} to serve as a state-dependent, low-energy avatar of the Lieb-Robinson velocity. Other models display a different form for the spatial structure of $\C(t,\r)$ \cite{BSDC17,DCon}.

We leave a detailed discussion of the results related to chaos in the models discussed in this paper for the future. However, we can make a few general observations for the single-band model below. The Lyapunov exponent has a scaling form,
\beq
\lambda_L &=& T~{\cal{L}}\bigg(\frac{T}{\Omega_c^*}\bigg),
\eeq
where ${\cal{L}}(y)$ is a function that describes the crossover of the Lyapunov exponent from the low-temperature Fermi liquid to the high temperature incoherent phase and is of the form,
\beq
{\cal{L}}(y) &\sim& y,~~y\ll1,\\
{\cal{L}}(y) &=& \tn{constant},~~y\gg1.
\eeq
In the low-temperature FL regime, it is clear that $\lambda_L\sim T^2/\Omega_c^*$. On the other hand, at high temperatures in the locally incoherent regime, $\lambda_L\sim T$. Furthermore, for operators defined on the same site, the constant above is simply $2\pi k_B/\hbar$, reflecting the maximally chaotic nature of each $(0+1)-$dimensional SYK island. 

\bibliographystyle{apsrev4-1_custom}

\bibliography{nfl}

\end{document}